\documentclass[journal=gmj]{CUP-JNL-DTM}%

\usepackage{graphicx}
\usepackage{multicol,multirow}
\usepackage{amsmath,amssymb,amsfonts}
\usepackage{mathrsfs}
\usepackage{amsthm}
\usepackage{rotating}
\usepackage{appendix}
\usepackage{ifpdf}
\usepackage[T1]{fontenc}
\usepackage{newtxtext}
\usepackage{newtxmath}
\usepackage{textcomp}
\usepackage{xcolor}
\usepackage{lipsum}
\usepackage[colorlinks,allcolors=blue]{hyperref}

\usepackage{subfig}
\usepackage{tikz}

\theoremstyle{definition}

\numberwithin{equation}{section}

\newenvironment{myitemize}
{\begin{list}{$\bullet$}{%
\setlength{\itemsep}{0pt}%
\setlength{\topsep}{-5pt}%
\setlength{\partopsep}{0pt}%
\setlength{\parsep}{7pt}%
\setlength{\labelwidth}{6pt}%
\setlength{\labelsep}{7pt}%
\setlength{\itemindent}{0pt}%
\setlength{\leftmargin}{15pt}}}%
{\end{list}}

\newcommand{\sfrac}[2]{{\textstyle\frac{#1}{#2}}}

\jname{Journal of Nonlinear Waves}
\articletype{RESEARCH ARTICLE}
\jyear{2026}
\begin{document}

\begin{Frontmatter}

\title[Article Title]{Small-time asymptotics and the emergence of complex \\ singularities for the KdV equation}

% There is no need to include ORCID IDs in your .pdf; this information is captured by the submission portal when a manuscript is submitted.
\author[1]{Scott~W. McCue}
\author[2]{Christopher~J. Lustri}
\author[3]{Daniel J. VandenHeuvel}
\author[2]{Jocelyn Zhang}
\author[4]{John~R. King}
\author[5]{S.~Jonathan Chapman}

\authormark{Scott W. McCue \textit{et al}.}

\address[1]{School of Mathematical Sciences, Queensland University of Technology, Brisbane QLD 4001, Australia}
\address[2]{School of Mathematics and Statistics, The University of Sydney, Sydney, New South Wales 2006, Australia}
\address[3]{Department of Mathematics, Imperial College London, London SW7 2AZ, UK}
\address[4]{School of Mathematical Sciences, University of Nottingham, Nottingham
NG7 2RD, UK}
\address[5]{Mathematical Institute, University of Oxford, Oxford OX2 6GG, UK
}

\keywords{KdV equation, complex-plane singularities, exponential asymptotics, matched asymptotic expansions, Stokes phenomenon, dispersive waves, solitary waves, Painlev\'{e} II, tritronqu\'{e}e solutions, transseries}

%\keywords[MSC Codes]{\codes[Primary]{CODE1}; \codes[Secondary]{CODE2, CODE3}}

\abstract{While real-valued solutions of the Korteweg--de Vries (KdV) equation have been studied extensively over the past 50 years,  much less attention has been devoted to solution behaviour in the complex plane.  Here we consider the analytic continuation of real solutions of KdV and investigate the role that complex-plane singularities play in early-time solutions on the real line.  We apply techniques of exponential asymptotics to derive the small-time behaviour for dispersive waves that propagate in one direction, and demonstrate how the amplitude, wavelength and speed of these waves depend on the strength and location of double-pole singularities of the initial condition in the complex plane.  Using matched asymptotic expansions in the limit $t\rightarrow 0^+$, we show how complex singularities of the time-dependent solution of the KdV equation emerge from these double-pole singularities.  Generically, their speed as they move from their initial position is of $\mathcal{O}(t^{-2/3})$, while the direction in which these singularities propagate initially is dictated by a Painlev\'{e} II (P$_{\mathrm{II}}$) problem with decreasing tritronqu\'{e}e solutions.  The well-known $N$-soliton solutions of KdV correspond to rational solutions of P$_{\mathrm{II}}$ with a finite number of singularities; otherwise, we postulate that infinitely many complex-plane singularities of KdV solutions are born at each double-pole singularity of the initial condition.  We also provide asymptotic results for some non-generic cases in which singularities propagate more slowly than in the generic case.  Our study makes progress towards the goal of providing a complete description of KdV solutions in the complex plane and, in turn, of relating this behaviour to the solution on the real line.}

\end{Frontmatter}

\section{Introduction} \label{intro}

The Korteweg--de Vries (KdV) equation is an exceptionally well-studied third-order nonlinear dispersive equation that is used to model weakly nonlinear water waves and other wave motion in physics \cite{KdV1895,Miles1981,Zabusky1965}, as well being a prototype model for a competition between nonlinear advection and linear dispersion.
Through inverse-scattering and related techniques \cite{Gardner1967}, much is known about this integrable model, including exact descriptions for interacting solitons \cite{Hirota1971}.
Alternatively, asymptotic methods provide detailed descriptions of long-time behaviour \cite{Ablowitz1977,Grunert2009} or the limit of vanishing dispersion \cite{Claeys2010,Deng2016},
and, of course, the form of solitary and dispersive waves that arise from the KdV model can be explored via numerical computation \cite{Grava2007,Trogdon2012}.

A powerful approach for studying the KdV equation on the real line, namely
\begin{equation}
u_t+6uu_x+u_{xxx}=0,
\quad u(x,0)=u_0(x),\quad x\in\mathbb R,
\label{eq:kdv}
\end{equation}
is to treat the analytic continuation of real solutions of (\ref{eq:kdv}) to the complex-$x$ plane, in part because the complex singularities of the solution are intimately related to the propagation of real-valued dispersive waves and solitons.  In this spirit, Kruskal \cite{Kruskal1974}, Thickstun \cite{Thickstun1976} and Bona \& Weissler \cite{Bona2009} study the role of complex-plane singularities of solutions in soliton interactions, while Airault et al. \cite{Airault1977} and Deconinck \& Segur \cite{Deconinck2000} conduct related studies for elliptic solutions of the KdV equation.  Other more recent research focuses on applying numerical methods to track singularities of solutions of the KdV equation, especially for periodic initial conditions or small dispersion \cite{Bona2023,Caflisch2015,Gargano2016,Weideman2022}.

Following these studies, we are also interested in the evolution of complex-variable singularities of (the analytic continuation of) solutions to (\ref{eq:kdv}), in our case with a focus on the limit $t\rightarrow 0^+$.  The small-time limit is important because singularities cannot be born at finite times; therefore, the initial number, type and motion of singularities can provide a solid indication of their behaviour for later times.  Apart from being of interest in their own right, our study of singularity behaviour in the complex plane, especially those singularities closest to the real axis, has facilitated new results for the real-valued problem, including a short-time asymptotic description of the dispersive waves.

It is worth acknowledging there is a significant body of research concerned with using asymptotic methods and numerical computation to characterise and track complex singularities of solutions of partial differential equations (pdes), including for Burgers' equation or a semilinear heat equation
\cite{Baker1996,Bessis1984,Chapman2007,Deconinck2007,Fasondini2023,Senouf1997,VandenHeuvel2023,Weideman2003,Weideman2022},
various problems in fluid mechanics \cite{Costin2008,Cowley1999,Gargano2014,Matsumoto2008,Siegel2009,Tanveer1993a,Tanveer1993b} and third-order pdes~\cite{Costin2000,Costin2006}.  A relevant summary of many of the key issues is provided by Costin \& Tanveer~\cite{Costin2004}, for example.  For many of the studies in this extensive list, there is a focus on small-time behaviour of singularities since, as we mention above, the singularity structure for early times can shed light on the subsequent behaviour (including the number of singularities, their type and their initial trajectories).  Our study follows the same strategy.

Returning to our KdV problem, we shall be studying the analytic continuation of solutions of (\ref{eq:kdv}).  These complex-valued solutions will have singularities in the complex-$x$ plane that are all double poles with principal part $-2$.  For each such complex singularity $x=s(t)$, the local behaviour must be
\begin{equation}
u\sim -\frac{2}{(x-s(t))^2}\quad\mbox{as}\quad x\rightarrow s(t),
\label{eq:doublepoleleadingorder}
\end{equation}
being associated with the balance $u_{zzz}\sim -6uu_z$, where $z\equiv x-s(t)$.
With this in mind, we shall focus on initial conditions that are analytic functions and also have double poles off the real axis, that is, initial conditions with
\begin{equation}
u_0\sim \frac{A_0}{(x-x_0)^2} \quad\mbox{as}\quad x\rightarrow x_0,
\label{eq:ICA}
\end{equation}
where $A_0$ may be complex (although in this paper the examples we include have real values of $A_0$).  Here, each $x_0\notin \mathbb{R}$ will be one of a complex-conjugate pair, since we want $u_0$ to be real for $x\in\mathbb R$.  We shall address the broad question of how solutions with double poles like (\ref{eq:doublepoleleadingorder}) emerge from initial conditions with (\ref{eq:ICA}), especially in terms of the small-time dynamics.  The analytical tools we employ are based on formal asymptotics, including exponential asymptotics (using the approach in \cite{chapman1998,OldeDaalhuis1995}) and matched asymptotic expansions in the limit $t\rightarrow 0^+$ (cf.~\cite{Cowley1999,Fasondini2024,VandenHeuvel2023} and others), analyses of the Stokes phenomenon and transseries to locate poles of the inner problem near $x_0$ (like \cite{Costin2001,Lustri2023}), while the computational tools include direct numerical solution of the real problem, and a pole-solver algorithm to analyse the inner problem numerically \cite{Fornberg2011,Fornberg2014}.  Note that analytic initial conditions for (\ref{eq:kdv}) guarantee solutions that are analytic for $t>0$ \cite{Grujic2002,Hayashi1991}.

\begin{figure}
\centering
\subfloat[$A_0 = -1/4$]{
\includegraphics[width=0.47\textwidth]{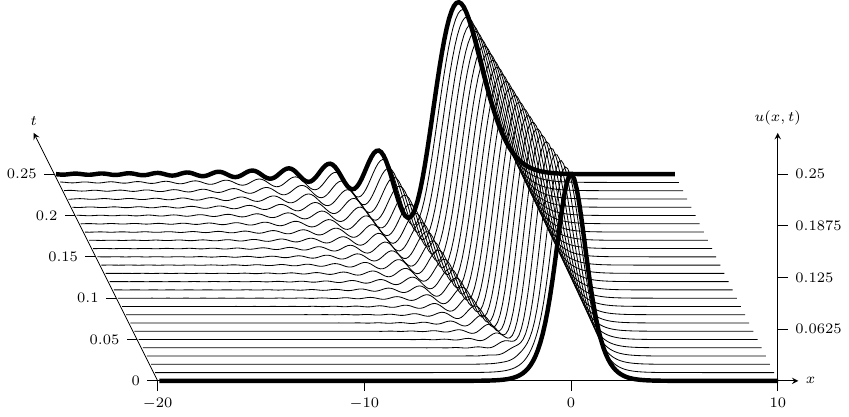}
}
\subfloat[$A_0 = -1$]{
\includegraphics[width=0.47\textwidth]{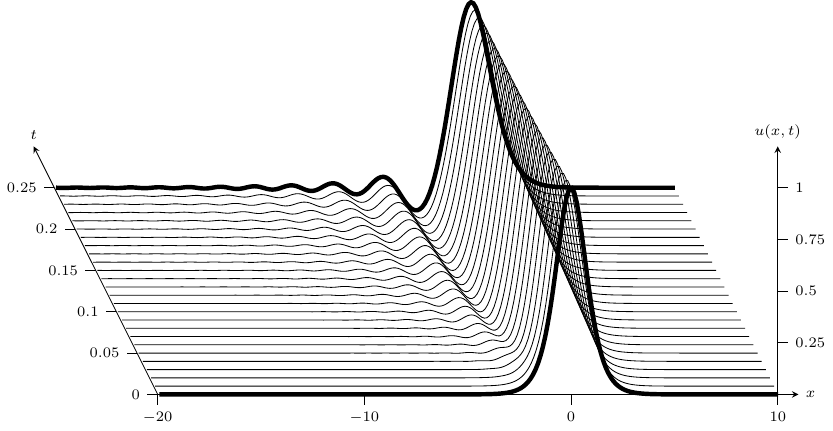}
}

\subfloat[$A_0 = -4$]{
\includegraphics[width=0.47\textwidth]{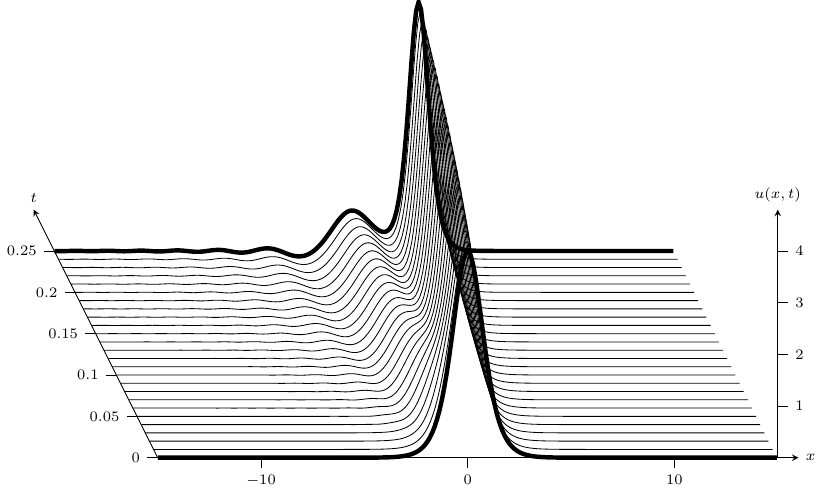}
}
\subfloat[$A_0 = -8$]{
\includegraphics[width=0.47\textwidth]{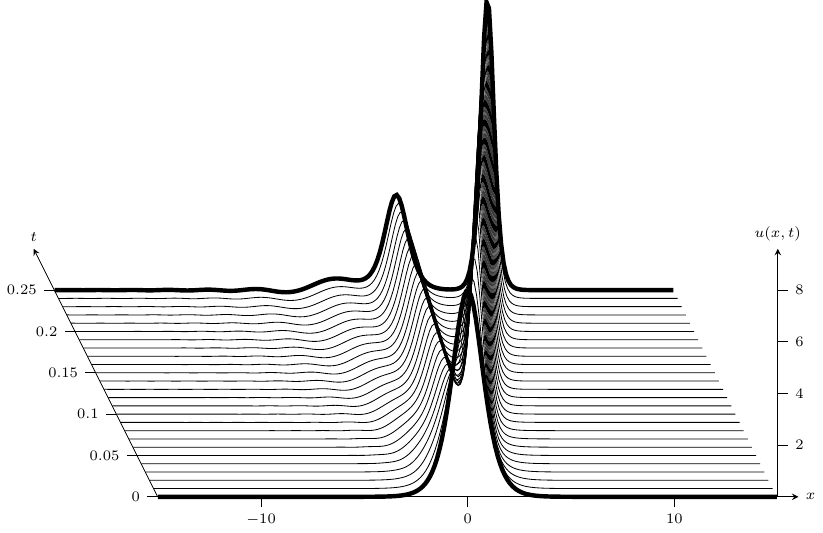}
}

\caption{Numerical solutions of the KdV model (\ref{eq:kdv}) on the real line using the $\mathrm{sech}^2$-type initial condition (\ref{eq:ICsech}), showing dispersive waves propagating to the left.  Calculations are performed using
the spin command \cite{Montanelli2020} in Chebfun \cite{Driscoll2014}.
For (a),(b) since $0< -A_0 <2$, there are no solitons moving to the right; (c) since $2<-A_0<6$, there is one soliton; (d) since $6<-A_0<12$, there are two solitons.  Note, although hard to see on this scale, the dispersive waves are roughly the same size for each of these four examples, while the height of the initial hump $u_0(0)=-A_0$ of course increases as $-A_0$ increases.}
\label{fig:waterfallplots}
\end{figure}

One reason for starting with analytic initial conditions with (\ref{eq:ICA}) is that it turns out the matched asymptotic expansions for the limit $t\rightarrow 0^+$ near each $x_0$ are much easier to deal with than if the singularity of $u_0$ is a different type (i.e., other than a double pole).  Another reason for focussing on initial conditions with (\ref{eq:ICA}) is to observe the difference between having initial double poles with $A_0=-2$ and the more general case $A_0\neq -2$.  Finally, the very well-studied $\mathrm{sech}^2$-type initial conditions have double poles as in (\ref{eq:ICA}), namely
\begin{equation}
u_0=-A_0\,\mathrm{sech}^2x, \quad x_0=\left(n+\sfrac{1}{2}\right)\pi\mathrm{i}, \quad A_0\in\mathbb R, \quad n\in\mathbb Z,
\label{eq:ICsech}
\end{equation}
providing further motivation for (\ref{eq:ICA}).  Properties of solutions of (\ref{eq:kdv}) with (\ref{eq:ICsech}) are well known, mostly via the application of inverse-scattering techniques \cite{Ablowitz2011,Drazin1989}.
For $A_0<0$, the profile $u_0$ in (\ref{eq:ICsech}) on the real line is a ``hump'' centred at $x=0$, and the subsequent solution of (\ref{eq:kdv}) will in general involve a finite number of solitons propagating to the right, together with dispersive radiation characterised by waves propagating very quickly to the left, as illustrated in figure~\ref{fig:waterfallplots}.  For the special cases $A_0=-N(N+1)$, $N\in\mathbb N$, there are exact $N$-soliton solutions of (\ref{eq:kdv}) with (\ref{eq:ICsech}), which will be of interest as test cases.  The 1-soliton solution, $N=1$ ($A_0=-2$), is an exceptional case that falls outside our analysis for $A_0\neq -2$, as we shall explain.  On the other hand, for $A_0>0$ the shape is a ``dip'', with solutions of (\ref{eq:kdv}) involving dispersive waves only.  One of the goals of our study is to attempt to identify links between complex-plane singularities of solutions and the qualitative behaviour of the solutions on the real line.

The outline of our paper is as follows.  Various aspects of the small-time analysis for the generic case (\ref{eq:kdv}) with (\ref{eq:ICA}) and $A_0\neq -2$ are provided in section~\ref{sec:A0neqminus2}.  The starting point in subsection~\ref{sec:outerexpansion} is an ``outer'' expansion in powers of $t$, which is not able to describe the dispersive waves on the real line, as their amplitudes are exponentially small in $t$ and hence are formally smaller than each term $t^nu_n(x)$ in the algebraic series.  This is reflected in the outer expansion being divergent, with the leading-order term $u_0(x)$ having singularities at $x=x_0$ off the real axis (of the form (\ref{eq:ICA})).  Thus, we require techniques in exponential asymptotics and the Stokes phenomenon to extract details of the dispersive waves in the small-time limit, with our analysis in subsection~\ref{sec:expasympt_main} predicting that the dispersive waves behave, to leading order, as
\begin{equation}
U_{\mbox{\footnotesize dis}}\sim
-\frac{2}{3^{3/4}\pi^{1/2}}
\cos\left(\frac{\pi}{2}\sqrt{1-4A_0}\right)
\frac{(-x)^{1/4}\,\mathrm{e}^{-y_0(-x/3t)^{1/2}}}{t^{3/4}}
\,\cos\left(\frac{2(-x)^{3/2}}{3(3t)^{1/2}}-\frac{\pi}{4}
\right)
\label{eq:farfielddispersivewaves}
\end{equation}
as $x\rightarrow -\infty$, $t\rightarrow 0^+$, where $y_0=\mathrm{Im}(x_0)$
(further details are provided in appendix~\ref{sec:expasympt_app}).  It is noteworthy that the cosine term out the front of the expression in (\ref{eq:farfielddispersivewaves}) has an explicit dependence on $A_0$, linking the amplitude of the dispersive waves to the strength of the double pole at $x=x_0$, with this term vanishing for the special cases $A_0=-N(N+1)$, where $N$ is a natural number.  We test our exponential asymptotics in subsection~\ref{sec:numericalexpo} by comparing with numerical solutions of (\ref{eq:kdv}) for the initial conditions (\ref{eq:ICsech}) and
\begin{equation}
u_0=-\frac{4A_0}{(1+x^2)^2}, \quad x_0=\pm\mathrm{i}, \quad A_0\in\mathbb R,
\label{eq:ICxsquared}
\end{equation}
(\ref{eq:ICxsquared}) being worth including in part because it is not of the special class (\ref{eq:ICsech}) and so therefore does not give rise to known exact solutions.  Further, (\ref{eq:ICxsquared}) is instructive because this initial condition has only one singularity, $x_0=\mathrm{i}$, in the upper half plane, so we avoid any possible complications from singularities born at multiple points in each half plane.  An additional, carefully constructed, initial condition with a vanishing residue (namely (\ref{eq:newIC})) is also used as a comparison.

We continue our study of the small-time limit in section~\ref{sec:PII}, where we consider an ``inner'' problem near $x=x_0$ in subsection~\ref{sec:innernearx0}.  We show how the initial dynamics of the singularities of our KdV problem are governed by the inhomogeneous Painlev\'{e} II (P$_{\mathrm{II}}$) equation (subsection~\ref{sec:reductionPII})
with decreasing tritronqu\'{e}e solutions (subsection~\ref{sec:WKB}).  In general, there are infinitely many singularities that emerge from each $x=x_0$; exceptional cases arise for $A_0=-N(N+1)$, whereby well-known rational solutions of P$_{\mathrm{II}}$ (subsection~\ref{sec:exactsolns}) correspond to a finite number of singularities for KdV.  (A summary of the effects of higher-order corrections to (\ref{eq:ICA}) is provided in appendix~\ref{sec:higherorder}, including some special cases.)
To illustrate the singularity structure of our P$_{\mathrm{II}}$ solutions, some numerical results are presented in subsection~\ref{sec:numericalPII}.
To close section~\ref{sec:PII}, we use a standard transseries approach in subsection~\ref{sec:trans} to derive approximations to the locations of the most important of these singularities.

As concrete examples of KdV solutions, the widely-studied initial conditions (\ref{eq:ICsech}) are used in section~\ref{sec:sechsquaredIC} to illustrate some of the global features of our small-time analysis from section~\ref{sec:A0neqminus2}, focussing on the role of rational solutions of P$_{\mathrm{II}}$ in the small-time dynamics of $N$-soliton solutions.  It is worth emphasising that, while the integrability of the KdV equation makes some of the calculations in various parts of our study more analytically amenable, our methodology should be broadly applicable to other dispersive wave equations.  We allude to these observations in section~\ref{sec:discussion}, where we also summarise our findings more generally and present a discussion about a significant number of unresolved issues and open problems.

\section{Small-time analysis for (\ref{eq:kdv}) with (\ref{eq:ICA})}\label{sec:A0neqminus2}

As mentioned in Section~\ref{intro}, one of our main motivations is to understand the role of complex-plane singularities of KdV solutions, which for $t>0$ must be double poles of the form (\ref{eq:doublepoleleadingorder}).  In this section and in section~\ref{sec:PII}, we concentrate on analytic initial conditions that also have double poles, as in (\ref{eq:ICA}), but with a different principal part to  (\ref{eq:doublepoleleadingorder}), namely $A_0\neq -2$.  The methodology we employ is based in part on matched asymptotic expansions in the limit $t\rightarrow 0^+$, including an outer region away from complex singularities of the initial condition $u_0(x)$ and inner regions near these singularities, as well as exponential asymptotics that involves analysis from both the outer and inner regions.  The present section summarises the exponential asymptotics, while a more detailed study of the inner regions is deferred until section~\ref{sec:PII}.

\subsection{Outer expansion away from $x=x_0$}\label{sec:outerexpansion}

Consider the KdV equation (\ref{eq:kdv}) together with an initial condition $u_0$ with the property (\ref{eq:ICA}), where $x_0$ (with
$\mathrm{Im}(x_0)=y_0>0$) is one of a complex-conjugate pair.  Assuming that $u_0$ is real for $x\in\mathbb R$, then a consequence of this problem formulation is that we need consider the solution only on the real line and the upper half plane, with the understanding that the behaviour of the singularities in the lower half plane will be an appropriate reflection about the real-$x$ axis.

To begin, we consider a straightforward power series expansion in time, the first two terms of which give
\begin{equation}
u\sim u_0(x)+tu_1(x)
\quad\mbox{as}\quad t\rightarrow 0^+.
\label{eq:outer}
\end{equation}
By substituting into (\ref{eq:kdv}), we find
\begin{equation}
u_1=-6u_0u_0'-u_0''',
\label{eq:u1}
\end{equation}
where the primes (here and throughout the document) denote differentiation with respect to $x$.  Therefore, we have the local behaviour
\begin{equation}
u_0\sim \frac{A_0}{(x-x_0)^2},
\quad
u_1\sim \frac{12A_0(A_0+2)}{(x-x_0)^5}
\quad\mbox{as}\quad x\rightarrow x_0,
\label{eq:local}
\end{equation}
noting that $u_1$ is three orders more singular than $u_0$ at $x=x_0$ (due to the third derivative in (\ref{eq:u1})).  Thus, we see immediately that the outer expansion (\ref{eq:outer}) applies along the real axis and in parts of the upper half complex plane away from $x=x_0$, but breaks down in regions where $u_0=\mathcal{O}(tu_1)$, or, in other words, where
$$
x-x_0=\mathcal{O}(t^{1/3}).
$$
This reasoning suggests there will be an inner problem near $x=x_0$, which we consider in subsection~\ref{sec:innernearx0}.  (Note that the power-series expansion will also break down in a sector in the upper half plane bounded by anti-Stokes lines, as discussed below in subsections~\ref{sec:expasympt_main} and \ref{sec:WKB}.)

\subsection{Exponential asymptotics argument for limit $t\rightarrow 0^+$}\label{sec:expasympt_main}

The terms $u_0$ and $u_1$ in (\ref{eq:outer}) are the first two in a divergent asymptotic expansion of the form
\begin{equation}
u\sim \sum_{n=0}^\infty t^n u_n(x)
\quad\mbox{as}\quad t\rightarrow 0^+,
\label{eq:powerseries}
\end{equation}
whose divergence is caused by the singularities of the leading-order term $u_0(x)$ in the complex-$x$ plane.  As such, the $u_n$ will be of the familiar factorial-over-power form for large $n$, due to repeatedly differentiating (three times) to obtain the next-order term.  Thus, we expect there to be an exponential term in our asymptotic representation for $u$ in the limit $t\rightarrow 0^+$ that appears ``beyond all orders'' of the original power series (\ref{eq:powerseries}) \cite{chapman1998,OldeDaalhuis1995}.  This term will ``switch on'' across Stokes lines in the $x$ plane and, in particular, will affect the solution on the real line by providing an exponentially small correction term to (\ref{eq:powerseries}), which we call $U_{\mbox{\footnotesize dis}}$.

Importantly, the emergence of dispersive waves that travel in the negative-$x$ direction can never be described by a power series (\ref{eq:powerseries}), as the amplitude of these waves turns out to be exponentially small in time compared to each term $t^n u_n$.  Thus, in order to approximate the dispersive wavetrain in the limit $t\rightarrow 0^+$, we must consider this exponential contribution and observe where it switches on across the real-$x$ axis.
To demonstrate how this works, we follow the framework in \cite{chapman1998}, summarise the main results here, and provide further details in appendix~\ref{sec:expasympt_app} (see \cite{Chapman2002,Lustri2013a,Lustri2012,Lustri2013b,Trinh2011}, for example, for other studies of wave motion using this framework of exponential asymptotics).

A crucial step is to analyse the late terms in (\ref{eq:powerseries}), which satisfy
\begin{equation}
(n+1)u_{n+1}=-6\sum_{j=0}^n u_j u'_{n-j}-u_n'''.
\label{eq:recurrence}
\end{equation}
Due to the factorial over power divergence of (\ref{eq:powerseries}), following Dingle~\cite{Dingle1973} we apply the ansatz
\begin{equation}
u_n\sim \frac{\mathcal{A}(x)\Gamma(2n+\gamma)}{\chi(x)^{2n+\gamma}}
\quad \mbox{as}\quad n\rightarrow\infty,
\label{eq:dingleansatz}
\end{equation}
which, after substituting into (\ref{eq:recurrence}), leads to
\begin{equation}
\frac{1}{2}\chi=\left(\chi'\right)^3,
\quad
\frac{\gamma}{2}\mathcal{A}
=3\mathcal{A}\chi'\chi''
+3\mathcal{A}'\left(\chi'\right)^2.
\label{eq:chiG}
\end{equation}
Here $\chi$ is the so-called singulant, which must vanish at singularities of $u_0$, and it suffices to consider the location $x=x_0$ (the full expansion of $u_n$ for large $n$ comprises a sum of terms of the form (\ref{eq:dingleansatz}) associated with each singularity $x_0$).  Thus, given $x_0$ is assumed to lie in the upper half plane, we explain in appendix \ref{sec:expasympt_lateorder} that the appropriate solutions are
\begin{equation}
\chi=-\frac{2}{3^{3/2}}(x-x_0)^{3/2},
\quad \mathcal{A}=\Lambda(x-x_0)^{(\gamma-1)/2},
\label{eq:chiandG}
\end{equation}
where $\Lambda$ is a constant, and therefore
\begin{equation}
u_n\sim \frac{\Lambda(-3^{3/2}/2)^{2n+\gamma}
\Gamma(2n+\gamma)}
{(x-x_0)^{3n+\gamma+1/2}}
\quad \mbox{as}\quad n\rightarrow\infty.
\label{eq:lateorder}
\end{equation}
For the initial conditions we are concerned with in this study, we have (\ref{eq:ICA}).  To be consistent with this local behaviour near $x=x_0$, we must choose $\gamma=3/2$.

With $\chi$ and $\mathcal{A}$ determined, we show in appendix~\ref{sec:expasympt_switch} that a consequence of this type of late-order behaviour for $u_n$ is that the exponentially-small quantity
\begin{equation}
\pi\mathrm{i}\,\mathcal{A}\, t^{-\gamma/2}\mathrm{e}^{-\chi/t^{1/2}}
=\pi\mathrm{i}\,\Lambda(x-x_0)^{1/4}t^{-3/4}
\mathrm{e}^{2(x-x_0)^{3/2}/3(3t)^{1/2}}
\label{eq:expsmall}
\end{equation}
switches on as we cross the Stokes line $\mathrm{arg}(x-x_0)=-2\pi/3$ from right to left (given that $x_0$ is in the upper half $x$ plane).  This Stokes line is found by setting the singulant term $\chi$ to be real and positive.  There will be another term like (\ref{eq:expsmall}) switched on across a Stokes line born in the lower half plane, and together the sum of these provides the asymptotic behaviour of the dispersive waves on the real line in the small-time limit.

To determine the constant $\Lambda$ in (\ref{eq:expsmall}), we need to match into an inner region near $x=x_0$, which we study below in subsection~\ref{sec:innernearx0}.  The details for this matching are provided in appendix~\ref{sec:expasympt_match}, with the key result that
\begin{equation}
\Lambda=
\frac{\mathrm{i}}{3^{3/4}\pi^{3/2}}
\cos\left(\frac{\pi}{2}\sqrt{1-4A_0}\right).
\label{eq:Lambda}
\end{equation}
A striking property of $\Lambda$ is that it vanishes for $A_0=-N(N+1)$, where $N\in \mathbb{N}$.   Therefore, this leading-order result predicts that a
train of dispersive waves is associated with each complex conjugate pair of singularities of the initial condition $u_0$, with the amplitude vanishing for $A_0=-N(N+1)$.  This prediction is consistent with the $\mathrm{sech}^2$-type initial condition (\ref{eq:ICsech}), for which it is known that there are no dispersive waves when $A_0$ takes on these triangular values (these are the $N$-soliton solutions, having flat tails).  (More generally, if $A_0=-N(N+1)$ for an initial condition that is not of the special form (\ref{eq:ICsech}), then parts of the required asymptotics to describe the dispersive waves will be different; we include a discussion on such cases in appendix~\ref{sec:expasympt_special}.)

To take an example, consider an initial condition that has a singularity lying on the imaginary axis, which we write as $x_0=\mathrm{i}y_0$.  Our analysis predicts that the term (\ref{eq:expsmall}) switches on across the Stokes line $\mathrm{arg}(x-\mathrm{i}y_0)=-2\pi/3$ in the upper half $x$ plane.  We would also observe an analogous term that switches on across $\mathrm{arg}(x+\mathrm{i}y_0)=2\pi/3$ in the lower half plane.  These Stokes lines both hit the real axis at $x=-y_0/\sqrt{3}$.
Thus, putting it together, we find the exponential contribution to our asymptotic solution on the real-$x$ axis is of the form
\begin{align}
U_{\mbox{\footnotesize dis}}\sim
&
\,\,\pi\mathrm{i}\Lambda t^{-3/4}\left(
(x-\mathrm{i}y_0)^{1/4}
\mathrm{e}^{2(x-\mathrm{i}y_0)^{3/2}/3(3t)^{1/2}}
+
(x+\mathrm{i}y_0)^{1/4}
\mathrm{e}^{2(x+\mathrm{i}y_0)^{3/2}/3(3t)^{1/2}}
\right)
\nonumber \\
=
&
-(2/3^{3/4}\pi^{1/2})
\cos\left(
\frac{\pi}{2}\sqrt{1-4A_0}
\right)
t^{-3/4}
(x^2+y_0^2)^{1/8}
\mathrm{e}^{
2(x^2+y_0^2)^{3/4}\cos (3\phi/2)/(3(3t)^{1/2})
}
\nonumber \\
& \times
\cos\left(
\frac{2}{3(3t)^{1/2}}
(x^2+y_0^2)^{3/4}
\sin \sfrac{3}{2}\phi +\sfrac{1}{4}\phi
\right),
\label{eq:expdispersive}
\end{align}
$x<-y_0/\sqrt{3}$, $t\rightarrow 0^+$, where $\phi=\arctan(y_0/x)$ (note $\cos (3\phi/2)<0$ for this interval in $x$).
For large negative $x$, noting that
$$
(x-\mathrm{i}y_0)^{3/2}\sim \mathrm{i}(-x)^{3/2}-\sfrac{3}{2}y_0(-x)^{1/2}+
\mathcal{O}((-x)^{-1/2}),
$$
these waves take the form (\ref{eq:farfielddispersivewaves}).
Thus, we see dispersive waves whose amplitude is exponentially small in both limits $x\rightarrow -\infty$ and $t\rightarrow 0^+$
(compared to the initial condition $u_0(x)$, which is assumed to be $\mathcal{O}(1)$), scaling as
$$
\mbox{const}\,(-x)^{1/4}t^{-3/4}\mathrm{e}^{-y_0(-x/t)^{1/2}}.
$$
Defining the locations of the crests of the dispersive waves to be $x_m$, with $m$ increasing from right to left, for large $m$ these propagate to the left as
\begin{equation}
x_m\sim -3\pi^{2/3}m^{2/3}t^{1/3},
\quad m\rightarrow\infty,
\quad t\rightarrow 0^+,
\label{eq:crests}
\end{equation}
so that the speed of the waves decreases algebraically as $\dot{x}_m\sim -\pi^{2/3}(t/m)^{-2/3}$ and the wavelength increases as $-x_{m+1}+x_m\sim 2\pi^{2/3}(t/m)^{1/3}$ as $t$ increases from zero.  These scalings explain why it is so difficult to compute numerical solutions of the KdV model (\ref{eq:kdv}) accurately for small time using elementary techniques (such as finite differences), given the challenge of capturing fast-moving, small-amplitude waves that would immediately reflect off the boundary of any truncated domain.  Such issues can be resolved, for example, via an inverse-scattering formulation, but such an approach is highly nontrivial \cite{Trogdon2012} (and, of course, limited to integrable systems); we instead use the spin (stiff pde integrator) command \cite{Montanelli2020} in Chebfun \cite{Driscoll2014}, which is good enough for our real-valued numerical solutions.

\subsection{Numerical tests for the exponential asymptotics}\label{sec:numericalexpo}

To test (\ref{eq:expdispersive}) against numerics, we consider the small-time expansion
$$
u\sim u_0+tu_1+U_{\mbox{\footnotesize dis}}
\quad\mbox{as}\quad t\rightarrow 0^+,
$$
i.e., we retain only the first two terms of the algebraic expansion (\ref{eq:outer}) but include the leading-order exponential correction (\ref{eq:expdispersive}) (in order to capture the dominant oscillatory contribution).  For the initial condition (\ref{eq:ICsech}), we have
$$
u_1=4A_0(3(A_0+2)-2\,\mathrm{cosh}^2x)\,\mathrm{sinh}\,x
\,\mathrm{sech}^5x
$$
and $y_0=\pi/2$, while for (\ref{eq:ICxsquared}) we have
$$
u_1=\frac{96A_0x(-5x^2+4A_0+3)}{(1+x^2)^5}
$$
and $y_0=1$. We also make use of a scaled version of (\ref{eq:expdispersive}), namely $U_{\mbox{\footnotesize scaled}}=U_{\mbox{\footnotesize dis}}/K$, where
\begin{equation}
K=t^{-3/4}
(x^2+y_0^2)^{1/8}
\mathrm{e}^{
2(x^2+y_0^2)^{3/4}\cos (3\phi/2)/(3(3t)^{1/2})}
\label{eq:defK}
\end{equation}
is chosen so that $U_{\mbox{\footnotesize scaled}}$ does not decay as $x\rightarrow -\infty$, in the comparison with the numerics.

\begin{figure}
\centering
\subfloat[$A_0 = -1/4$]{
\includegraphics[width=0.34\textwidth]{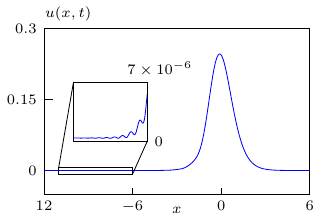}
\includegraphics[width=0.34\textwidth]{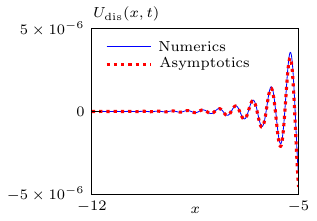}
\includegraphics[width=0.3\textwidth]{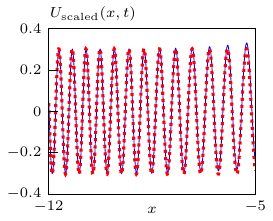}
}

\subfloat[$A_0 = -3/4$]{
\includegraphics[width=0.34\textwidth]{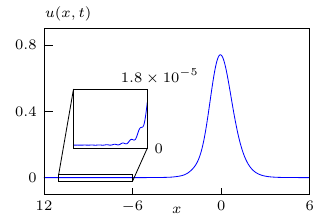}
\includegraphics[width=0.34\textwidth]{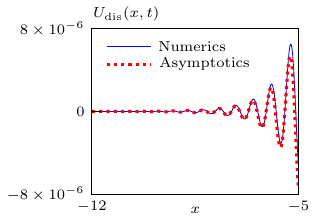}
\includegraphics[width=0.3\textwidth]{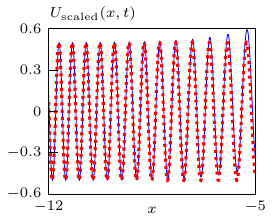}
}
\caption{(a) [left panel] Numerical solution of the KdV model (\ref{eq:kdv}) on the real line, computed at $t=0.02$ using the $\mathrm{sech}^2$-type initial condition (\ref{eq:ICsech}) with $A_0=-1/4$; [middle panel] plots of $u_{\mathrm{num}}-(u_0+tu_1)$ (blue) and $U_{\mbox{\footnotesize dis}}$ (red circles) versus $x$ using same parameters as in left panel; [right panel] plots of $\left(u_{\mathrm{num}}-(u_0+tu_1)\right)/K$ (blue) and $U_{\mbox{\footnotesize scaled}}$, again with same parameters as left panel.
(b) Same as (a), except that $A_0=-3/4$.}
\label{fig:testUdisp_1}
\end{figure}

For example, in figure~\ref{fig:testUdisp_1}(a), we show a numerical solution of (\ref{eq:kdv}), $u_{\mathrm{num}}$, computed with the $\mathrm{sech}^2$-type initial condition (\ref{eq:ICsech}), with $A_0=-1/4$, for a representative early time, $t=0.02$.  The very small dispersive waves we see in the inset on the left panel are propagating to the left.  In the middle panel, we plot $u_{\mathrm{num}}-(u_0+tu_1)$ at the specific time $t=0.02$, and compare with $U_{\mbox{\footnotesize dis}}$ from (\ref{eq:expdispersive}).  We see for these parameter values, the comparison is very good, which is a strong test for both our numerics and asymptotics.  As another test, we plot in the right panel the scaled versions $u_{\mathrm{num}}-(u_0+tu_1)/K$ and $U_{\mbox{\footnotesize scaled}}=U_{\mbox{\footnotesize dis}}/K$, where $K$ is defined in (\ref{eq:defK}).  The agreement is excellent.

\begin{figure}
\centering
\subfloat[initial condition $u_0=-A_0\,\mathrm{sech}^2x$]{
\includegraphics{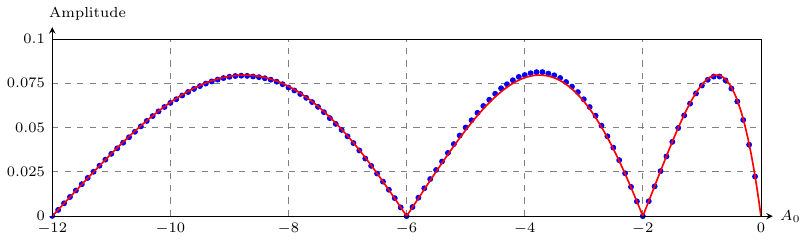}
}

\subfloat[initial condition $u_0=-4A_0/(1+x^2)^2$]{
\includegraphics{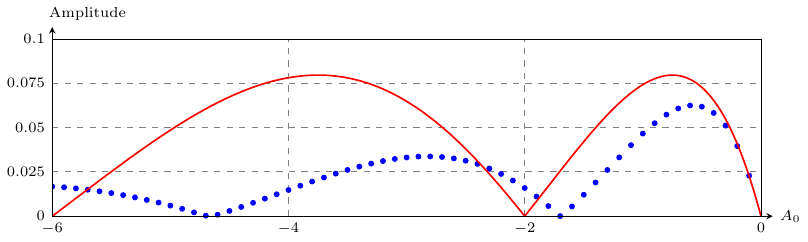}
}

\subfloat[initial condition $u_0=-4A_0/(1+x^2)^2+2A_0/(1+x^2)$]{
\includegraphics{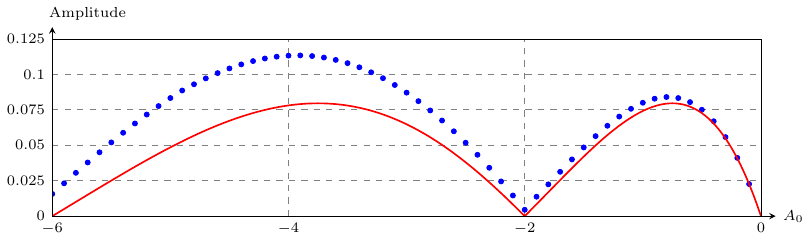}
}

\caption{A numerically computed amplitude of the scaled dispersive waves $\left(u_{\mathrm{num}}-(u_0+tu_1)\right)/K$ (blue dots) plotted for various values of the parameter $A_0$ at a fixed time $t=0.03$, compared with the asymptotic prediction $(2/3^{3/2}\pi^{1/2})\cos\left(
\frac{\pi}{2}\sqrt{1-4A_0}\right)$ (red solid).  (a) the $\mathrm{sech}^2$-type initial condition (\ref{eq:ICsech}); (b) the initial condition (\ref{eq:ICxsquared}); (c) the refined initial condition    (\ref{eq:newIC}).
}
\label{fig:testUdisp_3}
\end{figure}

We have generated a number of other examples like that presented in figure~\ref{fig:testUdisp_1}(a) using the initial condition (\ref{eq:ICsech}), computed for other values of $A_0<0$, and the agreement between numerics and asymptotics is also excellent.  One such example is for $A_0=-3/4$, is shown in figure~\ref{fig:testUdisp_1}(b).  This sweep of parameters includes the special cases $A_0=-N(N+1)$, where $N\in\mathbb N$, for which we have exact $N$-soliton solutions with no dispersive waves at all.  For those special choices, the amplitude of the waves is predicted by (\ref{eq:expdispersive}) to vanish, since
$$
\cos\left(\frac{\pi}{2}\sqrt{1-4A_0}\right)=
\cos\left(\frac{\pi}{2}\sqrt{(2N+1)^2}\right)=
0
$$
for all $N\in\mathbb N$.  Putting it together, we have presented in figure~\ref{fig:testUdisp_3}(a) a comparison of our numerical estimate of the amplitude of the wavelike term $\left(u_{\mathrm{num}}-(u_0+tu_1)\right)/K$ (blue dots) with the asymptotic prediction $(2/3^{3/4}\pi^{1/2})
\cos\left(\frac{\pi}{2}\sqrt{1-4A_0}\right)$ (red solid) for simulations that take the $\mathrm{sech}^2$-type initial condition (\ref{eq:ICsech}).  Given the very good agreement in this plot over a range of values of $A_0$, we are confident the approximation (\ref{eq:expdispersive}) is correctly describing the dispersive waves, at least for the initial condition (\ref{eq:ICsech}), in the small-time limit.

\begin{figure}
\centering
\subfloat[$A_0 = -1/4$]{
\includegraphics[width=0.34\textwidth]{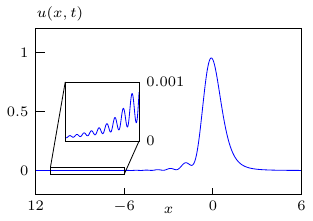}
\includegraphics[width=0.34\textwidth]{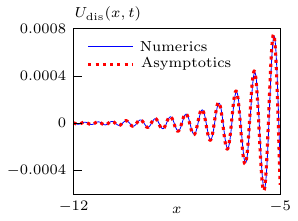}
\includegraphics[width=0.3\textwidth]{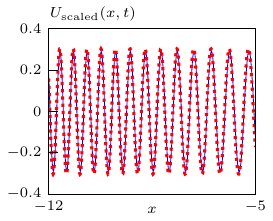}
}

\subfloat[$A_0 = -3/4$]{
\includegraphics[width=0.34\textwidth]{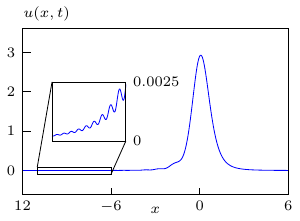}
\includegraphics[width=0.34\textwidth]{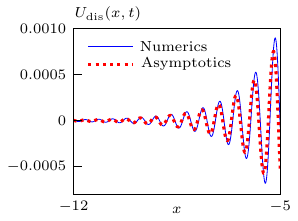}
\includegraphics[width=0.3\textwidth]{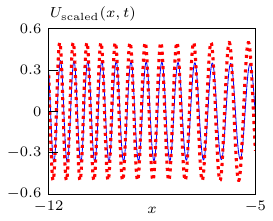}
}

\subfloat[$A_0 = -2$]{
\includegraphics[width=0.34\textwidth]{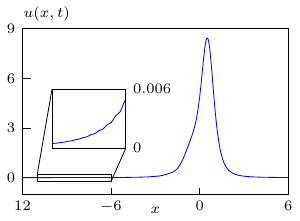}
\includegraphics[width=0.34\textwidth]{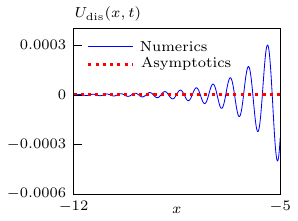}
\includegraphics[width=0.3\textwidth]{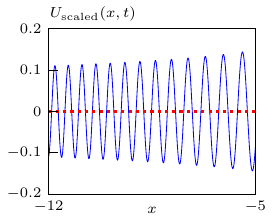}
}
\caption{(a) [left panel] Numerical solution of the KdV model (\ref{eq:kdv}) on the real line, computed at $t=0.02$ using the generic initial condition (\ref{eq:ICxsquared}) with $A_0=-1/4$; [middle panel] plots of $u_{\mathrm{num}}-(u_0+tu_1)$ (blue) and $U_{\mbox{\footnotesize dis}}$ (red circles) versus $x$ using same parameters as in left panel; [right panel] plots of $\left(u_{\mathrm{num}}-(u_0+tu_1)\right)/K$ (blue) and $U_{\mbox{\footnotesize scaled}}$, again with same parameters as left panel.
(b) Same as (a), except that $A_0=-3/4$.  (c) Same as (a), except that $A_0=-2$.}
\label{fig:testUdisp_2}
\end{figure}

These types of numerical tests turn out to be slightly more subtle for the initial condition (\ref{eq:ICxsquared}).  For example, in figure~\ref{fig:testUdisp_2} we show plots that are analogous to  figure~\ref{fig:testUdisp_1}, except that in figure~\ref{fig:testUdisp_2} we use the initial condition (\ref{eq:ICxsquared}).  Broadly speaking, we see very good agreement between numerics and asymptotics for $A_0=-1/4$, but the agreement for $A_0=-3/4$ is not quite as good.  Further, for the case $A_0=-2$ there is no agreement at all, since our asymptotic description (\ref{eq:expdispersive}) predicts that the dispersive waves should not appear for the precise value $A_0=-2$, whereas the numerical solutions clearly have waves (albeit of very small amplitude).  Of course, since (\ref{eq:ICxsquared}) is not of the special class of initial conditions that give rise to $N$-soliton solutions, we know that with (\ref{eq:ICxsquared}) there cannot be solutions without dispersive waves.  The explanation for the discrepancy is that the exponential asymptotics that led to (\ref{eq:expdispersive}) is only a first approximation.  As we vary the parameter $A_0$ so that it takes values closer and closer to the triangular numbers $-N(N+1)$, higher-order correction terms will inevitably come into play and eventually dominate.

Indeed, extending the local behaviour (\ref{eq:ICA}) of the initial condition about its singularity to be
\begin{equation}
u_0\sim \frac{A_0}{(x-x_0)^2}+\frac{A_1}{x-x_0}+A_2+A_3(x-x_0)+\ldots
\quad\mbox{as}\quad x\rightarrow x_0,
\label{eq:ICAmore}
\end{equation}
the full asymptotics for the dispersive waves will depend not only on $A_0$, but also (linearly) on $A_1$, $A_2$ and $A_3$, and so on, in an increasingly complicated manner.  For our special initial condition (\ref{eq:ICsech}), an expansion about $x_0=\pi\mathrm{i}/2$ shows that $A_1=0$, so the first-order correction terms vanish, while for (\ref{eq:ICxsquared}) an expansion about $x_0=\mathrm{i}$ shows that $A_1=\mathrm{i}A_0$, so the first-order correction terms in this case do not vanish.  To see the effect of the difference between these two cases, we show in figure~(\ref{fig:testUdisp_3})(b) a plot of the scaled amplitude versus $A_0$ for the initial condition (\ref{eq:ICxsquared}).  The agreement between the asymptotic prediction (in red) and the numerics (blue dots) appears to be good only for small values of $|A_0|$.  Compared with figure~(\ref{fig:testUdisp_3})(a) (for which $A_1=0$), it is clear that higher-order corrections (that depend linearly on $A_1$) are significant for (\ref{eq:ICxsquared}) (for which $A_1=\mathrm{i}A_0$) as $|A_0|$ increases.  To support this conclusion, we test a further initial condition, namely
\begin{equation}
u_0=-\frac{4A_0}{(1+x^2)^2}+\frac{2A_0}{1+x^2},
\label{eq:newIC}
\end{equation}
which has the property that
$$
u_0\sim \frac{A_0}{(x-\mathrm{i})^2}-\frac{A_0}{4}
-\frac{\mathrm{i}A_0}{4}(x-\mathrm{i})
\quad\mbox{as}\quad x\rightarrow\mathrm{i}.
$$
This is a carefully constructed initial condition that has the same leading-order behaviour near $x_0=\mathrm{i}$ as (\ref{eq:ICxsquared}), but has an additional term added so that $A_1=0$. Thus, locally near $x_0=\mathrm{i}$, the initial condition (\ref{eq:newIC}) is acting more like the special case (\ref{eq:ICsech}) near $x_0=\pi\mathrm{i}/2$ (which also has $A_1=0$). We plot in figure~(\ref{fig:testUdisp_3})(c) the scaled amplitude versus $A_0$ for (\ref{eq:newIC}), which shows much better agreement between numerics and asymptotics than in figure~\ref{fig:testUdisp_3}(b), again reinforcing the conclusion that there are important correction terms that scale with $A_1$.  While the treatment of correction terms in the context of our exponential asymptotics is rather complicated, we shall summarise how the analysis works for the special case $A_0=-N(N+1)$ in appendix~\ref{sec:expasympt_special}.

\begin{figure}
\centering
\subfloat[Solution at $t=0.3$, indicating peaks at $x_n$]{
\includegraphics[width=0.45\textwidth]{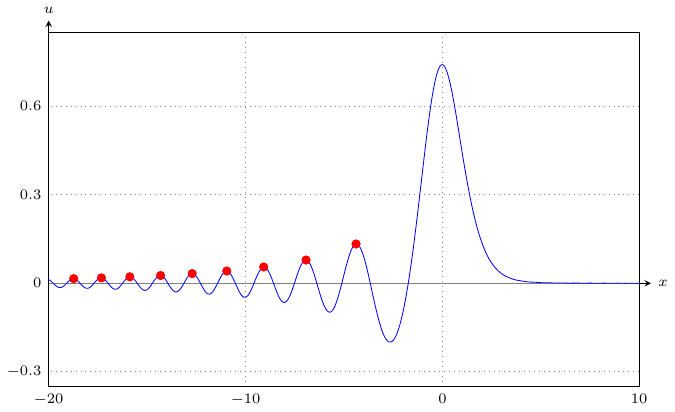}
}
\subfloat[plot of $x_n$ versus $t$]{
\includegraphics[width=0.45\textwidth]{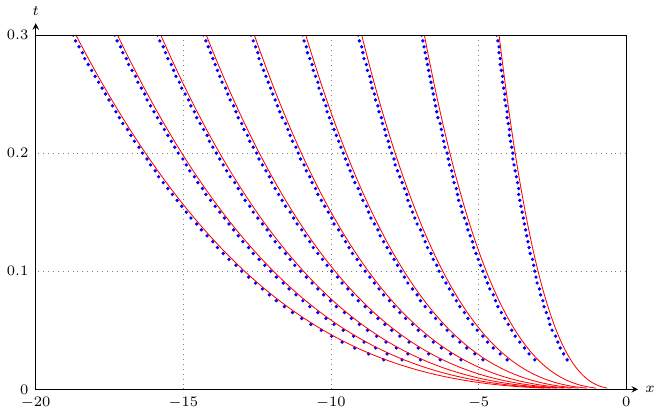}
}

\subfloat[plot of $\ln(-x_n)$ versus $\ln t$]{
\includegraphics[width=0.5\textwidth]{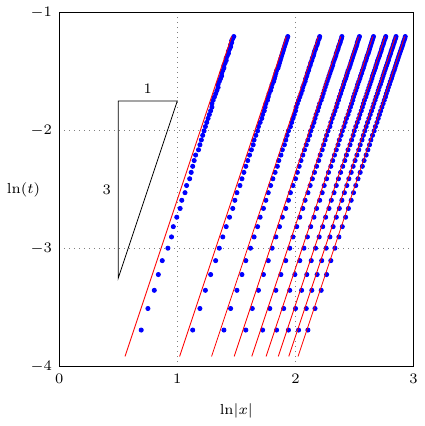}
}
\caption{(a) A snapshot of a numerical solution of (\ref{eq:kdv}) with the initial condition (\ref{eq:ICxsquared}), plotted for $t=0.3$ with the first nine wave crests indicated by red dots.  (b)--(c) Numerically-determined crest locations as a function of time (solid red) together with the asymptotic prediction (\ref{eq:crests}) (blue dots) for $m=1,\ldots,9$.  The slope of the hypotenuse of the triangle in the log-log plot indicates the scaling
$|x_m|\sim \,\mathrm{constant}\, t^{1/3}$.}
\label{fig:crests}
\end{figure}

The plots in figures~\ref{fig:testUdisp_1}--\ref{fig:testUdisp_2} are generated for a fixed small time.  As a simple check of the temporal scaling, we track the first nine dispersive wave crests for a specific example in figure~\ref{fig:crests}.  An image of the wave profile is shown in part (a) and the numerically-determined crest location as a function of time is plotted in (b) for the first nine crests.  The latter results reformatted on a log-log plot in (c) supports the $t^{1/3}$ scaling.  Further, in part (d) we have plotted the asymptotic prediction (\ref{eq:crests}) for the first nine crests $m=1,\ldots,9$.  This prediction shows good agreement with the numerics, especially given it holds formally for large $m$ (and small $t$) whereas these numerical results are for small $m$.

\begin{figure}
\centering
\includegraphics[width=0.52\textwidth]{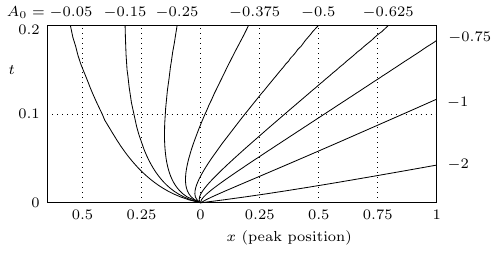}
\includegraphics[width=0.47\textwidth]{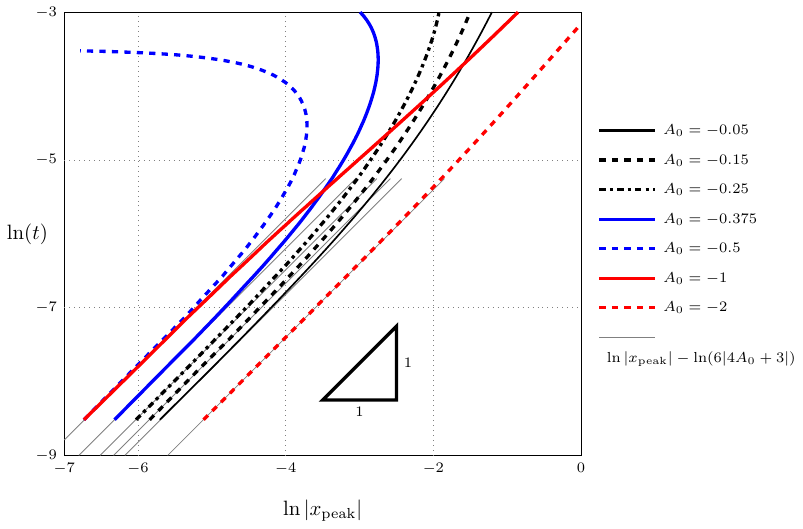}
\caption{(left) The location of the main peak in real solutions of (\ref{eq:kdv}) (horizontal axis) plotted against time (vertical axis).  In black, numerical results are shown for the initial condition (\ref{eq:ICxsquared}) plotted for $A_0=-0.05$, $-0.15$, $-0.25$, $-0.5$, $-0.625$, $-0.75$, $-1$ and $-2$.
(right) Main peak versus time on a log-log plot for the same initial condition, which shows how the numerical results (thick curves) approach the predicted limiting behaviour that comes from analysing the first two terms in the power series (\ref{eq:outer}), namely $\ln t \sim \ln |x_{\mathrm{peak}}| - \ln|6(4A_0+3)|$ as $t\rightarrow 0^+$ (thin solid lines).  Importantly (and as might be expected), the small-time behaviour of the main peak does {\em not} come from the exponential terms that appear beyond all order of the power series.}
\label{fig:maincrests}
\end{figure}

For the initial conditions we are considering here, there is a maximum (a ``peak'') at $x=0$.  This main peak can initially move to either the left or the right, depending on initial condition and the value $A_0$.  For example, we show in figure~\ref{fig:maincrests} the small-time evolution of the main peak for the initial condition (\ref{eq:ICxsquared}).  Note the behaviour of the solution in the neighbourhood of this peak is not driven by the exponentially-small terms (\ref{eq:expdispersive}), as these are only relevant to the left of $x=-y_0/\sqrt{3}=-1/\sqrt{3}$ (since the singularity is at $x=\mathrm{i}$ in this case) in the limit $t\rightarrow 0^+$.  Instead, for very small times, we expect the peak to move according to the first couple of terms in the power series, namely (\ref{eq:outer}), which suggests the peak location behaves like $x_{\mathrm{peak}}\sim -6(4A_0+3)t$.  That is, we expect the peak to initially move to the left for $A_0<-3/4$ and to the right for $-3/4<A_0<0$.  We can see this behaviour in the left panel of figure~\ref{fig:maincrests}, where the borderline case $A_0=-3/4$ is evident.  In the right panel, these numerical results on a log-log plot support the asymptotic prediction by approaching the appropriate straight line with slope unity for large negative values of $\ln t$.  Thus, an interesting observation is that the drift of the main peak of the solution scales like $t$ in the small time limit, found by tracking the local maximum of the first couple of terms in the power series expansion (\ref{eq:outer}), while the crests of the dispersive waves scale like $t^{1/3}$ (via (\ref{eq:crests})), driven by the exponentially-small terms that appear beyond all orders of the power series.

In summary, for our KdV problem (\ref{eq:kdv}), we see that the emergence of dispersive waves for small time can be explained by observing how exponentially-small terms are switched on across the point at which the Stokes lines intersect the real-$x$ axis.  As far as we are aware, this is the first asymptotic description of how dispersive waves propagate for the KdV equation in the limit that $t\rightarrow 0^+$.
While these results are interesting in their own right, we also use the above information about the Stokes line structure to inform how we apply far-field conditions in the inner problem in subsection~\ref{sec:innernearx0}.  Further, our analysis here locates an anti-Stokes lines at $\mathrm{arg}(x-x_0)=-\pi$, which is (near) where we expect there to be complex-plane singularities propagating out from each singularity of the initial condition, $x=x_0$.  One of our goals for the remainder of this paper is to track the initial dynamics of the singularities born at $x=x_0$, including the important string of singularities that align (close to) $\mathrm{arg}(x-x_0)=-\pi$.

\section{Inner problem governed by P$_{\mathrm{II}}$ with tritronqu\'{e}e  solutions}\label{sec:PII}

The inner regions near each of the double-pole singularities $x=x_0$ of the initial condition $u_0(x)$ are governed by decreasing tritronqu\'{e}e solutions of  Painlev\'{e} II (P$_{\mathrm{II}}$), as we explain in this section.

\subsection{Inner region near $x=x_0$}\label{sec:innernearx0}

Following on from subsection~\ref{sec:outerexpansion}, there is an inner problem as $t\rightarrow 0^+$, which holds for
\begin{equation}
\xi\equiv\frac{x-x_0}{(3t)^{1/3}}=\mathcal{O}(1).
\label{eq:xi}
\end{equation}
Using this new similarity-type variable, we can rewrite (\ref{eq:local}) as
\begin{equation}
u_0\sim \frac{A_0}{(3t)^{2/3}\xi^2},
\quad
u_1\sim \frac{12A_0(A_0+2)}{(3t)^{5/3}\xi^5},
\label{eq:localxi}
\end{equation}
which suggests the inner problem has
\begin{equation}
u=\frac{1}{(3t)^{2/3}}f(\xi,t).
\label{eq:inner}
\end{equation}
By substituting (\ref{eq:inner}) into (\ref{eq:kdv}), we find that $f$ satisfies the pde
\begin{equation}
3tf_t-2f-\xi f_\xi+6ff_\xi+f_{\xi\xi\xi}=0.
\label{eq:pdef}
\end{equation}
To leading order we write $f\sim f_0(\xi)$, so that (\ref{eq:pdef}) and (\ref{eq:localxi}) combine to give our inner problem
\begin{align}
&-2f_0-\xi \frac{\mathrm{d}f_0}{\mathrm{d}\xi} +6f_0\frac{\mathrm{d}f_0}{\mathrm{d}\xi}+\frac{\mathrm{d}^3f_0}{\mathrm{d}\xi^3}=0,
\label{eq:odef0}
\\
&f_0\sim \frac{A_0}{\xi^2}+\frac{4A_0(A_0+2)}{\xi^5}
\quad\mbox{as}\quad \xi\rightarrow -\mathrm{i}\infty.
\label{eq:farfieldf0}
\end{align}
Note the direction of the far-field condition (\ref{eq:farfieldf0}) comes from matching back down towards the real axis.  We expect (\ref{eq:farfieldf0}) to apply in a broader sector of the $\xi$ plane, as determined by our subsequent analysis.  Further, the extent to which (\ref{eq:farfieldf0}) may need to be supported by other far-field conditions will be explored later.

We comment here that the similarity reduction (\ref{eq:inner}) and the corresponding differential equations (\ref{eq:pdef}) and (\ref{eq:odef0}) are also used to derive the large-time asymptotics for the KdV problem (\ref{eq:kdv}) \cite{Ablowitz1977}.  In that case, the similarity variable is $\xi\equiv x/(3t)^{1/3}$ (instead of (\ref{eq:xi})), and the appropriate solution of (\ref{eq:odef0}) is valid on the real line for $x=\mathcal{O}(t^{1/3})$ as $t\rightarrow\infty$.  The relevant boundary conditions for (\ref{eq:odef0}) come from matching to an outer region as $\xi\rightarrow +\infty$.  All of these variables for the large-time asymptotics are real valued.
In contrast, our study of (\ref{eq:inner})-(\ref{eq:odef0}) is relevant for the region of the complex plane $x-x_0=\mathcal{O}(t^{1/3})$ as $t\rightarrow 0^+$.  The boundary condition (\ref{eq:farfieldf0}) is completely different to the matching condition used for the large-time asymptotics.  The variables $f$ and $\xi$ in our study of (\ref{eq:inner})-(\ref{eq:odef0}) are complex valued.

The third-order nonlinear ordinary differential equation (ode) problem (\ref{eq:odef0})--(\ref{eq:farfieldf0}) is difficult to analyse in this form, but much progress can be made by converting (\ref{eq:odef0}) to P$_{\mathrm{II}}$ in the usual way \cite{Joshi2004} (see the following subsection).  Ultimately, one of the goals of this exercise is to determine the singularities of $f_0(\xi)$ in the $\xi$ plane.  Our analysis then predicts that for any given singularity of $f_0$, which we call $\xi_0$, there is a double pole of $u(x,t)$ at $x=s(t)$ that emerges from $x_0$ like
\begin{equation}
s(t)\sim x_0+(3t)^{1/3}\xi_0
\quad\mbox{as}\quad t\rightarrow 0^+.
\label{eq:scalingfors}
\end{equation}
Before we proceed, there is a simple exact solution of (\ref{eq:odef0})--(\ref{eq:farfieldf0}), namely
\begin{equation}
f_0=-\frac{2}{\xi^2},
\quad A_0=-2.
\label{eq:exactA0minus2}
\end{equation}
The analysis in the present section is devoted to $A_0\neq -2$, while the special case $A_0=-2$ (for which there is this trivial exact solution for $f_0$) is treated separately in appendix~\ref{eq:A0m2}.

\subsection{Reduction to Painlev\'{e} II}\label{sec:reductionPII}

Applying the standard (Miura-transformation) reduction \cite{Boiti1979,Rosales1978}
\begin{equation}
f_0=\frac{\mathrm{d}F}{\mathrm{d}\xi}-F^2,
\label{eq:PIIreduction}
\end{equation}
we find that (\ref{eq:odef0}) is transformed to
\begin{equation}
\frac{\mathrm{d}^2}{\mathrm{d}\xi^2}
\left(\frac{\mathrm{d}^2F}{\mathrm{d}\xi^2}-\xi F-2F^3\right)-2F\frac{\mathrm{d}}{\mathrm{d}\xi}
\left(\frac{\mathrm{d}^2F}{\mathrm{d}\xi^2}-\xi F-2F^3\right)=0.
\label{eq:reduced}
\end{equation}
Before we consider (\ref{eq:reduced}) further, note that given the far-field condition (\ref{eq:farfieldf0}) and the change of variable (\ref{eq:PIIreduction}), we must have
\begin{equation}
F\sim \frac{\alpha}{\xi}
+\frac{2\alpha(1-\alpha^2)}{\xi^4}
\quad\mbox{as}\quad \xi\rightarrow -\mathrm{i}\infty,
\label{eq:farfieldF2term}
\end{equation}
where
\begin{equation}
\alpha=\sfrac{1}{2}(-1\pm\sqrt{1-4A_0}).
\label{eq:BA01}
\end{equation}
This observation is crucial for what follows.

Returning to (\ref{eq:reduced}), clearly one option is
\begin{equation}
\frac{\mathrm{d}^2F}{\mathrm{d}\xi^2}-\xi F-2F^3=\,\mbox{constant},
\label{eq:PIIconst}
\end{equation}
which is P$_{\mathrm{II}}$.  To determine the constant, we apply the far-field condition (\ref{eq:farfieldF2term}), which implies it is $-\alpha$.  We are free to proceed with either sign in (\ref{eq:BA01}); each choice will involve different solutions for $F$ but the same result for $f_0$ via (\ref{eq:PIIreduction}).  We choose the plus sign.
The other option, namely $F''-\xi F-2F^3\neq\,\mbox{constant}$, is not applicable as it is inconsistent with (\ref{eq:farfieldF2term}).

In summary, we conclude that (\ref{eq:odef0})--(\ref{eq:farfieldf0}) reduces to
\begin{align}
& \frac{\mathrm{d}^2F}{\mathrm{d}\xi^2}=2F^3+\xi F-\alpha,
\label{eq:pii}
\\
& F\sim \frac{\alpha}{\xi}+\frac{2\alpha(1-\alpha^2)}{\xi^4}
\quad\mbox{as}\quad \xi\rightarrow -\mathrm{i}\infty,
\label{eq:farfieldF}
\end{align}
where
\begin{equation}
\alpha=\sfrac{1}{2}(-1+\sqrt{1-4A_0}).
\label{eq:BA02}
\end{equation}
As mentioned above, (\ref{eq:pii}) is P$_{\mathrm{II}}$ with the constant $\alpha$ related to our key parameter $A_0$ (sometimes (\ref{eq:pii}) is referred to as the inhomogeneous P$_{\mathrm{II}}$ equation, with the homogenous version arising from $\alpha=0$).
Motivated by the examples discussed in the Introduction, we are mostly focused on real and negative values of $A_0$, which correspond to $\alpha>0$.  For example, the $N$-soliton solutions with $A_0=-N(N+1)$ correspond to $\alpha=N$ (via the rational solutions summarised in subsection~\ref{sec:exactsolns}).  For $0\leq A_0 \leq 1/4$, we have $-1/2\leq \alpha\leq 0$, and for $A_0>1/4$ we have complex values of $\alpha$.  These parameter intervals are also of some interest, although we shall restrict our exploration to real values of $\alpha$.  Finally, without further investigation (or a background in studying P$_{\mathrm{II}}$) it is not obvious whether or not (\ref{eq:farfieldF}) is sufficient to determine our solution uniquely, which is an issue we explore in subsection~\ref{sec:WKB}.

\subsection{Reduction to Painlev\'{e} 34}

In our work, we have applied the Miura-transformation (\ref{eq:PIIreduction}) to (\ref{eq:odef0})--(\ref{eq:farfieldf0}) so that our inner problem is described by the P$_{\mathrm{II}}$ problem (\ref{eq:pii})--(\ref{eq:BA02}).  Alternatively, we can set
\begin{equation}
f_0=g_0+\sfrac{1}{2}\xi,
\label{eq:P34reduction}
\end{equation}
so that
$$
g_0+2\xi \frac{\mathrm{d}g_0}{\mathrm{d}\xi} +6g_0\frac{\mathrm{d}g_0}{\mathrm{d}\xi}+\frac{\mathrm{d}^3g_0}{\mathrm{d}\xi^3}=0.
$$
Multiplying by $g_0$ and integrating, we find
\begin{equation}
g_0\frac{\mathrm{d}^2g_0}{\mathrm{d}\xi^2}=\frac{1}{2}\left(
\frac{\mathrm{d}g_0}{\mathrm{d}\xi}\right)^2
-2g_0^3-\xi g_0^2+\frac{1}{2}A_0-\frac{1}{8},
\label{eq:P34}
\end{equation}
where the constant has been determined by enforcing the far-field condition (\ref{eq:farfieldf0}).
Equation (\ref{eq:P34}) is known as Painlev\'{e} 34 (P$_{34}$), since an equivalent version was labelled XXXIV in section 14.33 of Ince~\cite{Ince1956}.

Clearly, given a solution $F$ of (\ref{eq:pii})--(\ref{eq:BA02}), we can recover a solution of P$_{34}$ by combining (\ref{eq:PIIreduction}) and (\ref{eq:P34reduction}) to give $g_0=\mathrm{d}F/\mathrm{d}\xi-F^2-\xi/2$.  Therefore all of our results for $F$ could be framed in terms of $g_0$.  Note the one-to-one correspondence between solutions of P$_{\mathrm{II}}$ and P$_{34}$ was given in Ref.~\cite{Fokas1982}, for example.  Equivalent links between P$_{\mathrm{II}}$ and P$_{34}$ can be established via the associate Hamiltonian, as described in section 32.6(iii) in the DLMF~\cite{DLMF} (equation 32.6.12 is equivalent to our P$_{34}$ in (\ref{eq:P34})).  See section 2.4 of Clarkson~\cite{Clarkson2003a} for further details.

\subsection{Liouville–Green (WKB) analysis and Stokes phenomenon}\label{sec:WKB}

For simplicity, we assume for the moment that the initial condition $u_0(x)$ has a single double pole in the upper half plane at $x=x_0$ (like (\ref{eq:ICxsquared}) does).  We wish to study (\ref{eq:pii}) subject to (\ref{eq:farfieldF}).   The solution we are after will satisfy this far-field condition in a broader sector that includes the negative imaginary direction.  To determine this sector, and also to determine whether (\ref{eq:farfieldF}) is sufficient to specify our solution uniquely,
we shall linearise about (\ref{eq:farfieldF}).

To proceed with this linearisation, we write
$$
F= \frac{\alpha}{\xi}+\frac{2\alpha(1-\alpha^2)}{\xi^4}+\ldots+J,
$$
where here the ellipsis denote higher terms in the algebraic expansion and $J$ represents the exponential correction.  Linearising about the algebraic series, to leading order we find that
$$
\frac{\mathrm{d}^2J}{\mathrm{d}\xi^2}=\left(\xi+\frac{6\alpha^2}{\xi^2}\right) J,
$$
which is related to the modified Bessel's equation via a change of variables.
The two possible linearly independent solutions, namely
$$
\xi^{1/2}I_{(1+24\alpha^2)^{1/2}/3}(\sfrac{2}{3}\xi^{3/2})
\quad\mbox{and}\quad
\xi^{1/2}K_{(1+24\alpha^2)^{1/2}/3}(\sfrac{2}{3}\xi^{3/2}),
$$
where $I_{\nu}(z)$ and $K_{\nu}(z)$ are modified Bessel functions of the first and second kind,
have the far-field behaviour $J\sim \mathrm{const}\,\xi^{-1/4}\,\mathrm{e}^{\pm 2\xi^{3/2}/3}$.  Therefore we have
$$
F\sim \frac{\alpha}{\xi}+\frac{2\alpha(1-\alpha^2)}{\xi^4}+\ldots
+\frac{\sigma_1}{\xi^{1/4}}\,\mathrm{e}^{2\xi^{3/2}/3}
+\frac{\sigma_2}{\xi^{1/4}}\,\mathrm{e}^{-2\xi^{3/2}/3}
\quad\mbox{as}\quad \xi\rightarrow -\mathrm{i}\infty,
$$
where $\sigma_1$ and $\sigma_2$ are important (Stokes) constants.
However, since $\mathrm{e}^{-2\xi^{3/2}/3}$ grows exponentially as $\xi\rightarrow -\mathrm{i}\infty$, we must set $\sigma_2=0$ in this asymptotic expression, leaving
\begin{equation}
F\sim \frac{\alpha}{\xi}+\frac{2\alpha(1-\alpha^2)}{\xi^4}+\ldots+\frac{\sigma_1}{\xi^{1/4}}\,\mathrm{e}^{2\xi^{3/2}/3}
\quad\mbox{as}\quad \xi\rightarrow -\mathrm{i}\infty.
\label{eq:farfieldwithK1}
\end{equation}
This exercise demonstrates that (\ref{eq:farfieldF}) is acting as one boundary condition only and to specify the solution of (\ref{eq:pii})--(\ref{eq:farfieldF}) uniquely we need to fix $\sigma_1$ in (\ref{eq:farfieldwithK1}), which is hidden beyond all orders of the algebraic expansion.

At this stage, we emphasise that (\ref{eq:pii}) with (\ref{eq:farfieldwithK1}) combine to give a one-parameter family of solutions of the type studied at length by Boutroux~\cite{Boutroux1913} and by others, including in Refs~\cite{Fokas2006,Fornberg2014,Its2003,Joshi1988,Joshi1992}.  We note that we are able to write out the full algebraic series in (\ref{eq:farfieldwithK1}) as
\begin{equation}
F\sim \frac{\alpha}{\xi}\sum_{n=0}^\infty \frac{b_n}{\xi^{3n}}
\quad\mbox{as}\quad \xi\rightarrow -\mathrm{i}\infty,
\label{eq:divergent}
\end{equation}
where the $b_n$ satisfy the recurrence relations \cite{Its2003}
\begin{equation}
b_{n+1}=(3n+1)(3n+2)b_n-2\alpha^2\sum_{\substack{i,j,k=0\\i+j+k=n}}^{n} b_ib_jb_k,\quad b_0=1.
\label{eq:its}
\end{equation}
Another way to write these coefficients is via $b_n=(-2)^n(\alpha^2-1)p_n(\alpha)$, where $p_n$ is a polynomial of order $2n-2$ of the form
$$
p_n=a^{(n)}_1 \alpha^{2n-2} - a^{(n)}_2 \alpha^{2n-4} + a^{(n)}_3 \alpha^{2n-6} - a^{(n)}_4 \alpha^{2n-8}+\ldots+(-1)^n a^{(n)}_{n-1}\alpha^2+(-1)^{n+1} a^{(n)}_{n}.
$$
As examples, the first and last coefficients $a^{(n)}_1$ and $a^{(n)}_{n}$ are A001764 and A025035 in the online encyclopedia of integer sequences \cite{OEIS}, respectively, with large $n$ behaviour
$$
a^{(n)}_1=\frac{(3n)!}{(2n+1)(2n)!n!}\sim \frac{3^{n+1/2}}{4{\pi}^{1/2}n^{3/2}}
\left(\frac{3}{2}\right)^{2n},
\quad
a^{(n)}_{n}=\frac{(3n)!}{6^n n!}\sim \frac{\sqrt{3}}{2^n}\left(\frac{3n}{\mathrm{e}}\right)^{2n}.
$$
Its \& Kapaev~\cite{Its2003} show that the coefficients $b_n$ have the large $n$ behaviour
\begin{equation}
b_n \sim
\frac{\sin\pi\alpha}{\alpha\pi^{3/2}}\left(\frac{3}{2}\right)^{2n+1/2}
\Gamma\left(2n+\sfrac{1}{2}\right)
\quad\mbox{as}\quad n\rightarrow\infty
\label{eq:largebn}
\end{equation}
(see also \cite{Fokas2006}).  A key point is that for $\alpha\in \mathbb{N}$, the series (\ref{eq:divergent}) is convergent (see section~\ref{sec:exactsolns} below), while otherwise it is divergent.

\begin{figure}
\centering
\includegraphics[width=0.7\textwidth]{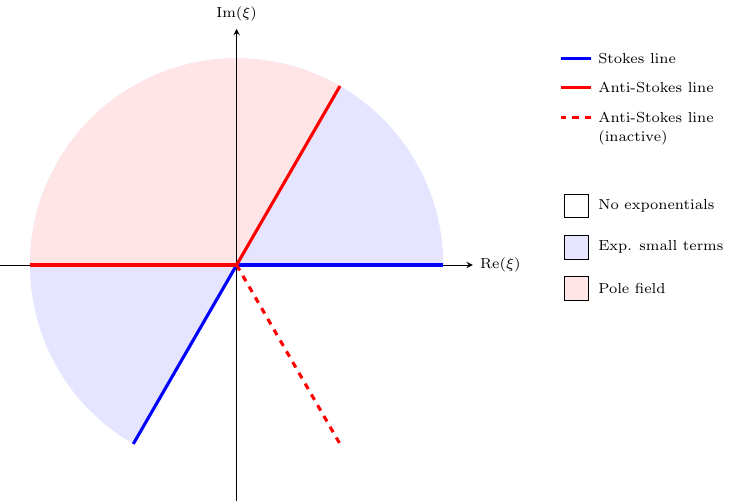}
\caption{A schematic of the $\xi$ plane, where $\xi=(x-x_0)/(3t)^{1/3}$, indicating the
Stokes structure for our decreasing tritronqu\'{e}e solution of P$_{\mathrm{II}}$.}
\label{fig:xiplane}
\end{figure}

To proceed with our argument about the one-parameter family of solutions to (\ref{eq:pii}) with (\ref{eq:farfieldwithK1}), we explore broader far-field sectors in the $\xi$ plane, referring to figure~\ref{fig:xiplane}.  With $\xi=\rho\mathrm{e}^{\mathrm{i}\theta}$, then let the direction $\xi\rightarrow -\mathrm{i}\infty$ be $\theta=-\pi/2$.  For any $\sigma_1\neq 0$, then we would have $\sigma_1\xi^{-1/4}\mathrm{e}^{2\xi^{3/2}/3}$ decaying along the ray $\theta=-\pi/2$.  Along a path $\rho=\,\mathrm{const}$ (with $\rho\gg 1$), as $\theta$ increases from $\theta=-\pi/2$ this term would increase in size until it was $\mathcal{O}(1)$ at the anti-Stokes line $\theta=-\pi/3$ (dashed line in figure~\ref{fig:xiplane}).  For $\theta>-\pi/3$, the term would be exponentially growing.

In order to eliminate the inclusion of an exponentially growing term in the sector $-\pi<\theta<0$, we need to force $\sigma_1=0$ in (\ref{eq:farfieldwithK1}).  In that case, both exponentials $\xi^{-1/4}\mathrm{e}^{2\xi^{3/2}/3}$ and $\xi^{-1/4}\mathrm{e}^{-2\xi^{3/2}/3}$ are absent along $\theta=-\pi/2$.  As $\theta$ increases from $\theta=-\pi/2$, we see that $\xi^{-1/4}\mathrm{e}^{-2\xi^{3/2}/3}$ is switched on at the Stokes line $\theta=0$ (solid blue line) and becomes of $\mathcal{O}(1)$ at $\theta=\pi/3$.  Transversing the other way, as $\theta$ decreases from $\theta=-\pi/2$, the exponential $\xi^{-1/4}\mathrm{e}^{2\xi^{3/2}/3}$ is switched on at the Stokes line $\theta=-2\pi/3$ (solid blue line) and becomes of $\mathcal{O}(1)$ at $\theta=-\pi$.  Putting it together, a complete description of our inner problem is
\begin{align}
& \frac{\mathrm{d}^2F}{\mathrm{d}\xi^2}=2F^3+\xi F-\alpha,
\label{eq:pii_2}
\\
& F\sim \frac{\alpha}{\xi}+\frac{2\alpha(1-\alpha^2)}{\xi^4}+\ldots
+\frac{\sigma_1}{\xi^{1/4}}\,\mathrm{e}^{2\xi^{3/2}/3}
\quad\mbox{as}\quad |\xi|\rightarrow \infty, \quad -2\pi/3<\mathrm{arg}(\xi)<0
\label{eq:farfieldF_2}
\\
& \sigma_1=0, \quad -2\pi/3<\mathrm{arg}(\xi)<0,
\label{eq:K1equals0}
\end{align}
where it is understood that the ellipsis represent further terms in the divergent power series (\ref{eq:divergent}); that is, there is a one-parameter family of solutions that satisfies (\ref{eq:pii_2})--(\ref{eq:farfieldF_2}), but the specific solution we are after is selected by enforcing (\ref{eq:K1equals0}).  Furthermore, the unique solution to (\ref{eq:pii_2})--(\ref{eq:K1equals0}) should be pole-free region in the far field of the sector $-\pi<\theta<\pi/3$ (bounded by active anti-Stokes lines at $\theta=-\pi$ and $\theta=\pi/3$) and the power-series part of the far-field condition, namely
\begin{equation}
F\sim \frac{\alpha}{\xi}+\frac{2\alpha(1-\alpha^2)}{\xi^4}
\quad\mbox{as}\quad |\xi|\rightarrow \infty,
\label{eq:farfieldsector}
\end{equation}
applies in the sector $-\pi<\theta<\pi/3$.  This type of solution with a $4\pi/3$ pole-free sector has been referred to as a decreasing tritronqu\'{e}e solution \cite{Fokas2006} (not to be confused with increasing tritronqu\'{e}e solutions, which behave to leading order like $F\sim \pm (-\xi/2)^{1/2}$ as $|\xi|\rightarrow \infty$ \cite{Miller2018}).  Note that our solution $F(\xi)$ is equivalent to the function $y_3(x,\alpha)$ in \cite{Its2003} or $u_3(x|\alpha)$ in \cite{Fokas2006}, where the Stokes multipliers $s_1=s_2=0$ and $s_3=-2\sin\pi\alpha$, using their notation.

\begin{figure}
\centering
\resizebox{0.8\textwidth}{!}{
\begin{tikzpicture}
    \node[anchor=south west,inner sep=0] (image) at (0,0) {\includegraphics[width=0.9\textwidth]{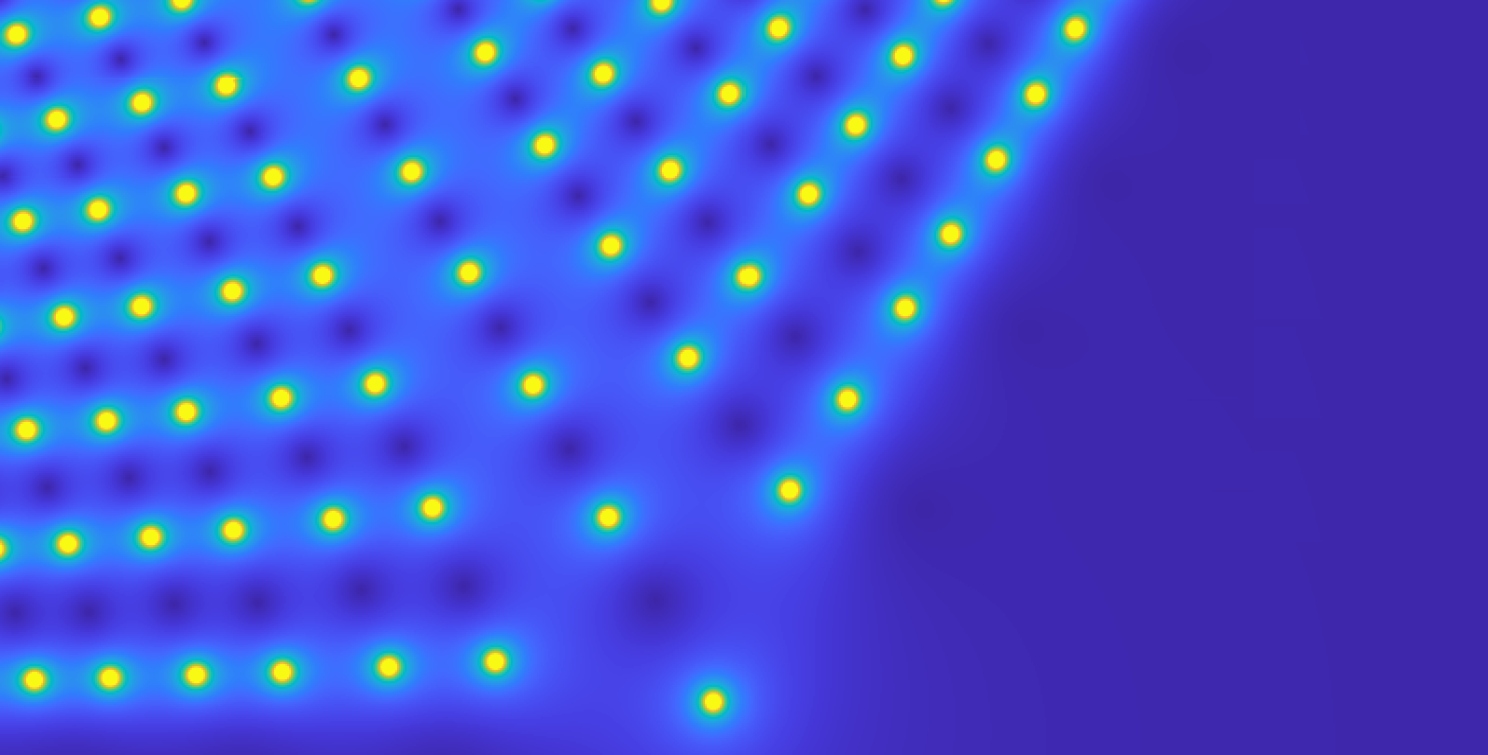}};
    \begin{scope}[x={(image.south east)},y={(image.north west)}]
        \node [below] at (0,0) {\scriptsize{$-10$}};
        \node [below] at (1,0) {\scriptsize{$10$}};
        \node [below] at (0.25,0) {\scriptsize{$-5$}};
        \node [below] at (0.75,0) {\scriptsize{$5$}};
        \node [below] at (0.5,0) {\scriptsize{$0$}};
        \draw [yshift=0.25pt, ->] (0,0) -- (1.02,0) node[right] {\scriptsize{$\mathrm{Re}(\xi)$}};
        \draw [yshift=0.25pt, ->] (0,0) -- (0,1.03) node[above] {\scriptsize{$\mathrm{Im}(\xi)$}};
        \node [left] at (0,0) {\scriptsize{$0$}};
        \node [left] at (0,0.5) {\scriptsize{$5$}};
        \node [left] at (0,1) {\scriptsize{$10$}};
        \draw[line width=1pt,red] (0,0) -- (0.5,0) -- (0.788675,1);
    \end{scope}
\end{tikzpicture}
}
\caption{An image showing $|F|$ for a numerical solution of Painlev\'{e} II with $\alpha = \sfrac{1}{2}(-1 + \sqrt{5})$ using the algorithm from Fornberg \& Weideman~\cite{Fornberg2014}.  The yellow dots represent the pole field for this solution.  The two active anti-Stokes lines are shown in red.  While solutions of P$_{\mathrm{II}}$ can be uniquely specified by the ``initial conditions'' $F(0)$ and $F'(0)$, our scheme provides this pair as outputs.  For this example, the numerically obtained values are $F(0) \approx 0.5941 + 1.0289\mathrm{i}$ and $F'(0) \approx 0.7995 - 1.3848\mathrm{i}$.}
\label{fig:PIIexample}
\end{figure}

Finally, from a practical (numerical) perspective, perhaps the best way to enforce (\ref{eq:farfieldF_2})--(\ref{eq:K1equals0}) (which count as two boundary conditions) is to apply the far-field condition (\ref{eq:farfieldsector}) along the ray $\theta=-\pi/3$ (the inactive anti-Stokes line).  The reason is that for any $\sigma_1\neq 0$, the condition  (\ref{eq:farfieldsector}) would not apply along $\theta=-\pi/3$ (since, as already explained, the term $\sigma_1\xi^{-1/4}\mathrm{e}^{2\xi^{3/2}/3}$ would be $\mathcal{O}(1)$ along this ray).  Therefore, applying (\ref{eq:farfieldsector}) along $\theta=-\pi/3$ has the effect of killing off $\sigma_1\xi^{-1/4}\mathrm{e}^{2\xi^{3/2}/3}$ (as well as the other exponential $\sigma_2\xi^{-1/4}\mathrm{e}^{-2\xi^{3/2}/3}$).
As an example of a numerical solution of (\ref{eq:pii_2})--(\ref{eq:K1equals0}) that is found by enforcing (\ref{eq:farfieldsector}) along $\theta=-\pi/3$, we show in figure~\ref{fig:PIIexample} an image computed for the case $\alpha= \sfrac{1}{2}(-1 + \sqrt{5})$ using the algorithm outlined by Fornberg \& Weideman~\cite{Fornberg2014}.  Here the yellow dots indicate poles of the solution, which are aligned in arrays.  The red lines indicate anti-Stokes lines.  Note that for this value of $\alpha$, there are no poles in the third or fourth quadrant.
We discuss other numerical solutions of (\ref{eq:pii_2})--(\ref{eq:K1equals0}) below in subsection~\ref{sec:numericalPII}.

\subsection{Rational solutions of P$_{\mathrm{II}}$ for $\alpha\in \mathbb{N}$}\label{sec:exactsolns}

There are well-known exact rational solutions of (\ref{eq:pii_2})--(\ref{eq:K1equals0}), which will be important for our study.  For a start, when $\alpha=1$, the solution of (\ref{eq:pii_2})--(\ref{eq:K1equals0}) is simply $F=1/\xi$.  This scenario comes from $A_0=-2$, which happens to apply for the single-soliton solution.  Here, the inner problem (\ref{eq:pii_2})--(\ref{eq:K1equals0}) is insufficient to describe the early time behaviour (as a singularity $\xi_0=0$ does not correspond to a moving singularity in the original variables) and higher-order terms are required.  This case is considered in appendix~\ref{eq:A0m2}.

More generally, there are rational solutions of P$_{\mathrm{II}}$ for all $\alpha\in \mathbb{N}$, all of which satisfy (\ref{eq:pii_2})--(\ref{eq:K1equals0}).  Each of these solutions is of the form
\begin{equation}
F=-\frac{\mathrm{d}}{\mathrm{d}\xi}\ln\left(\frac{Q_{\alpha-1}(\xi)}{Q_\alpha(\xi)}\right),
\quad
\alpha\in \mathbb{N},
\label{eq:rationalF}
\end{equation}
where the $Q_n$ satisfy the recursion relation \cite{Clarkson2006,Clarkson2003}
$$
Q_{n+1}Q_{n-1}=\xi Q_n^2-4\left(Q_n\frac{\mathrm{d}^2 Q_n}{\mathrm{d}\xi^2}-\left(\frac{\mathrm{d} Q_n}{\mathrm{d}\xi}\right)^2\right),
\quad
Q_0=1,
\quad
Q_1=\xi.
$$
These $Q_n$ are referred to as Yablonskii-Vorob'ev polynomials \cite{Vorobev1965,Yablonskii1959}.  They are special cases of the Adler-Moser polynomials, which are associated with rational solutions of the KdV equation \cite{Adler1978,Airault1977}; various other properties of the Yablonskii-Vorob'ev polynomials are summarised in Refs~\cite{Bertola2015,Clarkson2003b,Kanedo2003,Roffelsen2012,Taneda2000}.

Using the recursion relation (or, equivalently, a determinant representation derived in \cite{Kajiwara1996}), we find
\begin{equation}
Q_2=\xi^3+4, \quad Q_3=\xi^6+20\xi^3-80, \quad Q_4=\xi(\xi^9 + 60\xi^6 + 11200), \quad \ldots.
\label{eq:moreQ}
\end{equation}
For example, the first few rational solutions are (see table 7.5.1 of \cite{Ablowitz1991})
\begin{align}
\alpha=1:\quad  F= & -\frac{\mathrm{d}}{\mathrm{d}\xi}\ln\left(\frac{1}{\xi}\right)=\frac{1}{\xi},
\label{eq:F1}
\\
\alpha=2:\quad  F= & -\frac{\mathrm{d}}{\mathrm{d}\xi}\ln\left(\frac{\xi}{\xi^3+4}\right)=-\frac{1}{\xi}
+\frac{3\xi^2}{\xi^3+4},
\label{eq:F2}
\\
\alpha=3:\quad  F= & -\frac{\mathrm{d}}{\mathrm{d}\xi}\ln\left(\frac{\xi^3+4}{\xi^6+20\xi^3-80}\right)=
-\frac{3\xi^2}{\xi^3+4}+\frac{6\xi^2(\xi^3+10)}{\xi^6+20\xi^3-80},
\label{eq:F3}
\\
\alpha=4:\quad  F= & -\frac{\mathrm{d}}{\mathrm{d}\xi}\ln\left(\frac{\xi^6+20\xi^3-80}{\xi(\xi^9+60\xi^6+11200)}\right)
\nonumber
\\
=  & \, \frac{1}{\xi}-\frac{6\xi^2(\xi^3+10)}{\xi^6+20\xi^3-80}
+\frac{9\xi^5(\xi^3+40)}{\xi^9+60\xi^6+11200}.
\label{eq:F4}
\end{align}
We can easily check that each of these satisfies the far-field condition (\ref{eq:farfieldF_2}) with (\ref{eq:K1equals0}).  Further details of rational solutions of P$_{\mathrm{II}}$ are provided in Refs~\cite{Buckingham2014,Miller2017}

In terms of the inner problem for KdV, (\ref{eq:odef0})--(\ref{eq:farfieldf0}), applying the formula (\ref{eq:PIIreduction}), these rational solutions correspond to
\begin{align}
\alpha=1:\quad & f_0=-\frac{2}{\xi^2},
\label{eq:B1f0}
\\
\alpha=2:\quad & f_0=-\frac{6\xi(\xi^3-8)}{(\xi^3+4)^2},
\label{eq:B2f0}
\\
\alpha=3:\quad & f_0=-\frac{12\xi(\xi^9+600\xi^3+1600)}{(\xi^6+20\xi^3-80)^2},
\label{eq:B3f0}
\\
\alpha=4:\quad & f_0=-\frac{20(\xi^{18} + 48\xi^{15} + 2520\xi^{12} - 78400\xi^9 - 1881600\xi^6 + 12544000)}
{\xi^2(\xi^9+60\xi^6+11200)^2}.
\label{eq:B4f0}
\end{align}
One conclusion is that the double poles of $f_0$ for each $\alpha\in \mathbb{N}$ are located at the zeros of the polynomials $Q_\alpha(\xi)$.  These occur in an almost triangular structure \cite{Bertola2015,Buckingham2014,Clarkson2006,Clarkson2003,Miller2017}.  Note that, while solutions $F$ of P$_{\mathrm{II}}$ have simple poles of residue either $+1$ or $-1$, the application of (\ref{eq:PIIreduction}) shows that only those with residue $+1$ correspond to double poles of $f_0$ (for simple poles with residue $-1$, say at $\xi=\bar{\xi}$, the function $f_0$ takes the value $\bar{\xi}/2$).  This is why, for $\alpha\in \mathbb{N}$, there are $\alpha^2$ simple poles of $F$, but only $\sfrac{1}{2}\alpha(\alpha+1)$ double poles of $f_0$.

A more direct way of deriving the rational solutions of $f_0$ is via
$$
f_0=2\frac{\mathrm{d}^2}{\mathrm{d}\xi^2}\ln Q_\alpha(\xi),
\quad
\alpha\in \mathbb{N}
$$
(see Clarkson~\cite{Clarkson2003b}, for example, where $f_0$ is denoted by $W$ in equation (46) in Clarkson's paper).  This formula has the advantage of requiring only one Yablonskii-Vorob'ev polynomial $Q_n$, whereas (\ref{eq:rationalF}) involves two consecutive polynomials.

For later, we shall check the result here for $\alpha=2$, which will be used to compare with the well-known 2-soliton solution of (\ref{eq:kdv}).  We see above that the poles of $f_0$ for $\alpha=2$ (the zeros of $Q_2=\xi^3+4$) are at
\begin{equation}
\xi_0=-2^{2/3}, \quad 2^{2/3}\left(\frac{1}{2}+\frac{\sqrt{3}}{2}\mathrm{i}\right),
\quad 2^{2/3}\left(\frac{1}{2}-\frac{\sqrt{3}}{2}\mathrm{i}\right),
\label{eq:xi0forB2}
\end{equation}
where $2^{2/3}\approx 1.587$.
Thus, when $\alpha=2$ (that is, $A_0=-6$), our small-time analysis (see (\ref{eq:scalingfors})) suggests there are three singularities that move like
\begin{equation}
s(t)\sim x_0 - 2^{2/3}(3t)^{1/3},
\,\, s(t)\sim x_0+ 2^{2/3}\left(\frac{1}{2}+\frac{\sqrt{3}}{2}\mathrm{i}\right)
(3t)^{1/3},
\,\, s(t)\sim x_0+ 2^{2/3}\left(\frac{1}{2}-\frac{\sqrt{3}}{2}\mathrm{i}\right)
(3t)^{1/3}
\label{eq:checkB2}
\end{equation}
as $t\rightarrow 0^+$.  That is, they initially propagate out from $x_0$ in equispaced directions $-\pi$, $\pi/3$ and $-\pi/3$ with asymptotic speed $(2/3t)^{2/3}\approx 0.763 \,t^{-2/3}$.

Similarly, we shall later check our results here for $\alpha=3$ against the 3-soliton solution of (\ref{eq:kdv}).  In this case, the poles of $f_0$ for $\alpha=3$ (the zeros of $Q_3=\xi^6+20\xi^3-80$) are at
$$
\xi_0=(-10+6\sqrt{5})^{1/3},
\quad
(-10+6\sqrt{5})^{1/3} \left(-\frac{1}{2}+\frac{\sqrt{3}}{2}\mathrm{i}\right),
\quad
(-10+6\sqrt{5})^{1/3} \left(-\frac{1}{2}-\frac{\sqrt{3}}{2}\mathrm{i}\right),
$$
\begin{equation}
-(10+6\sqrt{5})^{1/3},
\quad
(10+6\sqrt{5})^{1/3}
\left(\frac{1}{2}+\frac{\sqrt{3}}{2}\mathrm{i}\right),
\quad
(10+6\sqrt{5})^{1/3}
\left(\frac{1}{2}-\frac{\sqrt{3}}{2}\mathrm{i}\right),
\label{eq:checkB3}
\end{equation}
where $(-10+6\sqrt{5})^{1/3}\approx 1.506$ and $10+6\sqrt{5}\approx 2.861$.
Thus, for $\alpha=3$ ($A_0=-12$), our analysis predicts there are six singularities that emerge from each $x_0$, and they propagate like (see (\ref{eq:scalingfors}))
\begin{equation}
s(t)\sim x_0+ (-10+6\sqrt{5})^{1/3}(3t)^{1/3},
\quad s(t)\sim x_0+
(-10+6\sqrt{5})^{1/3} \left(-\frac{1}{2}+\frac{\sqrt{3}}{2}\mathrm{i}\right)
(3t)^{1/3},
\label{eq:checkB31}
\end{equation}
\begin{equation}
s(t)\sim x_0+
(-10+6\sqrt{5})^{1/3} \left(-\frac{1}{2}-\frac{\sqrt{3}}{2}\mathrm{i}\right)
(3t)^{1/3},
\quad
s(t)\sim x_0-(10+6\sqrt{5})^{1/3}(3t)^{1/3},
\end{equation}
\begin{equation}
s(t)\sim x_0+
(10+6\sqrt{5})^{1/3} \left(\frac{1}{2}+\frac{\sqrt{3}}{2}\mathrm{i}\right)
(3t)^{1/3},
\quad s(t)\sim x_0+
(10+6\sqrt{5})^{1/3} \left(\frac{1}{2}-\frac{\sqrt{3}}{2}\mathrm{i}\right)
(3t)^{1/3},
\label{eq:checkB32}
\end{equation}
as $t\rightarrow 0^+$.  This time, three singularities propagate out from $x_0$ in equispaced directions $0$, $2\pi/3$ and $-2\pi/3$ with asymptotic speed $(-10+6\sqrt{5})^{1/3}/(3t)^{2/3}
\approx 0.724 \,t^{-2/3}$, while the other three propagate out in directions $-\pi$, $\pi/3$ and $-\pi/3$ with asymptotic speed
$(10+6\sqrt{5})^{1/3}/(3t)^{2/3}
\approx 1.375 \,t^{-2/3}$.

\subsection{Numerical solutions of (\ref{eq:pii_2})--(\ref{eq:K1equals0})}\label{sec:numericalPII}

\begin{figure}
\centering
\includegraphics[width=0.9\textwidth]{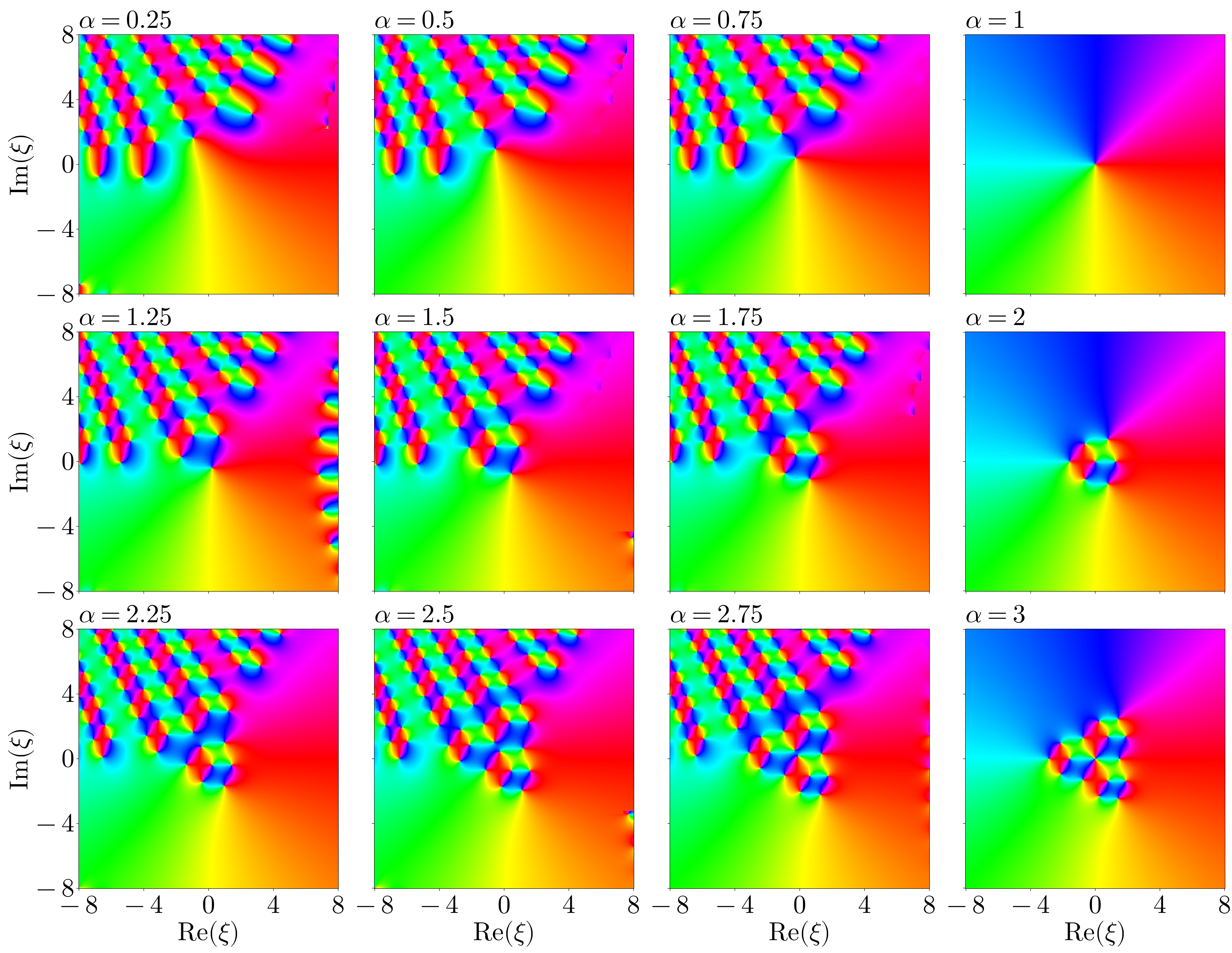}
\caption{Phase portraits of solutions of P$_{\mathrm{II}}$, (\ref{eq:pii_2})--(\ref{eq:K1equals0}), computed for various values of $\alpha$, namely $0.25$, $0.5$, $0.75$ and $1$ in the first row, $1.25$, $1.5$, $1.75$, and $2$ in the second row, and $2.25$, $2.5$, $2.75$ and $3$ in the third row.  The colour denotes the phase of $F$, with red denoting real and positive, yellow imaginary and positive, light blue real and negative, and dark blue negative and imaginary.  Note the images for $\alpha=1$, $2$ and $3$ in this figure are generated using the exact solutions (\ref{eq:F1})--(\ref{eq:F3}), while the rest are computed using the pole field solver~\cite{Fornberg2014}.  In the latter case, we can see some numerical error in some images around $\mathrm{Re}(\xi)\approx 8$.
}
\label{fig:daniel}
\end{figure}

By adapting the code used to run numerical simulations of P$_{\mathrm{II}}$ in Fornberg \& Weideman~\cite{Fornberg2014} (referred to as the ``pole field solver''; see also \cite{Fornberg2011,Fornberg2015}), we are able to generate numerical solutions of (\ref{eq:pii_2})--(\ref{eq:K1equals0}) for a selection of values of the parameter $\alpha$.  Some of these results are presented in figure~\ref{fig:daniel}.
To obtain these results, we treat (\ref{eq:pii_2}) as an initial-value problem with $F(L\mathrm{e}^{-\pi\mathrm{i}/3})$ determined by an optimal truncation of the far-field condition (\ref{eq:farfieldF_2}) using the recurrence relation in (\ref{eq:its}) to obtain as many terms as needed in (\ref{eq:divergent}), where $L$ is some moderately large value of $L$ ($L=10$, say).  Using the pole field solver~\cite{Fornberg2014}, we integrate from $\xi = L\mathrm{e}^{-\pi\mathrm{i}/3}$ towards $\xi = 0$ in order to estimate $F(0)$ and $F'(0)$. Once these estimates are obtained, the pole field solver algorithm is again used, now integrating along the entire square computational domain. In cases where $F(\xi)$ is singular at or near $\xi = 0$, instead of integrating until $\xi = 0$ we integrate until $|\xi| < h$ for some small $h$, and then use that stopping location as an initial value for evaluating the solution on the complete domain. Lastly, we note that if there a singularity on or near the ray $\xi = \rho\mathrm{e}^{-\pi\mathrm{i}/3}$, the pole field solver algorithm is able to adjust the path of integration automatically, thus no adjustment is needed to handle this situation.

Returning to figure~\ref{fig:daniel}, the most obvious distinction between some of these examples and others is that the solutions for $\alpha\in\mathbb{N}$ do not have a lattice of poles in the far field, as these are rational solutions with a finite number of poles.  Note that the exact solutions for $\alpha=1$, $2$ and $3$ are given by (\ref{eq:F1})--(\ref{eq:F3}), and the images in figure~\ref{fig:daniel} are generated from these exact results (numerical solutions for $\alpha=1$, $2$ and $3$ are visually indistinguishable from the exact results for $-7\lesssim\mathrm{Re}(\xi)\lesssim7$, $-7\lesssim\mathrm{Im}(\xi)\lesssim7$, but show some numerical error outside of this region).  Recall that our solutions of P$_{\mathrm{II}}$ have simple poles with reside either $+1$ or $-1$.  Using the solution for $\alpha=1$ in figure~\ref{fig:daniel} (right panel of first row), namely $F=1/\xi$, as a reference, it should be clear how the colour scheme in the phase portrait displays a simple pole with residue $+1$ (with colouring red, dark blue, light blue, yellow, then red again, as we traverse once around the pole in the positive direction, starting at $\theta=0$).  For example, we see the solution for $\alpha=2$ in figure~\ref{fig:daniel} (right panel of second row), which has simple poles with residue $+1$ at the three points given by (\ref{eq:xi0forB2}).  The other pole for $\alpha=2$, located at the origin, has residue $-1$ (with colouring light blue, yellow, red, dark blue and light blue as we traverse around the pole starting at $\theta=0$).  For $\alpha=3$ in figure~\ref{fig:daniel} (right panel of third row), the six poles with residue $+1$ (\ref{eq:checkB3}) form an almost triangular shape, while the three remaining poles with residue $-1$ lie inside this almost-triangle, and so on.  Clear images of the pole fields for these and other rational solutions, together with the location of the simple zeros, can be found in Fornberg \& Weideman~\cite{Fornberg2014}, for example.

\begin{figure}
\centering
\includegraphics[width=0.33\textwidth]{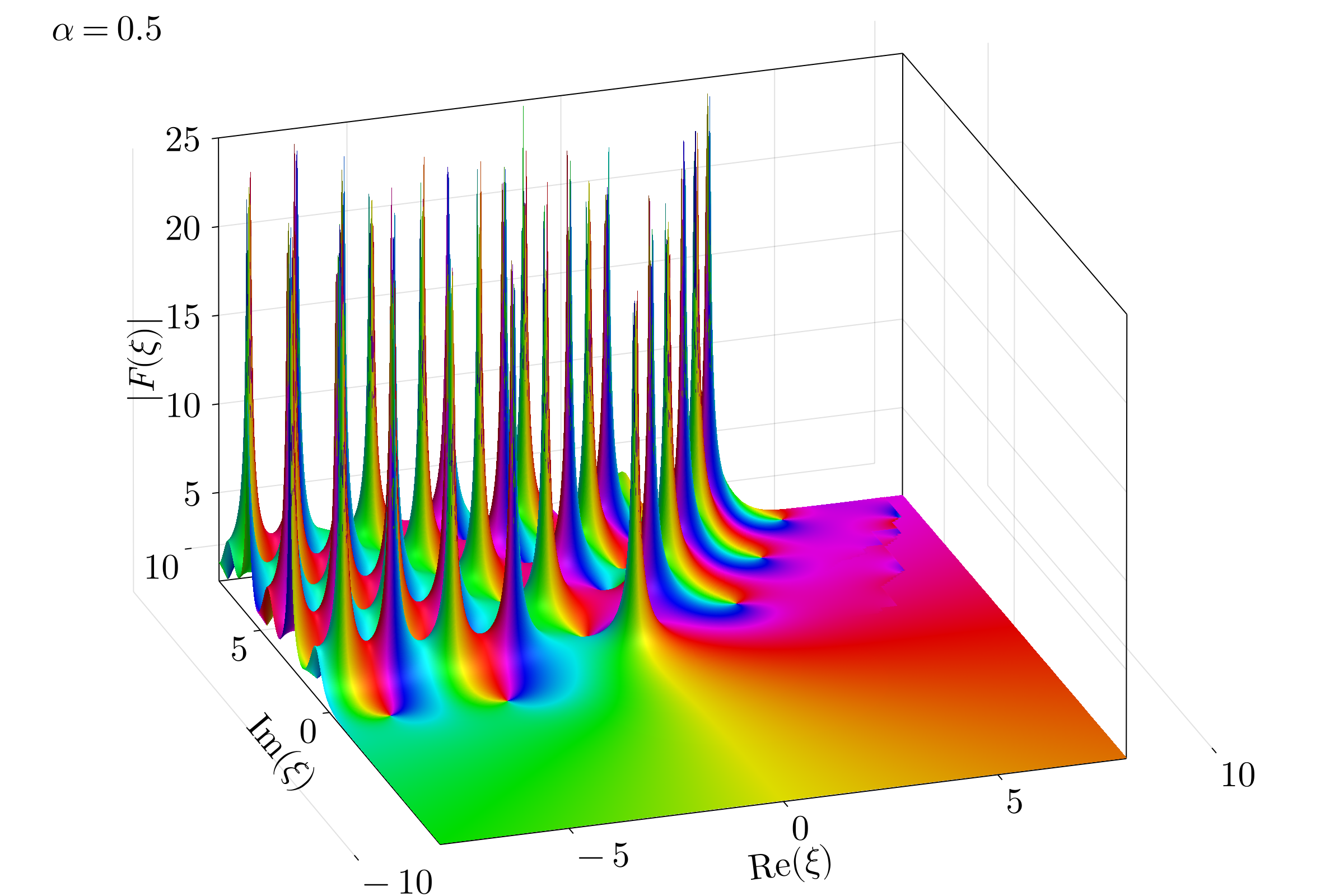}
\includegraphics[width=0.33\textwidth]{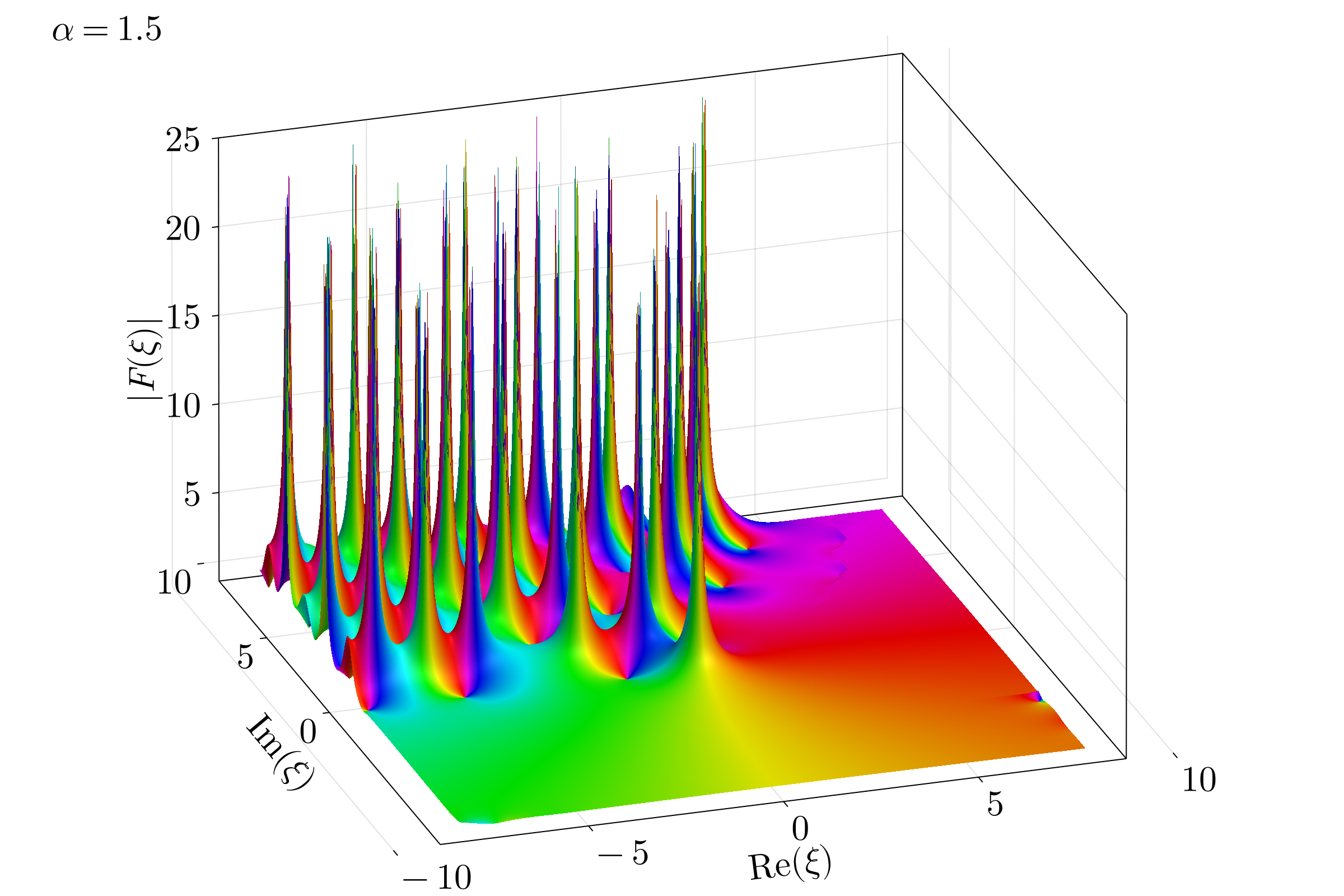}
\includegraphics[width=0.33\textwidth]{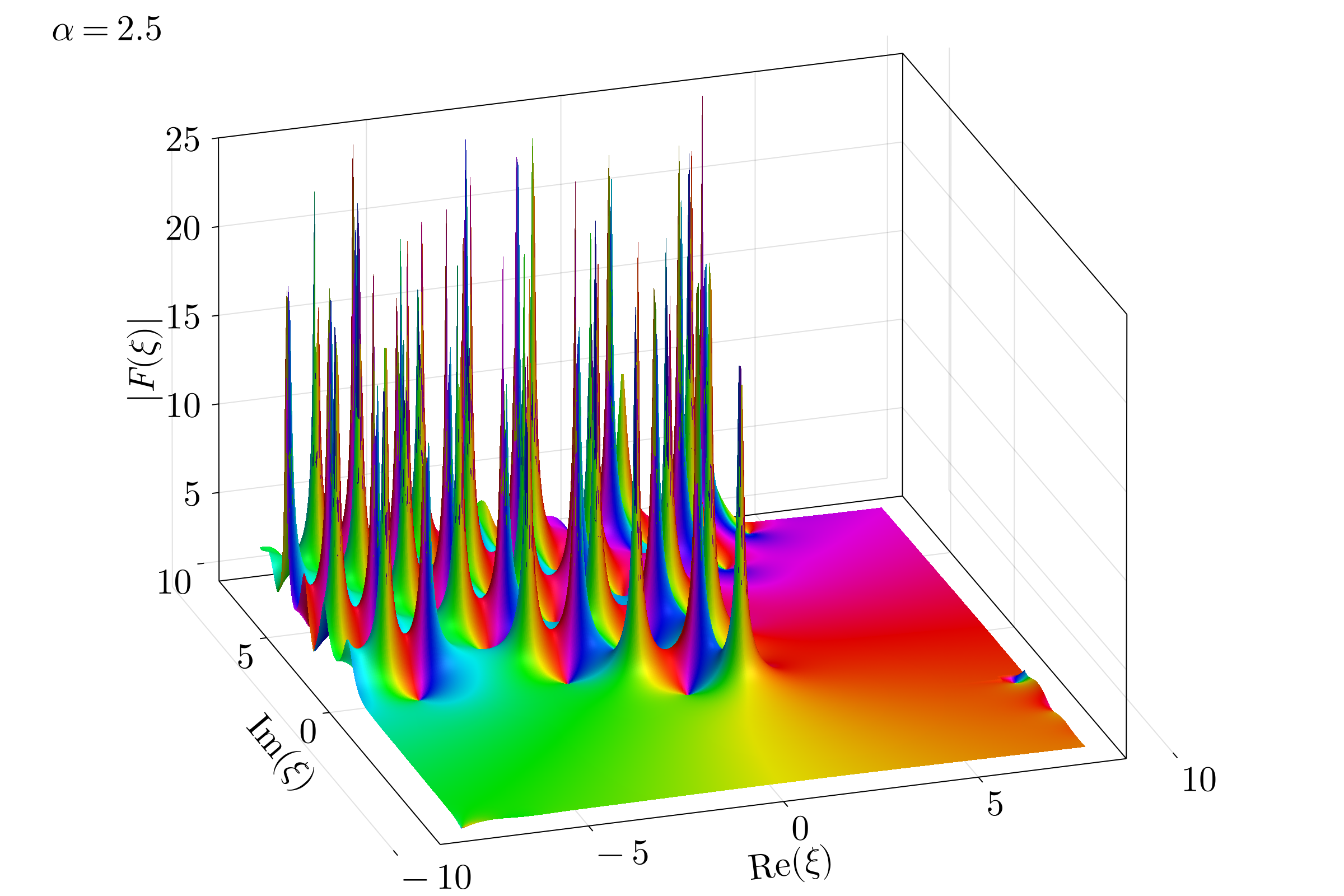}

\caption{Analytic landscape plots for numerical solutions of P$_{\mathrm{II}}$, (\ref{eq:pii_2})--(\ref{eq:K1equals0}), computed with $\alpha=0.5$, $1.5$ and $2.5$.  The surfaces are $|F(\xi)|$, while the colour scheme is the same as in figure~\ref{fig:daniel}.  The clear spikes in $|F(\xi)|$ correspond to simple poles.
}
\label{fig:daniel2}
\end{figure}

\begin{figure}
\centering
\includegraphics[width=0.6\textwidth]{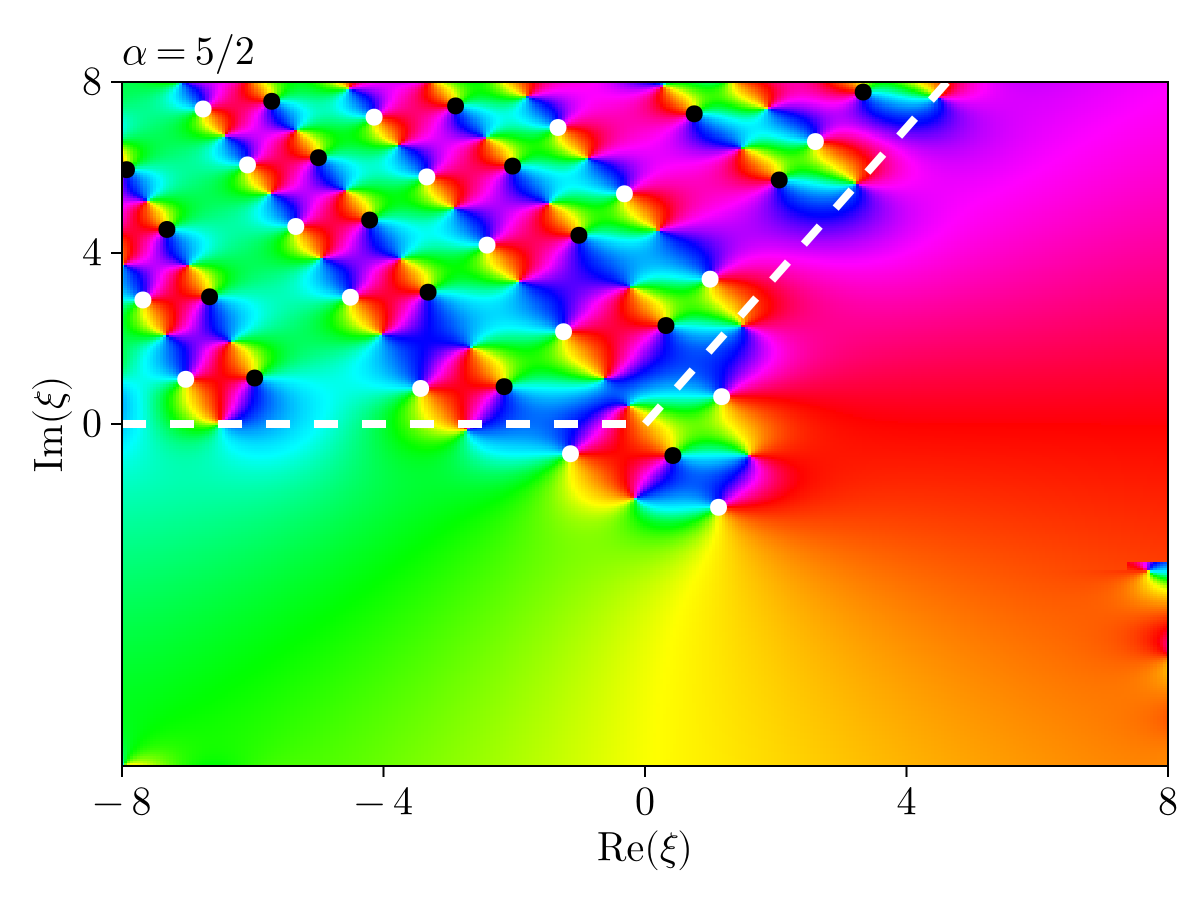}
\caption{Phase portrait of P$_{\mathrm{II}}$ solution for $\alpha=5/2$.  The white dashed lines indicate the boundary of the pole-free sector in the far field.  The white and black dots denote poles with residues $+1$ and $-1$, respectively.}
\label{fig:scott}
\end{figure}

For the solutions in figure~\ref{fig:daniel} for $\alpha\notin\mathbb{N}$, the lattice of poles in the far field is restricted to a sector of angle $2\pi/3$, which is why these are called tritronqu\'{e}e solutions.  The pole-free sector in the far field is $-\pi<\mathrm{arg}(\xi)<\pi/3$.  Another perspective of this lattice of poles is shown in figure~\ref{fig:daniel2}, where there are a few analytic landscape plots, from which it is easy to appreciate where the poles are located.  To help distinguish between poles with residue $+1$ or $-1$, we have added white and black dots to a phase portrait in figure~\ref{fig:scott}.  For this solution, drawn for $\alpha=5/2$, there are four poles (three with residue $+1$ and one with residue $-1$) that sit inside $-\pi<\mathrm{arg}(\xi)<\pi/3$.  As $\alpha$ increases, these ``move'' in the general direction $-\pi/3$ and ultimately are located at known positions when $\alpha=3$ (the three with residue $+1$ are the first, third and sixth poles in equation (\ref{eq:checkB3})).  Thus, we see that the property of a pole-free sector only holds in the far field, not the near field.  To be clear, while in this section we are discussing solutions of (\ref{eq:pii_2})--(\ref{eq:K1equals0}), when we apply the change of variables (\ref{eq:PIIreduction}), it is only the poles with residue $+1$ that are important for our solution $f_0$, and therefore it is only these poles that are relevant for the original KdV problem.

\subsection{Locating singularities of Painlev\'{e} II using transseries}\label{sec:trans}

As we see in figures~\ref{fig:PIIexample}--\ref{fig:scott}, for each solution of (\ref{eq:pii_2})--(\ref{eq:K1equals0}) with $\alpha\notin\mathbb{N}$, there is a lattice of simple-pole singularities in the upper half $\xi$ plane.
We are interested here in the string of singularities that align themselves in an almost-horizontal direction and tend towards the anti-Stokes line $\mathrm{Im}(\xi)=0$, $\mathrm{Re}(\xi)<0$ ($\mathrm{arg}(\xi)=-\pi$) in the far field.  These are important as they will have the strongest influence on the real-line solution $u(x,t)$ of (\ref{eq:kdv}) for $x<0$ in the small-time limit.  Indeed, our working hypothesis is that these singularities link directly to the important properties of dispersive waves, such as their speed and wavelength.  In this subsection, we summarise how to apply a standard transseries expansion to estimate the location of these singularities in the far field.  The approach we take relies on transseries and is equivalent to that of Costin \& Costin~\cite{Costin2001}, who briefly use a version of P$_{\mathrm{II}}$ as an example.  Other studies that apply transseries methodology for locating singularities of P$_{\mathrm{II}}$ are for the homogeneous version with $\alpha=0$, for which some of the details are different \cite{Baldino2023,Marino2008,Schiappa2014}.

The starting point is to note that the solution to (\ref{eq:pii_2})--(\ref{eq:K1equals0}) has the behaviour
\begin{align}
& F\sim \frac{\alpha}{\xi}+\frac{2\alpha(1-\alpha^2)}{\xi^4}+\ldots
+\frac{\sigma_1}{\xi^{1/4}}\,\mathrm{e}^{2\xi^{3/2}/3}
\quad\mbox{as}\quad |\xi|\rightarrow \infty, \quad -\pi<\mathrm{arg}(\xi)<-2\pi/3
\label{eq:farfieldFantiS}
\\
& \sigma_1\neq 0, \quad -\pi<\mathrm{arg}(\xi)<-2\pi/3,
\label{eq:K1noequals0}
\end{align}
where the Stokes constant $\sigma_1$ in (\ref{eq:farfieldFantiS})--(\ref{eq:K1noequals0}) is given explicitly by
\begin{equation}
\sigma_1=\frac{\sin\pi\alpha}{\pi^{1/2}}
\label{eq:sigma1}
\end{equation}
\cite{Fokas2006,Its2003}.  Clearly, the exponential term in (\ref{eq:farfieldFantiS}) is exponentially small in the sector $-\pi<\mathrm{arg}(\xi)<-2\pi/3$, and can be interpreted as the leading-order behaviour of the non-perturbative part of the asymptotic expansion.
(An abbreviated argument for (\ref{eq:sigma1}) is to see from (\ref{eq:largebn}) that the terms in the algebraic terms in the asymptotic series for $F$ behave as
$$
\frac{\alpha b_n}{\xi^{3n+1}}
\sim
-\mathrm{i}  \frac{\sin\pi\alpha}{\pi^{3/2}} \xi^{-1/4} \frac{\Gamma(2n+1/2)}{(-2\xi^{3/2}/3)^{2n+1/2}}
\quad\mbox{as}\quad n\rightarrow\infty,
$$
which is of the form $\mathcal{A}(\xi)\Gamma(2n+1/2)/\chi(\xi)^{2n+1/2}$, where $\chi$ is a singulant.  Therefore, the exponentially small term
$$
\pi\mathrm{i}\mathcal{A}\mathrm{e}^{-\chi}=\frac{\sin\pi\alpha}{\pi^{1/2}}\xi^{-1/4} \mathrm{e}^{2\xi^{3/2}/3}
$$
is switched on as we cross the Stokes line $\mathrm{arg}(\xi)=-2\pi/3$ in the clockwise direction.)

As we explain in appendix~\ref{sec:stocktake}, we can extend (\ref{eq:farfieldFantiS}) to a transseries of the form
\begin{equation}
F\sim \sum_{n=0}^\infty  \sigma_1^n \,\mathrm{e}^{2n\xi^{3/2}/3}
\sum_{m=0}^\infty \frac{F_m^{(n)}}{\xi^{-1/2+3n/4+3m/2}}
=
\sum_{n=0}^\infty\left(\frac{\sigma_1\,\mathrm{e}^{2\xi^{3/2}/3}}
{\xi^{3/4}}\right)^n \sum_{m=0}^\infty \frac{F_m^{(n)}}{\xi^{-1/2+3m/2}},
\label{eq:transseries1}
\end{equation}
where $F_{2m}^{(0)}=0$, $F_0^{(2n)}=0$, $F_0^{(1)}=1$ and, comparing with the notation in (\ref{eq:divergent}), $F_{2m+1}^{(0)}=\alpha b_m$.
The transseries (\ref{eq:transseries1}) is valid in the sector $-\pi<\mathrm{arg}(\xi)<-2\pi/3$, as each term $\mathrm{e}^{2n\xi^{3/2}/3}$ is exponentially smaller than the previous one as $n$ increases.

To proceed, we define the transseries variable $\tau$ by
\begin{equation}
\tau = \frac{\sigma_1\,\mathrm{e}^{2\xi^{3/2}/3}}{\xi^{3/4}}.
\label{eq:transvariable}
\end{equation}
The representation (\ref{eq:transseries1}) remains well-ordered provided $|\tau|\ll 1$, which is certainly true for $|\xi|\gg 1$ in the
sector $-\pi\leq \mathrm{arg}(\xi)<-2\pi/3$.
Thus, in this region we can swap the order of summation, so that
\begin{equation}
F\sim \sum_{m=0}^\infty \frac{1}{\xi^{-1/2+3m/2}}
\sum_{n=0}^\infty \tau^n F_m^{(n)},
\label{eq:transseries2}
\end{equation}
and, furthermore, in the neighbourhood of the anti-Stokes line $\mathrm{Im}(\xi)=0$, $\mathrm{Re}(\xi)<0$, we have
\begin{equation}
F\sim \sum_{m=0}^\infty \frac{G_m(\tau)}{\xi^{-1/2+3m/2}},
\label{eq:transseries3}
\end{equation}
where the $G_m(\tau)$ are functions of $\tau$ that for $-\pi\leq \mathrm{arg}(\xi)<-2\pi/3$ are given by the second (convergent) summation in (\ref{eq:transseries2}).  Crucially, the asymptotic representation (\ref{eq:transseries3}) applies regardless of whether $|\tau|$ is small, which means it applies as we cross the negative real $\xi$ axis from below (where $|\tau|$ is exponentially small) to above (where $|\tau|$ is exponentially large).  In particular, we shall apply (\ref{eq:transseries3}) in the neighbourhood of the string of singularities that align themselves slightly above the negative real $\xi$ axis for large $|\xi|$.

Thus, near the negative real $\xi$ axis, we substitute (\ref{eq:transseries3}) into P$_{\mathrm{II}}$ (\ref{eq:pii}) to derive odes for $G_m$, which are subject to boundary conditions that come from matching into the lower half plane, namely
\begin{equation}
G_m\sim F_m^{(0)}+F_m^{(1)}\tau+F_m^{(2)}\tau^2+\ldots
\quad\mbox{as}\quad \tau\rightarrow 0.
\label{eq:Gmmatching}
\end{equation}
We show in appendix~\ref{sec:termsGm} that the leading-order solution is
$$
G_0=\frac{4\tau}{4-\tau^2},
$$
which has singularities at $\tau = \tau_0 = \pm 2$ (corresponding to poles of P$_{\mathrm{II}}$ with residues -1 and 1, respectively).  Therefore, armed with this leading-order solution only, we can see that
\begin{equation}
F\sim \frac{4\tau\xi^{1/2}}{4-\tau^2} = \frac{4\sigma_1\xi^{-3/4}\,\mathrm{e}^{2\xi^{3/2}/3}\xi^{1/2}}
{4-\sigma_1^2\xi^{-3/2}\,\mathrm{e}^{4\xi^{3/2}/3}}
\quad\mbox{as}\quad |\xi|\rightarrow\infty,
\label{eq:firstapprox}
\end{equation}
and the first approximation for the location of the singularities as $|\xi|\rightarrow\infty$ is determined by the transcendental equation
\begin{equation}
\sigma_1\xi_0^{-3/4}\,\mathrm{e}^{2\xi_0^{3/2}/3}=\tau_0^{(0)}=\pm 2.
\label{eq:transcend}
\end{equation}
Rearranging, we can write
$$
\xi_0^{3/2}=3n\pi\mathrm{i}+\frac{3}{4}\log \left(\xi_0^{3/2}\right)
+\frac{3}{2}\log \left(\frac{\tau_0^{(0)}}{\sigma_1}\right).
$$
By interpreting $\mathrm{i}=\mathrm{e}^{-3\pi\mathrm{i}/2}$, we solve this equation asymptotically in the limit $|\xi_0|\rightarrow\infty$ to give
$$
\xi_0^{3/2}\sim 3n\pi\mathrm{i}+\frac{3}{4}\ln (3n\pi) - \frac{9\pi\mathrm{i}}{8}
+\frac{3}{2}\log \left(\frac{\tau_0^{(0)}}{\sigma_1}\right)
\quad\mbox{as}\quad n\rightarrow\infty,
$$
or, alternatively,
\begin{equation}
\xi_0\sim -(3\pi n)^{2/3}+\frac{1}{(3\pi n)^{1/3}}
\left(
\frac{3\pi}{4}+\frac{\mathrm{i}}{2}\ln (3\pi n)
+\mathrm{i}\log\left(\frac{\tau_0^{(0)}}{\sigma_1}\right)
\right)
\quad\mbox{as}\quad n\rightarrow\infty,
\label{eq:xi0largen}
\end{equation}
where $\tau_0^{(0)}=\pm 2$ and $n\in\mathbb{N}$.

\begin{figure}
\centering
\includegraphics[width=0.7  \textwidth]{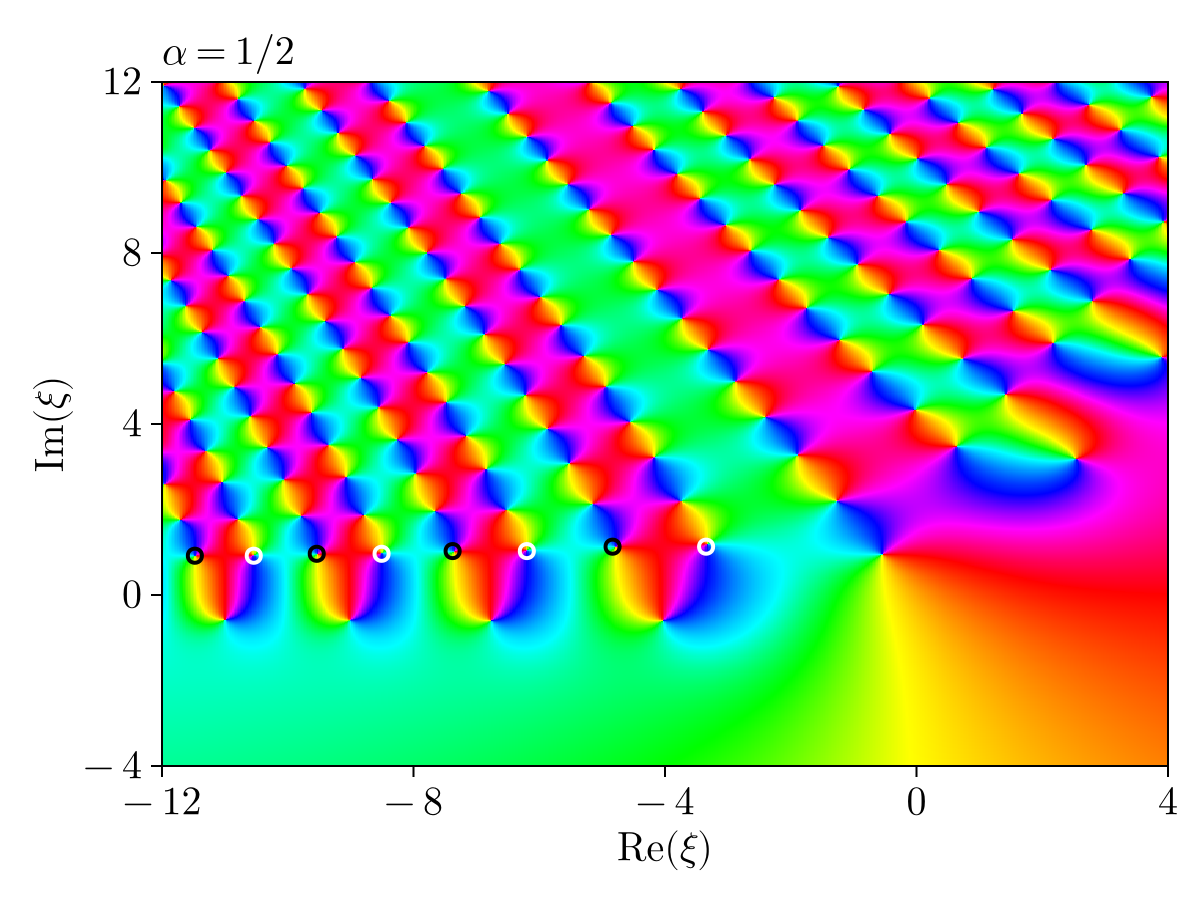}
\caption{Phase portrait  of a numerical solution of (\ref{eq:pii_2})--(\ref{eq:K1equals0}), computed for $\alpha=1/2$.  The black and white circles denote estimates of poles with residues $-1$ and $+1$ computed via (\ref{eq:xi0quotientpos}) and (\ref{eq:xi0quotientneg}), respectively.  The closest two circles to the origin are computed using $n=1$, the next closest pair with $n=2$, and so on.
}
\label{fig:polecheck}
\end{figure}

For our purposes, it is important to note that the Stokes constant $\sigma_1$ in (\ref{eq:sigma1}) is real, but may be positive or negative, depending on $\alpha$.  Further, $\sigma_1$ vanishes for $\alpha\in\mathbb{N}$.  Thus, given our two choices $\tau_0^{(0)}=\pm 2$, the term $\tau_0^{(0)}/\sigma_1$ can also be positive or negative, provided $\alpha\notin\mathbb{N}$.  With this in mind, we can rewrite (\ref{eq:xi0largen}) for the two separate cases as
\begin{equation}
\xi_0\sim -(3\pi n)^{2/3}+\frac{1}{(3\pi n)^{1/3}}
\left(\frac{3\pi}{4}
+\frac{\mathrm{i}}{2}\ln \left(\frac{12\pi^2 n}{\sin^2 \pi\alpha}\right)
\right)
\quad\mbox{as}\quad n\rightarrow\infty
\quad\mbox{for}\quad \tau_0^{(0)}/\sigma_1>0,
\label{eq:xi0quotientpos}
\end{equation}
\begin{equation}
\xi_0\sim -(3\pi n)^{2/3}+\frac{1}{(3\pi n)^{1/3}}
\left(-\frac{\pi}{4}
+\frac{\mathrm{i}}{2}\ln \left(\frac{12\pi^2 n}{\sin^2 \pi\alpha}\right)
\right)
\quad\mbox{as}\quad n\rightarrow\infty
\quad\mbox{for}\quad \tau_0^{(0)}/\sigma_1<0.
\label{eq:xi0quotientneg}
\end{equation}
For example, if we choose $\alpha=1/2$, then $\sigma_1>0$.  With this parameter choice, poles with $\tau_0^{(0)}=2$ (those with residue $-1$) are located using (\ref{eq:xi0quotientpos}), while poles with $\tau_0^{(0)}=-2$ (residue $+1$) are located using (\ref{eq:xi0quotientneg}).
We show this case in figure~\ref{fig:polecheck}, where we see the asymptotic formulae (\ref{eq:xi0quotientpos})--(\ref{eq:xi0quotientneg}) do an excellent job of estimating the pole locations for moderately small values of $n$ even though they hold formally in the limit $n\rightarrow\infty$.
(Note that by considering $G_1(\tau)$ in (\ref{eq:transseries3}), we can extend (\ref{eq:xi0quotientpos})--(\ref{eq:xi0quotientneg}) to include higher-order corrections.  See appendix~\ref{sec:C3}.)

We make two further observations about our asymptotic formulae (\ref{eq:xi0quotientpos})--(\ref{eq:xi0quotientneg}).  First, the poles alternate between residues $+1$ and $-1$ and, for very large $n$, are separated by an asymptotic distance $\pi^{2/3}/(3n)^{1/3}$.  Second, these formulae
assume that $\ln |\sigma_1|=\mathcal{O}(1)$.  For values of $\alpha$ extremely close to a natural number $N$, $\sigma_1$ becomes sufficiently small in magnitude that $\ln |\sigma_1|$ becomes large and these formulae need reconsidering.  Correspondingly, as $\alpha$ approaches $N$, the lattice of poles ``moves away'' from the origin, leaving behind a block of $N^2$ poles that are generated by the relevant rational solution for $\alpha=N$.

It is worth recalling that, in terms of our original KdV problem, double poles of solutions $u(x,t)$, located at $x=s(t)$, have the small-time behaviour (\ref{eq:scalingfors}).  Thus, we can use (\ref{eq:xi0quotientpos}) and (\ref{eq:xi0quotientneg}) for $\tau_0^{(0)}=-2$, to describe the leading-order location of double poles of $u(x,t)$ that propagate almost horizontally to the left like $s(t)\sim x_0-3\pi^{2/3}n^{2/3}t^{1/3}$.  Each of these can be associated with a crest of the dispersive waves on the real line, linking real-valued behaviour with poles in the complex plane.

\section{Check of early-time behaviour using $N$-soliton solutions} \label{sec:sechsquaredIC}

As a check on our small-time asymptotics, we consider the exact $N$-soliton solutions that comes from the initial condition (\ref{eq:ICsech}) with $A_0=-N(N+1)$.  For these examples, there are $N(N+1)/2$ double-pole singularities that emerge from each singularity of the initial condition at $x_0=(n+\frac{1}{2})\pi\mathrm{i}$.  While initially these move out from each $x_0$ in the directions and speeds given by the rational solutions of $f_0$ (see (\ref{eq:B1f0})--(\ref{eq:B4f0}), for example), they will ultimately rearrange themselves so that $N$ of them align with the fastest soliton, $N-1$ align with the next fastest, and so on, until only $1$ singularity from each $x_0$ is associated with the slowest soliton.

At this point it is worth repeating that we have used a change of variables (\ref{eq:PIIreduction}) to rewrite our inner problem for $f_0(\xi)$ in terms of solutions $F(\xi)$ of P$_{\mathrm{II}}$.  By inverting this transformation, we observe that, of all the simple poles of a solution for $F$, only those with residue $+1$ correspond to double poles of $f_0$; the simple poles of $F$ with reside $-1$ do not correspond to singularities of $f_0$ and therefore are not associated with singularities of our original KdV problem.  For the rational solutions of P$_{\mathrm{II}}$, there are $N^2$ simple poles, but only $N(N+1)/2$ of these are double poles of the corresponding rational solutions of $f_0$.

\subsection{$1$-soliton solution}\label{eq:1soliton}

An obvious example is a 1-soliton solution $u(x,t)=2\,\mathrm{sech}^2 (x-4t)$, which evolves from the initial condition $u_0=2\,\mathrm{sech}^2 x$, that is, from (\ref{eq:ICsech}) with $A_0=-2$.  The 1-soliton solution is simply a constant-shape travelling wave moving from left to right with speed $4$.  Clearly, the solution has poles where $\cosh(x-4t)=0$, which are located at
\begin{equation}
s(t)=\left(n+\sfrac{1}{2}\right)\pi\mathrm{i}+4t,
\label{eq:poles1soliton}
\end{equation}
where $n$ is an integer.  That is, a single pole emerges from each $x_0=\left(n+\sfrac{1}{2}\right)\pi\mathrm{i}$ at $t=0$ and travels horizontally in the complex plane with speed 4.  Another way to write the 1-soliton solution is
$$
u(x,t)=2\frac{\partial^2}{\partial x^2}\left(\ln U_1\right),
\quad
U_1=1+\mathrm{e}^{2x-8t}.
$$
With this representation, we see poles of $u$ can be determined via the zeros of $U_1$, giving the same result as (\ref{eq:poles1soliton}), as expected.

To compare with our small-time results, we refer to appendix~\ref{eq:A0m2}, since $A_0=-2$ is a special case.  Further, noting that this initial condition has poles at $x_0=\left(n+\sfrac{1}{2}\right)\pi\mathrm{i}$ with local behaviour
$$
u_0\sim \frac{-2}{(x-x_0)^2}+\frac{2}{3}
\quad\mbox{as}\quad x\rightarrow x_0,
$$
we see from (\ref{eq:ICAmore}) that $A_0=-2$, $A_1=0$ and $A_2=2/3$.  Following this very special case in appendix~\ref{sec:evenmorespecial}, our analysis predicts that the early time behaviour of singularity is described by (\ref{eq:veryspecialcase}), namely $s(t)\sim x_0 +6A_2t$ as $t\rightarrow 0^+$. Given $A_2=2/3$, this predicted speed is $6A_2=4$, which agrees with the exact solution for the 1-soliton solution.  Thus, while the 1-soliton solution itself is trivial in the sense that we can easily determine the small-time behaviour of the complex singularities without any of our asymptotic analysis, this check of (\ref{eq:veryspecialcase}) provides support for the details in appendix~\ref{eq:A0m2} (which apply for other, more complicated, initial conditions that also have $A_0=-2$).

\subsection{$2$-soliton solution}\label{sec:2soliton}

The simplest less trivial example with an exact solution is a $2$-soliton solution (see \cite{Ablowitz1991,Drazin1989} and the references therein)
\begin{equation}
u(x,t)=12\frac{3+4\cosh(2x-8t)+\cosh(4x-64t)}{(3\cosh(x-28t)+\cosh(3x-36t))^2},
\label{eq:2soliton}
\end{equation}
which evolves from the initial condition
\begin{equation}
u_0=6\,\mathrm{sech}^2x.
\label{eq:6sech}
\end{equation}
This solution consists of two solitons moving to the right, with the long-time limit \cite{Ablowitz1991,Drazin1989}
$$
u(x,t)\sim 2\,\mathrm{sech}^2(x-4t+\sfrac{1}{2}\ln 3)
+8\,\mathrm{sech}^2(2x-32t-\sfrac{1}{2}\ln 3)
\quad\mbox{as}\quad t\rightarrow\infty.
$$
That is, for sufficiently large time, the solution is dominated by a larger soliton moving with speed 16 and a smaller soliton moving with speed 4.  To compare with our small-time analysis in section~\ref{sec:A0neqminus2}, we note that the initial condition (\ref{eq:6sech}) has poles at $x_0=\left(n+\sfrac{1}{2}\right)\pi\mathrm{i}$ and is a member of (\ref{eq:ICsech}) with $A_0=-6$.

Another way of writing (\ref{eq:2soliton}) is
$$
u(x,t)=2\frac{\partial^2}{\partial x^2}\left(\ln U_2\right),
\quad U_2=1+3\mathrm{e}^{2x-8t}+3\mathrm{e}^{4x-64t}+\mathrm{e}^{6x-72t},
$$
so that clearly the singularities of $u$ can be determined by setting $x=s(t)$ and solving for zeros of $U_2$.  By letting $\gamma=\mathrm{e}^{2s}$, this exercise reduces to finding roots of the cubic \cite{Drazin1989}
$$
U_2=1+3\gamma\mathrm{e}^{-8t}+3\gamma^2\mathrm{e}^{-64t}+\gamma^3\mathrm{e}^{-72t}.
$$
By using a symbolic manipulation package (Maple, for example) or otherwise, we can use the cubic formula to derive complicated expressions for the roots $\gamma$ and then determine the small-time behaviour using the appropriate series command.  We find the three behaviours (one for each solution of the cubic)
\begin{align*}
\gamma\sim & -1+2^{5/3}(3t)^{1/3}+2^{7/3}(3t)^{2/3}-8t, \\
& -1-2^{5/3}\left(\frac{1}{2}+\frac{\sqrt{3}}{2}\mathrm{i}\right)(3t)^{1/3}
-2^{7/3}\left(\frac{1}{2}-\frac{\sqrt{3}}{2}\mathrm{i}\right)(3t)^{2/3}-8t, \\
& -1-2^{5/3}\left(\frac{1}{2}+\frac{\sqrt{3}}{2}\mathrm{i}\right)(3t)^{1/3}
-2^{7/3}\left(\frac{1}{2}+\frac{\sqrt{3}}{2}\mathrm{i}\right)(3t)^{2/3}-8t.
\end{align*}
For each of these we expand $s=(\ln\gamma)/2$ for small time to give the three options
\begin{equation}
s(t)\sim \left(n+\sfrac{1}{2}\right)\pi\mathrm{i}-2^{2/3}(3t)^{1/3}+12t,
\quad
\left(n+\sfrac{1}{2}\right)\pi\mathrm{i}+ 2^{2/3}\left(\frac{1}{2}\pm\frac{\sqrt{3}}{2}\mathrm{i}\right)(3t)^{1/3}+12t
\label{eq:2solitonsingularity}
\end{equation}
as $t\rightarrow 0^+$.

\begin{figure}
\centering
\subfloat[2-soliton solution]{
\includegraphics[width=0.9\textwidth]{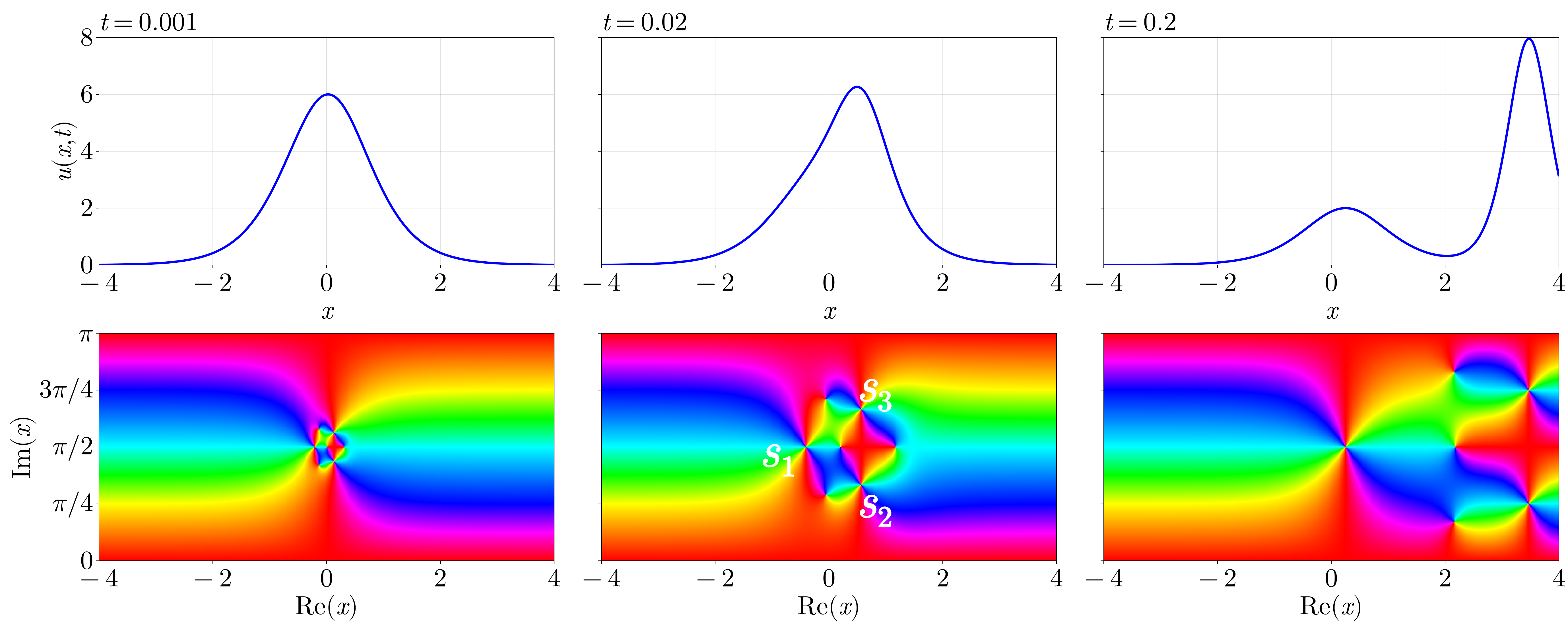}
}

\subfloat[3-soliton solution]{
\includegraphics[width=0.9\textwidth]{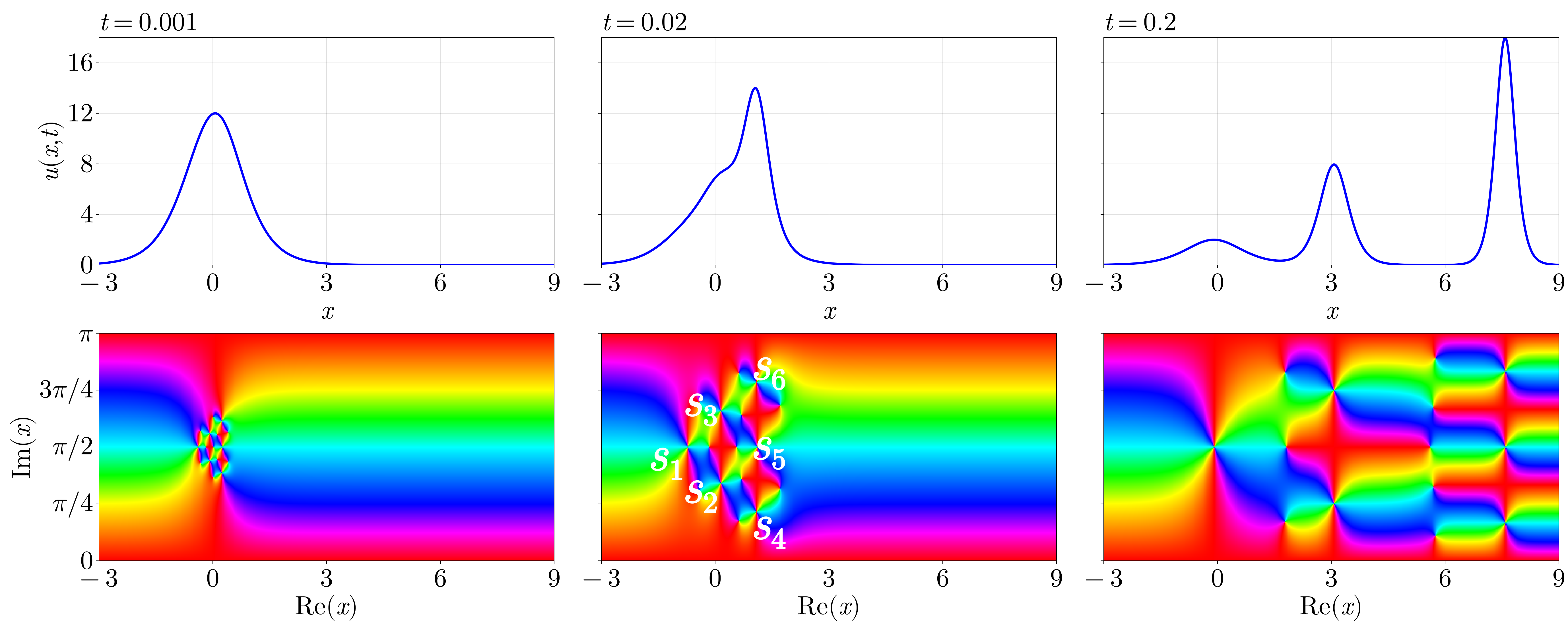}
}
\caption{Solution profiles and phase portraits for (a) the $2$-soliton solution (\ref{eq:2soliton}), which evolves from $u(x,0)=6\,\mathrm{sech}^2x$ and (b) the $3$-soliton solution (\ref{eq:3soliton}), which evolves from $u(x,0)=12\,\mathrm{sech}^2x$.}
\label{fig:twothreesoliton}
\end{figure}

The $\mathcal{O}(t^{1/3})$ part of (\ref{eq:2solitonsingularity}) agrees with our prediction (\ref{eq:checkB2}), that comes from our inner problem for the case $A_0=-6$ (and involves the rational solution of P$_{\mathrm{II}}$ for $\alpha=2$).  The $\mathcal{O}(t)$ terms in (\ref{eq:2solitonsingularity}) are all $12t$; this result agrees with our prediction (\ref{eq:predictA0min6A2equal2}) from appendix~\ref{eq:A0neqm2A1eq0} (found by considering the effect of higher-order terms in (\ref{eq:ICA})).  Further, these $\mathcal{O}(t)$ terms in (\ref{eq:2solitonsingularity}) are consistent with the well-known results that, for the 2-soliton solution, there are three singularities that initially move out of each of $x_0=\left(n+\sfrac{1}{2}\right)\pi\mathrm{i}$ in three equally spaced directions but then all begin to move to the right.  More specifically, taking the closest $x_0$ to the real-$x$ axis as an example ($n=0$), we can label
$$
s_1(t)\sim \sfrac{1}{2}\pi\mathrm{i}-2^{2/3}(3t)^{1/3}+12t
$$
as $t\rightarrow 0^+$, which is associated with the late-time soliton $2\,\mathrm{sech}^2(x-4t+\sfrac{1}{2}\ln 3)$ and therefore has the asymptotic behaviour $s_1(t)\sim \sfrac{1}{2}\pi\mathrm{i}
-\sfrac{1}{2}\ln 3 + 4t$ as $t\rightarrow\infty$.  Further, we can label the singularities
$$
s_2(t)\sim \sfrac{1}{2}\pi\mathrm{i}+ 2^{2/3}\left(\frac{1}{2}-\frac{\sqrt{3}}{2}\mathrm{i}\right)(3t)^{1/3}+12t,
\quad
s_3(t)\sim \sfrac{1}{2}\pi\mathrm{i}+ 2^{2/3}\left(\frac{1}{2}+\frac{\sqrt{3}}{2}\mathrm{i}\right)(3t)^{1/3}+12t,
$$
as $t\rightarrow 0^+$, so that both $s_2(t)$ and $s_3(t)$ are associated with the late-time soliton $8\,\mathrm{sech}^2(2x-32t-\sfrac{1}{2}\ln 3)$ and have the asymptotic behaviours $s_2(t)\sim \sfrac{1}{4}\pi\mathrm{i}
+\sfrac{1}{4}\ln 3 + 16t$, $s_3(t)\sim \sfrac{3}{4}\pi\mathrm{i}
+\sfrac{1}{4}\ln 3 + 16t$ as $t\rightarrow\infty$.  We illustrate this behaviour in figure~\ref{fig:twothreesoliton}(a), where snapshots of the $2$-soliton solution are provided at times $t=0.001$, $0.02$ and $0.2$.  The accompanying phase portraits show how the singularities first move away from $x=\pi\mathrm{i}/2$ in different directions, as described above, but ultimately all propagate to the right and align themselves with the peaks of the relevant soliton.

Another check on our 2-soliton solution is to put $x=\sfrac{1}{2}\pi\mathrm{i}+(3t)^{1/3}\xi$ into (\ref{eq:2soliton}) and then take the limit $t\rightarrow 0^+$.  The result is that
$$
u(x,t)\sim -\frac{1}{(3t)^{2/3}}\frac{6\xi(\xi^3-8)}{(\xi^3+4)^2}+\frac{2\xi^9-24\xi^6+1440\xi^3-640}{(\xi^3+4)^3}
\quad\mbox{as}\quad t\rightarrow 0^+.
$$
The leading-order part here agrees with the rational solution computed for $\alpha=2$, namely (\ref{eq:B2f0}), as expected.  The correction term can be checked by showing it is a solution of the appropriate ode problem (\ref{eq:odef2ifA10}) with (\ref{eq:farfieldf2more}), the details of which are discussed in appendix~\ref{eq:A0neqm2A1eq0}.  In summary, by carefully studying the exact solution for the $2$-soliton solution we are able to check our matched asymptotics, including the correction terms, providing additional confidence that our analysis is correct.

\subsection{$3$-soliton solution}\label{sec:3soliton}

A more complicated example is a $3$-soliton solution that evolves from the initial condition
\begin{equation}
u_0=12\,\mathrm{sech}^2x,
\label{eq:12sech}
\end{equation}
and can be represented by
$$
u(x,t)=2\frac{\partial^2}{\partial x^2}\left(\ln U_3\right),
$$
$$
U_3=1+6\mathrm{e}^{2x-8t}+15\mathrm{e}^{4x-64t}+10\mathrm{e}^{6x-72t}
+10\mathrm{e}^{6x-216t}+15\mathrm{e}^{8x-224t}+6\mathrm{e}^{10x-280t} +\mathrm{e}^{12x-288t}.
$$
By performing the two partial derivatives and simplifying, we can show that another version of this $3$-soliton solution is
\begin{equation}
u(x,t)=H_1/H_2^2,
\label{eq:3soliton}
\end{equation}
where
$$
H_1=24[126+50\cosh(2x-8t)+135\cosh(2x-56t)+25\cosh(2x-152t)
+80\cosh(4x-64t)+40\cosh(4x-208t)
$$
$$
+15\cosh(6x-72t)+30\cosh(6x-216t)+10\cosh(8x-224t)
+\cosh(10x-280t)],
$$
$$
H_2=15\cosh(2x-80t)+6\cosh(4x-136t)+\cosh(6x-144t)+10\cosh(72t)
$$
(see \cite{Shen2012}).  This solution involves three solitons moving to the right, with the long-time limit
$$
u(x,t)\sim 2\,\mathrm{sech}^2(x-4t+\sfrac{1}{2}\ln 6)
+8\,\mathrm{sech}^2(2x-32t+\sfrac{1}{2}\ln \sfrac{5}{3})
+18\,\mathrm{sech}^2(3x-108t-\sfrac{1}{2}\ln 10)
\quad\mbox{as}\quad t\rightarrow\infty.
$$
In other words, for late times, the smallest soliton is moving with speed $4$, the middle soliton is moving with speed $16$, while the largest soliton is moving with speed $36$.

To determine the small-time behaviour or complex singularities for the $3$-soliton solution, we set $x=s(t)$ and solve for the zeros of $U_3$, which becomes a sixth-order polynomial in terms of $\gamma=\mathrm{e}^{2s}$.  Leaving out the details, by concentrating on singularities that emerge from $x_0=\sfrac{1}{2}\pi\mathrm{i}$, we have the six small-time behaviours
$$
s_1(t)\sim \sfrac{1}{2}\pi\mathrm{i}-(10+6\sqrt{5})^{1/3}(3t)^{1/3}
+24t,
\quad
s_2(t)\sim \sfrac{1}{2}\pi\mathrm{i}+(-10+6\sqrt{5})^{1/3} \left(-\frac{1}{2}-\frac{\sqrt{3}}{2}\mathrm{i}\right)
(3t)^{1/3}
+24t,
$$
$$
s_3(t)\sim \sfrac{1}{2}\pi\mathrm{i}+(-10+6\sqrt{5})^{1/3} \left(-\frac{1}{2}+\frac{\sqrt{3}}{2}\mathrm{i}\right)
(3t)^{1/3}
+24t,
\,\,
s_4(t)\sim \sfrac{1}{2}\pi\mathrm{i}+(10+6\sqrt{5})^{1/3} \left(\frac{1}{2}-\frac{\sqrt{3}}{2}\mathrm{i}\right)
(3t)^{1/3}
+24t,
$$
$$
s_5(t)\sim \sfrac{1}{2}\pi\mathrm{i}+(-10+6\sqrt{5})^{1/3}(3t)^{1/3}
+24t,
\quad
s_6(t)\sim \sfrac{1}{2}\pi\mathrm{i}+(10+6\sqrt{5})^{1/3} \left(\frac{1}{2}+\frac{\sqrt{3}}{2}\mathrm{i}\right)
(3t)^{1/3}
+24t,
$$
as $t\rightarrow 0^+$.  With this labelling, one of the six singularities, namely $s_1$, is associated with the slowest soliton, with $s_1(t)\sim \sfrac{1}{2}\pi\mathrm{i}
-\sfrac{1}{2}\ln 6 + 4t$ as $t\rightarrow\infty$; two of the singularities, $s_2$ and $s_3$, are associated with the intermediate-speed soliton, with $s_2(t)\sim \sfrac{1}{4}\pi\mathrm{i} -\sfrac{1}{4}\ln \sfrac{5}{3} + 16t$ and $s_3(t)\sim \sfrac{3}{4}\pi\mathrm{i} -\sfrac{1}{4}\ln \sfrac{5}{3} + 16t$ as $t\rightarrow\infty$; and the remaining three singularities are associated with the fastest soliton, with
$s_4(t)\sim \sfrac{1}{6}\pi\mathrm{i} +\sfrac{1}{6}\ln 10 + 36t$,
$s_5(t)\sim \sfrac{1}{2}\pi\mathrm{i} +\sfrac{1}{6}\ln 10 + 36t$, and
$s_6(t)\sim \sfrac{5}{6}\pi\mathrm{i} +\sfrac{1}{6}\ln 10 + 36t$ as $t\rightarrow\infty$.  This $3$-soliton solution and the singularity dynamics are illustrated in figure~\ref{fig:twothreesoliton}(b).  As with the $2$-soliton example, we see in the phase portraits how the singularities initial spread out from $x=\pi\mathrm{i}/2$ before propagating to the right and aligning with a soliton.

To compare these results for the $3$-soliton solution with our analysis in section~\ref{sec:A0neqminus2}, we note that (\ref{eq:12sech}) is a member of (\ref{eq:ICsech}) with $A_0=-12$.  Thus, the relevant leading-order results are (\ref{eq:checkB31})--(\ref{eq:checkB32}), which come from the roots of $f_0$ when $\alpha=3$ (that are derived via the rational solution of P$_{\mathrm{II}}$ for $\alpha=3$).  The $\mathcal{O}(t^{1/3})$ parts of $s_1(t)$ to $s_6(t)$ above agree with  (\ref{eq:checkB31})--(\ref{eq:checkB32}), as expected, but nevertheless this comparison acts as an important check on our leading-order asymptotic analysis.   Further, the $\mathcal{O}(t)$ terms in $s_1(t)$ to $s_6(t)$ above are all $24t$.  These terms agree with the prediction (\ref{eq:predictA0min12A2equal4}) that comes from considering higher-order effects in appendix~\ref{eq:A0neqm2A1eq0}.

As an additional check, we set $x=\sfrac{1}{2}\pi\mathrm{i}+(3t)^{1/3}\xi$ in (\ref{eq:3soliton}) and take the limit $t\rightarrow 0^+$, to give
$$
u(x,t)\sim -\frac{1}{(3t)^{2/3}}\frac{12\xi(\xi^9+600\xi^3+1600)}{(\xi^6+20\xi^3-80)^2}
$$
$$
+\frac{4(\xi^{18} + 12\xi^{15} + 2880\xi^{12} - 136000\xi^9 - 1075200\xi^6 - 8064000\xi^3 - 3584000)}{(\xi^6+20\xi^3-80)^3}
$$
as $t\rightarrow 0^+$.  The leading-order part of this expression agrees with the rational solution of $f_0$ for $\alpha=3$, given by (\ref{eq:B3f0}), as required.  Further, the correction term can be shown to satisfy the appropriate problem (\ref{eq:odef2ifA10}) with (\ref{eq:farfieldf2more}), as mentioned in appendix~\ref{eq:A0neqm2A1eq0}.  These tests, including those for higher-order terms, again provide confidence that our matched asymptotics is correct.

\section{Discussion}\label{sec:discussion}

Solutions of the KdV equation (\ref{eq:kdv}) are well known to exhibit dispersive waves that travel in the negative-$x$ direction.  For initial conditions that are analytic functions of $x$, we expect the amplitude of these waves to be exponentially small in time.  Therefore, they cannot be captured by a naive algebraic expansion in powers of $t$, but instead can be described by considering exponentially-small corrections to the algebraic series and employing Stokes phenomenon in complex-$x$ plane to observe where the waves appear on the real-$x$ axis.  One of the goals of our study is to derive the asymptotic form of these waves in the small-time limit and establish connections between this real-valued behaviour and certain singularity structures in the complex-$x$ plane.

For the purposes of our study, a key observation is that (the analytic continuation of) solutions of the KdV equation (\ref{eq:kdv}) have singularities in the complex plane that are all double poles with a strength $-2$ (i.e., (\ref{eq:doublepoleleadingorder})).  Therefore, we have been motivated to study pole dynamics for the KdV equation with a class of initial conditions that also have double poles, each with a strength $A_0$, where in general $A_0\neq -2$ (see (\ref{eq:doublepoleleadingorder})).  By applying techniques in exponential asymptotics in the complex plane, we show that dispersive waves behave like (\ref{eq:expdispersive}) on the real line in the small-time limit (see also (\ref{eq:farfielddispersivewaves})).  The dependence of (\ref{eq:expdispersive}) on  $A_0$ and $x_0$ demonstrates how the strength and location of the double-pole singularities of the analytic continuation of the initial condition have an explicit influence on crucial real-valued wave-like behaviour.  Further, without any reference to inverse-scattering techniques or the Painlev\'{e} II equation, our exponential asymptotics is able to identify the exceptional nature of the triangular values $A_0=-N(N+1)$ in the $\mathrm{sech}^2$-type initial conditions (\ref{eq:ICsech}).  More generally, these asymptotic results highlight the role that applied complex analysis can play in analysing real-valued pde solutions.

Next, we have used matched asymptotic expansions to describe how the double poles of KdV emerge from the double poles of the initial condition, and how they propagate for small time.  In the neighbourhood of a double pole of the initial condition, $x=x_0\notin\mathbb{R}$, we show that the early-time dynamics is driven by a solution of the P$_{\mathrm{II}}$ equation (\ref{eq:pii}), written in terms of the similarity variable $\xi=(x-x_0)/(3t)^{1/3}$, whose constant $\alpha$ is related to $A_0$ via (\ref{eq:BA02}).  This P$_{\mathrm{II}}$ problem comes from writing $u=(3t)^{-2/3}f(\xi,t)$, taking the leading-order term $f\sim f_0(\xi)$ as $t\rightarrow 0^+$, and making the change of variable (\ref{eq:PIIreduction}).
While there are well-known links between KdV (and modified KdV) and P$_{\mathrm{II}}$, for example via large-time asymptotics or the small-dispersion limit, these typically involve tronqu\'{e}e solutions of
the homogeneous version of P$_{\mathrm{II}}$ (i.e., with $\alpha=0$) that are real for real values of the independent variable \cite{Claeys2009,Hastings1980,Segur1981}, whereas our situation is different because it involves (decreasing) {\em tri}\,tronqu\'{e}e solutions of the {\em in}homogeneous version of P$_{\mathrm{II}}$ (i.e., with $\alpha\neq 0$) that is complex on the real line.

Based on the inner problem centred on each $x=x_0$, we conjecture that, generically, an infinite number of double poles emerge from $x=x_0$.  Each one of these poles initially moves like $s(t)\sim x_0 + (3t)^{1/3}\xi_0$ as $t\rightarrow 0^+$, where $\xi_0$ is one of the poles of P$_{\mathrm{II}}$ with residue $+1$ (the poles of P$_{\mathrm{II}}$ with residue $-1$ do not correspond to double poles of KdV).  Given the tritronqu\'{e}e structure, this implies that there is a lattice of such poles in the $\xi$ plane, bounded by two strings of poles that tend towards the anti-Stokes lines at $\theta=\pi/3$ and $\theta=-\pi$. It appears that the poles that line up close to $\theta=-\pi$ will have the strongest effect on the dispersive waves in the original real-valued problem (\ref{eq:kdv}).  We can, in principle, compute more terms for the location of each singularity $x=s(t)$ in the small-time limit by considering higher-order correction terms $f\sim f_0(\xi)+(3t)^{1/3}f_1(\xi) +(3t)^{2/3}f_2(\xi)$ to give $s(t)\sim x_0 + (3t)^{1/3}\xi_0+(3t)^{2/3}\xi_1+(3t)\xi_2$ as $t\rightarrow 0^+$, where $\xi_1$ comes from solving the linear (but complicated) problem for $f_1$, $\xi_2$ comes from the (even more complicated) problem for $f_2$, and so on.  The problems for $f_1$, $f_2$, $\ldots$, depend on the parameters $A_1$, $A_2$, $\ldots$, in (\ref{eq:ICAmore}), which ultimately links all of these small-time results to the local expansion of the singularity of the initial condition (\ref{eq:ICAmore}).

We have applied some formal techniques in transseries asymptotics to approximate the location of the string of singularities for our P$_{\mathrm{II}}$ problem that tend to the negative real $\xi$ axis as $|\xi|\rightarrow\infty$.  To leading order, these results imply that corresponding poles of KdV move like $s(t)\sim x_0 - 3\pi^{2/3}n^{2/3}t^{1/3}$ as $t\rightarrow 0^+$, $n\rightarrow\infty$, where $n$ is a natural number.  On the other hand, our exponential asymptotics suggest that the crests of the dispersive waves also move like $\mathrm{Re}(x_0)- 3\pi^{2/3}n^{2/3}t^{1/3}$ on the real line (see (\ref{eq:crests})).  Clearly, there is a link between certain complex singularities of a solution of the KdV equation and the wavelike behaviour of that solution on the real line.

We have identified a number of special cases.  For example, when the parameter $\alpha=N$ for $N\in \mathbb{N}$, our tritronqu\'{e}e solutions of P$_{\mathrm{II}}$, with an infinite number of poles, reduce to rational solutions, which have a finite number of poles.  These solutions are relevant for $\mathrm{sech}^2$-type initial conditions (\ref{eq:ICsech}), where the prefactor $-A_0$ is a triangular number $N(N+1)$.  We have shown that for $2$- and $3$-soliton solutions, our asymptotics agree with the exact solutions.  Other special cases come from $A_0=-2$, for which the singularities move with a different temporal scaling.  For example, if an initial condition has a double pole at $x=x_0$ with local behaviour (\ref{eq:ICAmore}), where $A_0=-2$ but $A_1\neq 0$, then we expect a singularity to emerge from $x_0$ with the scaling $s(t)= x_0+\mathcal{O}(t^{2/3})$ (see (\ref{eq:finalresultA1neq0})).  Further, if $A_0=-2$, $A_1= 0$, $A_2\neq 0$, then the scaling changes to $s(t)= x_0+\mathcal{O}(t)$ (see (\ref{eq:veryspecialcase})).

We regard our work as a preliminary study that has left a number of unresolved issues and suggests numerous open problems worth attention.  We list some of these here in the remainder of this section.
\begin{myitemize}
\item The exponential asymptotics detailed in subsection~\ref{sec:expasympt_main} and appendix~\ref{sec:expasympt_app} provides leading order behaviour of the dispersive waves for the generic case $A_0\neq -N(N+1)$ as well as the special case $A_0 = -N(N+1)$, the latter depending on the coefficient of the correction term in (\ref{eq:ICAmore}), namely $A_1$.  It remains to derive the correction terms for the generic case $A_0\neq -N(N+1)$, which will involve both $A_0$ and $A_1$.  This type of higher-order analysis for this nonlinear problem within the context of exponential asymptotics is likely to be nontrivial.

\item Much of the methodology in subsection~\ref{sec:expasympt_main} and appendix~\ref{sec:expasympt_app} should carry over to the linear KdV problem
    $$
    u_t+u_{xxx}=0,
\quad u(x,0)=u_0(x),\quad x\in\mathbb R,
    $$
    which is not surprising as dispersive waves are exponentially small in the small-time limit and therefore the nonlinear term in the full KdV equation does not play a role in deriving the form of the late-order terms (\ref{eq:lateorder}) nor the exponentially-small contributions (\ref{eq:expsmall}).  The matching into the inner region near $x=x_0$ will be different for the linear problem, however, since the function $f_0$ will satisfy the linear ode
    $$
    -2f_0-\xi \frac{\mathrm{d}f_0}{\mathrm{d}\xi} +\frac{\mathrm{d}^3f_0}{\mathrm{d}\xi^3}=0
    $$
    instead of (\ref{eq:odef0}).  Therefore, applying the series (\ref{eq:seriesf0}), the recurrence relation (\ref{eq:an}) will be $a_n= (3n-1)(3n+1)a_{n-1}$, which can be solved easily to evaluate the limit (\ref{eq:lambda1}), suggesting
    $\Lambda=\mathrm{i}A_0/3^{3/4}\sqrt{\pi}$ and
\begin{equation}
U_{\mbox{\footnotesize dis}}\sim
-\frac{2\sqrt{\pi}A_0}{3^{3/4}}
\frac{(-x)^{1/4}\,\mathrm{e}^{-y_0(-x/3t)^{1/2}}}{t^{3/4}}
\,\cos\left(\frac{2(-x)^{3/2}}{3(3t)^{1/2}}-\frac{\pi}{4}
\right)
\label{eq:farfielddispersivewaveslinear}
\end{equation}
as $x\rightarrow -\infty$, $t\rightarrow 0^+$ (cf.~(\ref{eq:farfielddispersivewaves})).  The claim (\ref{eq:farfielddispersivewaveslinear}) could be checked using integral transforms and appropriate limits and then expanded upon to hold for different initial conditions other than those with the property (\ref{eq:ICA}).

\item For the well-studied initial conditions of the form (\ref{eq:ICsech}), it remains to explain how the singularity structure for (\ref{eq:pii_2})--(\ref{eq:K1equals0}) changes as the parameter $A_0$ varies from slightly greater than $A_0=-N(N+1)$ for a natural number $N$ ($\alpha<N$) to slightly less than $-N(N+1)$ ($\alpha>N$), and how these changes affect qualitative behaviour for KdV on the real line.  For example, in figure~\ref{fig:daniel3} we show three phase portraits illustrating solutions of (\ref{eq:pii_2})--(\ref{eq:K1equals0}) for $\alpha=2.99$, $3$ and $3.01$.  It is not clear how the lattice of infinitely many poles are arranged in the two limits $\alpha\rightarrow 3^-$ and $\alpha\rightarrow 3^+$, given only nine are present for the borderline case $\alpha=3$.

    Furthermore, with the special initial condition (\ref{eq:ICsech}), when $A_0$ is slightly greater than $-12$ ($\alpha\lesssim 3$), there will be two solitons that eventually evolve to the right, while for $A_0=-12$ ($\alpha=3$) or slightly less than $-12$ ($\alpha\gtrsim 3$), there will be three.  Therefore, while we can speculate that the pole dynamics for the solitons when $A_0$ is slightly less than $-12$ is qualitatively similar to that for $A_0=-12$ (discussed in subsection~\ref{sec:3soliton}), it is not at all clear what happens for $A_0$ slightly greater than $-12$.

\begin{figure}
\centering
\includegraphics[width=0.9  \textwidth]{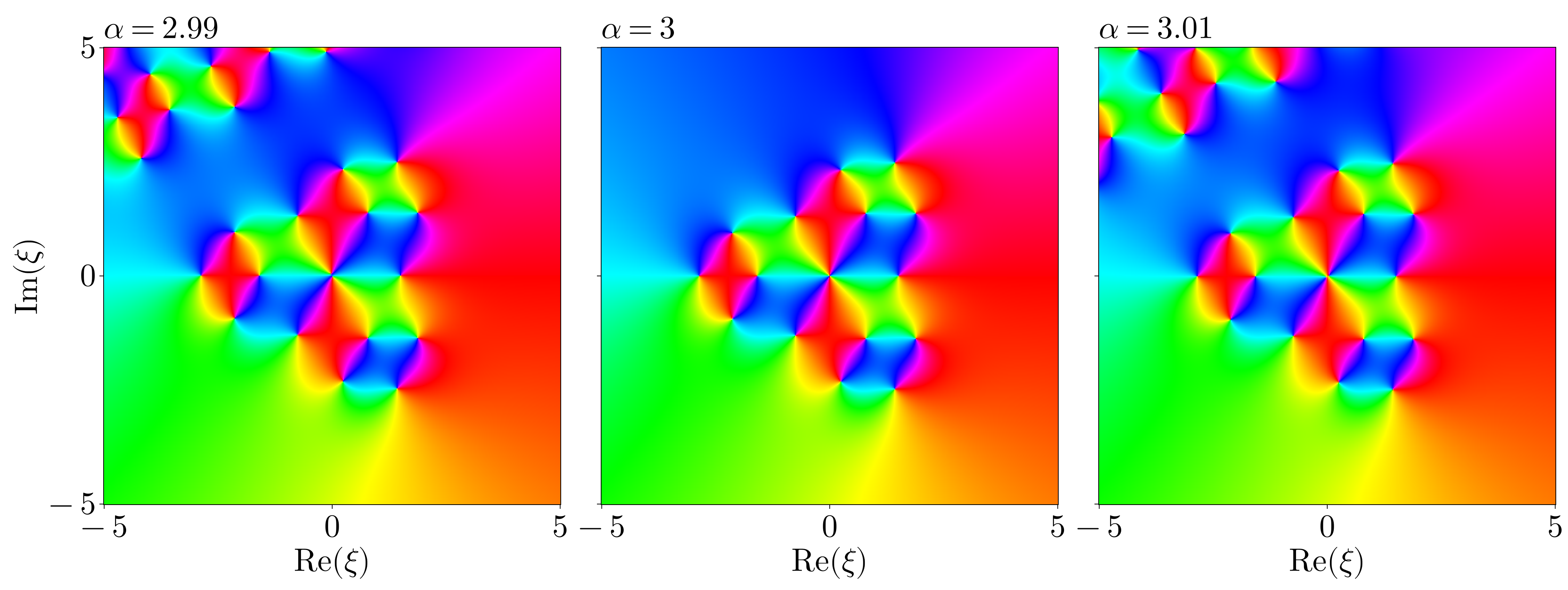}
\caption{Phase portraits of numerical solutions of (\ref{eq:pii_2})--(\ref{eq:K1equals0}) computed for  $\alpha=2.99$ (left), $3$ (centre) and $3.01$ (right).
}
\label{fig:daniel3}
\end{figure}

\item For initial conditions with singularities of the form (\ref{eq:ICA}), we have not explored the parameter range $A_0>1/4$, which corresponds to complex values of $\alpha$.  This parameter range would be relevant for ``dip''-type versions of (\ref{eq:ICsech}) or (\ref{eq:ICxsquared}).  It would be interesting to study relevant solutions of P$_{\mathrm{II}}$, which may behave differently to the cases in which $\alpha$ is real-valued.

\item While the asymptotic methods used in subsection~\ref{sec:trans} are able to estimate the location of the important singularities of (\ref{eq:pii_2})--(\ref{eq:K1equals0}) that align near the negative real $\xi$ axis as $|\xi|\rightarrow\infty$ (see (\ref{eq:xi0quotientpos})--(\ref{eq:xi0quotientneg})), a more refined version of these techniques, using a two-parameter transseries \cite{Aniceto2015,Schiappa2014}, could be developed to describe the full singularity structure between the two active anti-Stokes lines (the negative real $\xi$ axis and $\mathrm{arg}(\xi)=\pi/3$).

\item For a more general class of initial conditions with singularities of the form
    \begin{equation}
    u_0\sim \frac{A}{(x-x_0)^\beta} \quad\mbox{as}\quad x\rightarrow x_0,
    \label{eq:ICbeta}
    \end{equation}
    with $\beta\neq 2$, the details for the matched asymptotic expansions must be very different to the case we have considered (namely $\beta=2$).  For a start, the second equation in (\ref{eq:local}) must be replaced by
    \begin{equation}
    u_1\sim \frac{6\beta A^2}{(x-x_0)^{2\beta+1}}
    +\frac{\beta(\beta+1)(\beta+2)A}{(x-x_0)^{\beta+3}}
    \quad\mbox{as}\quad x\rightarrow x_0.
    \label{eq:u1beta}
    \end{equation}
    For $\beta>2$, the first term on the right-hand side of (\ref{eq:u1beta}) dominates the second, and thus the inner problem must have
    $$
    u=\frac{1}{(3t)^{\beta/(\beta+1)}}f(\xi,t),
    \quad \xi=\frac{x-x_0}{(3t)^{1/(\beta+1)}}=\mathcal{O}(1)
    $$
    (cf. (\ref{eq:xi}) and (\ref{eq:inner})).  It turns out that the third-order dispersive term does not appear in the leading-order problem,
    governed by
    $$
    -\frac{3\beta}{\beta+1}f_0-\frac{3}{\beta+1}\xi \frac{\mathrm{d}f_0}{\mathrm{d}\xi}+6f_0\frac{\mathrm{d}f_0}{\mathrm{d}\xi}=0
    $$
    (cf. (\ref{eq:odef0})), which suggests a further rescaling will be required in order to bring in the as-yet neglected higher-order term.
For $0<\beta<2$, the second term on the right-hand side of (\ref{eq:u1beta}) dominates the first, and so the inner problem has
    $$
    u=\frac{1}{(3t)^{\beta/3}}f(\xi,t),
    \quad \xi=\frac{x-x_0}{(3t)^{1/3}}=\mathcal{O}(1).
    $$
    This time the nonlinear term will not appear in the leading-order ode
    $$
    -\beta f_0-\xi \frac{\mathrm{d}f_0}{\mathrm{d}\xi}+\frac{\mathrm{d}^3 f_0}{\mathrm{d}\xi^3}=0,
    $$
    again suggesting there will need to be a further rescaling, this time to bring in the nonlinear term.  These are complicated issues that we leave for the future.  In addition, the application of our exponential asymptotics to describe the dispersive waves in the small-time limit (subsection~\ref{sec:expasympt_main}) is more challenging for $\beta\neq 2$, in part because of the complications just described for the inner region near $x=x_0$, and so we shall report on these results elsewhere.

    Of course, there are many other classes of initial conditions that do not have singularities of the type (\ref{eq:ICbeta}), and these are also of interest for future work. For example, entire initial conditions do not have any singularities in the plane at all and, thus, for such cases, the singularities of the time-dependent solution must be born at infinity.  Say we had a Gaussian initial condition $u_0=A\,\mathrm{e}^{-x^2}$, then from (\ref{eq:u1}) we have $u_1=12A^2x\,\mathrm{e}^{-2x^2}-4Ax(3-2x^2)\,\mathrm{e}^{-x^2}$.  As we move up into the upper half of the complex-$x$ plane, both terms of $u_1$ grow exponentially, but the first dominates the second.  Indeed, the first term in $u_1$ grows faster than $u_0$ does in this direction.  Thus, the expansion (\ref{eq:outer}) breaks down where $A\,\mathrm{e}^{-x^2}=\mathcal{O}(12A^2tx\,\mathrm{e}^{-2x^2})$, which suggests an inner problem with scalings
    $$
    u=-\frac{\mathrm{i}}{12t\ln^{1/2}(1/t)}U(X,t),
    \quad
    x=\mathrm{i}\ln^{1/2}(1/t)+
    \frac{-\frac{1}{4}\mathrm{i}\ln(\ln(1/t))-\frac{1}{2}\mathrm{i}
    \ln(12A)+(n+\frac{1}{4})\pi+X}{\ln^{1/2}(1/t)},
    $$
    where $X=\mathcal{O}(1)$ and $n$ is an integer whose magnitude is of $\mathcal{O}(1)$.  To leading order, we write $U\sim U_0(X)$ as $t\rightarrow 0^+$, where
    \begin{equation}
    -U_0+\sfrac{1}{2}\mathrm{i}\frac{\mathrm{d}U_0}{\mathrm{d}X}-
    \sfrac{1}{2}\mathrm{i}U_0\frac{\mathrm{d}U_0}{\mathrm{d}X}=0.
    \label{eq:Gaussianinner}
    \end{equation}
    As the third-order dispersive term in KdV does not appear in (\ref{eq:Gaussianinner}), the resulting singularities of $U_0$ will be of the wrong type (they will be square-root-type branch points instead of double poles).  Thus, there will need to be a further rescaling near these singularities to establish the appropriate balance between advection and dispersion.  This analysis will complicate the description of the early-time pole dynamics as well as the exponential asymptotics.  Other entire initial conditions are expected to involve similar challenges.

\item In order to better understand the time-dependent motion of complex singularities of KdV solutions for $\mathcal{O}(1)$ time, it would be beneficial to generate accurate numerical results for the analytic continuation of $u(x,t)$.  For example, for the singularities that emerges from a point $x=x_0$ according to a solution of our P$_{\mathrm{II}}$ problem, we would like to understand which end up propagating to the right and aligning with a soliton and, if they don't, how they arrange themselves for $t=\mathcal{O}(1)$.
    An example of this type of numerical approach was explored for the KdV equation by Weideman~\cite{Weideman2022}, although that study was for a periodic formulation, allowing for a Fourier-Pad\'{e} method, and the results were not extensive.  For our formulation on the real line, an attractive alternative could be to use numerical rational approximation, for example via the adaptive Antoulas-Anderson (AAA) algorithm  \cite{Nakatsukasa2018,Nakatsukasa2024}.  Given the ill-posed nature of numerical analytic continuation, such a task would be challenging; further tinkering of the AAA algorithm may be required to deal with the double pole singularities that arise from KdV, which are not ideal for the types of numerical rational approximation that cater for simple poles only.

\item As an integrable pde, the KdV equation (\ref{eq:kdv}) is well known to be exceptional, with various special properties.  Clearly, in our study, the integrability of the KdV equation plays a role in the analysis of the inner problem for $f_0(\xi)$, which is a change of variables away from P$_{\mathrm{II}}$, most notably in the absence of branch points but also reflected in the Stokes multipliers being explicit.  On the other hand, the same general asymptotic framework we outline here for studying the emergence of complex-plane singularities should apply to other (non-integrable) dispersive wave equations, albeit with a stronger reliance on nontrivial numerical analysis of inner problems.  Further, it appears that integrability does not significantly affect the application of exponential asymptotics to derive the form of the dispersive waves (\ref{eq:farfielddispersivewaves}), suggesting that these techniques are likely to be versatile in this context.  As such, we are motivated to expand our study to other dispersive wave equations, whether they be integrable or not.

\item An important point to emphasise is that various parts of our analysis (or the analysis for more general initial conditions with (\ref{eq:ICbeta}), for example) require that the initial condition $u_0(x)$ be an analytic function of $x$.  If we drop that restriction, then we are unable to apply our methodologies directly to determine the small-time behaviour of the dispersive waves or the singularities of the solution in the complex plane.  For example, for non-analytic initial conditions, we are not able by the above arguments to make any claim about the form of the dispersive waves for early times, or even whether the waves are exponentially small in time.  Further, for non-analytic initial conditions we do not expect the early-time singularity structure to be related to the decreasing tritronqu\'{e}e solutions of P$_{\mathrm{II}}$, or even necessarily related to P$_{\mathrm{II}}$ at all.  With this in mind, a challenge would be to construct an appropriate analysis of the small-time behaviour of KdV solutions on the real line and the complex plane for non-analytic initial conditions.

\item Finally, all the predictions outlined in this paper are based on formal asymptotics supported by some numerics; as such, there is scope for revisiting these ideas using more rigorous analysis.
\end{myitemize}

\begin{Backmatter}

\vspace{3ex}
\paragraph{Acknowledgments}
SWM is grateful for valuable discussions with
Bernard Deconinck, Sergey Dyachenko, Tom Trogdon and Catherine Johnston, as well as to Bengt Fornberg and Andre Weideman for kindly providing a copy of the code they developed for Painlev\'{e} II~\cite{Fornberg2014}.  SWM and CJL thank In\^{e}s Aniceto for discussions about transseries.
SWM, CJL, JRK and SJC would like to thank the Isaac
Newton Institute for Mathematical Sciences, Cambridge, for support and hospitality during the programmes `Complex analysis: techniques, applications and computations' and `Applicable resurgent asymptotics: towards a universal theory', where part of the work on this paper was undertaken. These programmes were supported by the EPSRC grant no. EP/R014604/1.  All authors thank the anonymous referees for their constructive comments and suggestions.  For the purpose of open access, the authors have applied a CC BY public copyright licence to any author accepted manuscript arising from this submission.

\paragraph{Funding Statement}
SWM, CJL and SJC acknowledge the support of Australian Research Council Discovery Projects DP250101095, DP240101666 and DP190101190.  JRK gratefully acknowledges a Royal Society Leverhulme Trust Senior Research Fellowship.

\paragraph{Competing Interests}
None declared.

\paragraph{Data Availability Statement}
The datasets used in this study are available from the authors upon request.

\paragraph{Ethical Standards}
The research meets all ethical guidelines, including adherence to the legal requirements of the study country.

\paragraph{Author Contributions}
Conceptualization: SWM, CJL, JRK, SJC.
Methodology: SWM, CJL, DJV, JZ, JRK, SJC.
Data curation: SWM, CJL, DJV, JZ, JRK, SJC.
Data visualisation: CJL, DJV, JZ.
Writing original draft: SWM.
All authors approved the final submitted draft.

%\bibliography{example}
%%\printbibliography

\end{Backmatter}

\begin{appendix}

\begin{large}
\vspace{3ex}
\noindent {\bf Appendices}
\end{large}

\vspace{-3ex}
\section{Exponential asymptotics to describe emergence of dispersive waves}\label{sec:expasympt_app}

In this appendix, we provide further details of the exponential asymptotics from subsection~\ref{sec:expasympt_main}.  As mentioned in that subsection, we believe these are the first analytical results to describe the small-time behaviour of dispersive waves for the KdV equation.

\subsection{Late-order terms in power-series expansion}\label{sec:expasympt_lateorder}

We begin by writing out $u(x,t)$ as a regular perturbation in powers of $t$, as in (\ref{eq:powerseries}), and then substituting into the KdV equation (\ref{eq:kdv}) to determine the correction term $u_1$ and subsequent terms via the recurrence relation (\ref{eq:recurrence}).  Such an asymptotic expansion is divergent due to the repeated differentiation required to compute $u_{n+1}$ from $u_n$.  Therefore, the terms in (\ref{eq:powerseries}) become increasingly singular in the neighbourhood of any singularity of the leading-order term $u_0$.  For our problem, this leading-order term is simply the initial condition for the original pde.

The divergence in (\ref{eq:powerseries}) can be characterised by analysing late-order terms $u_n$ as $n\rightarrow\infty$.  One way to do this is to pose the factorial-over-power ansatz (\ref{eq:dingleansatz}), where $\mathcal{A}$ and $\chi$ are functions to be determined and $\gamma$ is an unknown constant \cite{chapman1998}.  By differentiating (\ref{eq:dingleansatz}) we find
$$
(n+1)u_{n+1}\sim
\frac{(n+1)\mathcal{A}\Gamma(2n+\gamma+2)}{\chi^{2n+\gamma+2}}
=\frac{\mathcal{A}\chi\Gamma(2n+\gamma+3)}{2\chi^{2n+\gamma+3}}
-\frac{\gamma \mathcal{A}\Gamma(2n+\gamma+2)}{2\chi^{2n+\gamma+2}},
$$
$$
u_n'''\sim -\frac{\mathcal{A}(\chi')^3\Gamma(2n+\gamma+3)}{\chi^{2n+\gamma+3}}
+\frac{(3\mathcal{A}\chi'\chi''+3\mathcal{A}'(\chi')^2)
\Gamma(2n+\gamma+2)}{\chi^{2n+\gamma+2}},
$$
as $n\rightarrow\infty$, so that the recurrence relation (\ref{eq:recurrence}) can be used to show that $\chi$ and $\mathcal{A}$ satisfy (\ref{eq:chiG}).  The first of these equations suggests $\chi=\pm (2/3^{3/2})(x-x_0)^{3/2}$, where recall we are assuming that the singularity $x_0$ is in the upper half plane.  We choose the negative sign so that $\chi$ is real and positive on the Stokes line $\mathrm{arg}(x-x_0)=-2\pi/3$.  It is interesting to note that, in order to derive (\ref{eq:chiG}), we need only balance the first two terms of $(n+1)u_{n+1}$ with the first two terms of $u_n'''$ in the limit $n\rightarrow\infty$, which means the nonlinear terms on the right-hand side of (\ref{eq:recurrence}) do not play a role in this part of the analysis.

Given the form of the late-order terms, (\ref{eq:lateorder}), we can determine by $\gamma$ by a consistency check with the leading order term $u_0$.  As we are focussing on the initial conditions with the property (\ref{eq:ICA}), we see that by setting $n=0$ in the denominator of (\ref{eq:lateorder}), we have $\gamma+1/2=2$.
While we do not consider other possibilities here, it is clear that other types of initial conditions would lead to different values of $\gamma$.

\subsection{Switching on exponential across Stokes lines}\label{sec:expasympt_switch}

In this subsection, we show that the term (\ref{eq:expsmall}) is switched on across the Stokes line $\mathrm{arg}(x-x_0)=-2\pi/3$.  A starting point is to truncate out divergent series (\ref{eq:powerseries}) so that
\begin{equation}
u = \sum_{n=0}^{N-1} t^n u_n(x) + R_N(x,t).
\label{eq:powerseries2}
\end{equation}
Here, $R_N$ is the remainder term which, for fixed $N$, is of $\mathcal{O}(t^{-1})$ as $t\rightarrow 0^+$.  However, if we truncate optimally (at the smallest term), then the remainder is exponentially small.  This result motivates what follows.  To complicate matters, the optimal truncation point, $N=N_{\mathrm{opt}}$, depends on $x$ and $t$.  For a fixed $x$, as we decrease $t$ we need more terms in our optimally truncated series.  Thus, we have $N_{\mathrm{opt}}\rightarrow\infty$ as $t\rightarrow 0^+$, which means that all of the crucial information is in the tail of the series (\ref{eq:powerseries}).

Substituting (\ref{eq:powerseries2}) into the KdV equation (\ref{eq:kdv}) gives
$$
\sum_{n=0}^{N-2}(n+1)t^nu_{n+1}+\frac{\partial R_N}{\partial t}
+6\sum_{n=0}^{2N-2}t^n\sum_{m=0}^n  u_m u'_{n-m}
+6R_N\sum_{n=0}^{N-1} t^nu'_n
$$
$$
+6\frac{\partial R_N}{\partial x}\sum_{n=0}^{N-1} t^nu_n
+6R_N\frac{\partial R_N}{\partial x}
+\sum_{n=0}^{N-1} t^nu'''_n
+\frac{\partial^3 R_N}{\partial x^3}=0,
$$
which, after cancelling the first $N-1$ terms in each series using (\ref{eq:recurrence}), simplifies to
$$
\frac{\partial R_N}{\partial t}+6t^{N-1}\sum_{m=0}^{N-1} u_m u'_{N-1-m}
+6u'_0R_N+6u_0 \frac{\partial R_N}{\partial x} +t^{N-1}u'''_{N-1}
+\frac{\partial^3 R_N}{\partial x^3}+\ldots=0.
$$
The ellipsis denote terms that will be smaller in size as $t\rightarrow 0^+$.  Again, appealing to (\ref{eq:recurrence}), our equation for the remainder becomes
\begin{equation}
\frac{\partial R_N}{\partial t}+6\frac{\partial}{\partial x}\left(u_0 R_N\right)
+\frac{\partial^3 R_N}{\partial x^3}
\sim
\frac{t^{N-1}\chi \mathcal{A}\, \Gamma(2N+\gamma+1)}{2\chi^{2N+\gamma+1}}
\quad\mbox{as}\quad t\rightarrow 0^+,
\label{eq:remaindereqn}
\end{equation}
where, for the right-hand side, we have used (\ref{eq:dingleansatz})
to derive a leading order approximation for $Nt^{N-1}u_N$.

Note that in (\ref{eq:remaindereqn}), and in what follows, the amplitude function $\mathcal{A}$ and the singulant $\chi$ are given by (\ref{eq:chiandG}).  These functions also arise as ingredients of the solution to the homogeneous part of (\ref{eq:remaindereqn}).  Indeed, applying a WKB method reveals that the homogenous solution, $R_\mathrm{H}$, behaves like
$$
R_\mathrm{H} \sim \mbox{constant}\,\mathcal{A}\, t^{-\gamma/2}\mathrm{e}^{-\chi/t^{1/2}}
\quad\mbox{as}\quad t\rightarrow 0^+.
$$
To proceed, we apply a type of variation of parameters argument, and let
$$
R_N \sim \mathcal{S}(x,t) \,\mathcal{A}\, t^{-\gamma/2}\mathrm{e}^{-\chi/t^{1/2}},
$$
where $\mathcal{S}$ is a Stokes multiplier, so that, to leading order,
$$
(\mathcal{S}_t+3t^{-1}(\chi')^2\mathcal{S}_x)t^{-\gamma/2}\mathrm{e}^{-\chi/t^{1/2}}
\sim \frac{t^{N-1}\chi\, \Gamma(2N+\gamma+1)}{2\chi^{2N+\gamma+1}}.
$$
Using the similarity variable
$$
\bar{\chi}=\frac{\chi}{t^{1/2}}
$$
and the first equation in (\ref{eq:chiG}), we find
\begin{equation}
\frac{\mathrm{d}\mathcal{S}}{\mathrm{d}\bar{\chi}}\sim
\frac{\Gamma(2N+\gamma+1)}{2\bar{\chi}^{2N+\gamma+1}}\,\mathrm{e}^{\bar{\chi}}
\quad\mbox{as}\quad |\bar{\chi}|\rightarrow\infty.
\label{eq:Seqn}
\end{equation}
This equation with $\bar{\chi}$ as an independent variable is now in a convenient form.

In the above analysis, we wish to take $N=N_{\mathrm{opt}}$, where $N_{\mathrm{opt}}$ comes from truncating (\ref{eq:powerseries2}) optimally.  We follow a heuristic which says optimal truncation occurs at the least term or, in other words, when
$$
|t^N u_N|\sim |t^{N+1}u_{N+1}|
\quad\mbox{as}\quad t\rightarrow 0^+.
$$
Using (\ref{eq:dingleansatz}), we find that $N_{\mathrm{opt}}\sim |\chi|/2t^{1/2}$, so we write
$$
N_{\mathrm{opt}}=\frac{|\chi|}{2t^{1/2}}+\omega
=\sfrac{1}{2}|\bar{\chi}|+\omega,
$$
where $\omega\in [0,1)$ is an unimportant quantity chosen so that  $N_{\mathrm{opt}}$ is an integer.

Before we substitute in our optimal value of $N$ into (\ref{eq:Seqn}), we make a change of variables $\bar{\chi}=\bar{\rho}\mathrm{e}^{\mathrm{i}\vartheta}$.
Since $N_{\mathrm{opt}}$ depends on $|\bar{\chi}|=\bar{\rho}$ and not $\vartheta$, then we rewrite our $\bar{\chi}$ derivative as
$$
\frac{\mathrm{d}}{\mathrm{d}\bar{\chi}}=
-\frac{\mathrm{i}\mathrm{e}^{-\mathrm{i}\vartheta}} {\bar{\rho}}\,\frac{\mathrm{d}}{\mathrm{d}\vartheta}.
$$
Using Stirling's formula, our equation (\ref{eq:Seqn}) for the Stokes multiplier becomes
$$
\frac{\mathrm{d}\mathcal{S}}{\mathrm{d}\vartheta}\sim
\frac{\mathrm{i}\sqrt{2\pi}(2N)^{2N+\gamma+1/2}
\mathrm{e}^{-\mathrm{i}\vartheta\gamma}
\mathrm{e}^{2N(-1-\mathrm{i}\vartheta)+\bar{\rho}\mathrm{e}^{\mathrm{i}\vartheta}}}
{2\bar{\rho}^{2N+\gamma}}
\quad\mbox{as}\quad \bar{\rho}\rightarrow\infty.
$$
Letting $N=N_{\mathrm{opt}}$, we have
$$
\frac{\mathrm{d}\mathcal{S}}{\mathrm{d}\vartheta}\sim
\mathrm{i}\,\sqrt{\frac{\pi}{2}}
\,\mathrm{e}^{-\mathrm{i}\vartheta(\gamma+2\omega)}
\,\bar{\rho}^{1/2}
\,\mathrm{e}^{\bar{\rho}(-1-\mathrm{i}\vartheta+\mathrm{e}^{\mathrm{i}\vartheta})}
\quad\mbox{as}\quad \bar{\rho}\rightarrow\infty
$$
The argument $-1-\mathrm{i}\vartheta+\mathrm{e}^{\mathrm{i}\vartheta}$ has a negative real part, except when $\vartheta=0$.  Thus, we see that the rate of change of the Stokes multiplier, $\mathrm{d}\mathcal{S}/\mathrm{d}\vartheta$, is exponentially small for $\bar{\rho}\gg 1$, except at $\vartheta=0$ where it is of $\mathcal{O}(\bar{\rho}^{1/2})$, suggesting a boundary layer near $\vartheta=0$.

Writing $\vartheta=\bar{\theta}/\bar{\rho}^{1/2}$, we have
$$
\frac{\mathrm{d}\mathcal{S}}{\mathrm{d}\bar{\theta}}\sim
\mathrm{i}\,\sqrt{\frac{\pi}{2}}
\,\mathrm{e}^{-\bar{\theta}^2/2}.
$$
By integrating to give an error function to leading order, we see the jump in $\mathcal{S}$ between $\bar{\theta}\rightarrow -\infty$ and $\bar{\theta}\rightarrow \infty$, denoted by $[\mathcal{S}]_-^+$, is
$$
[\mathcal{S}]_-^+ \sim \mathrm{i}\pi.
$$
In terms of the variable $\bar{\theta}$, the jump in $\mathcal{S}$ across $\bar{\theta}=0$ is smooth, approximated by an error function \cite{Berry1989}.  However, from the perspective of the original spatial variable $x$, the transition appears to be sharp and the observation is that we pick up a contribution $[\mathcal{S}]_-^+$ multiplied by $\mathcal{A}\, t^{-\gamma/2}\mathrm{e}^{-\chi/t^{1/2}}$ as we cross the Stokes line $\bar{\theta}=0$.  Now, the function $\chi=-(2/3^{3/2})(x-x_0)^{3/2}$ is real and positive when $\sin(\frac{3}{2}\mathrm{arg}(x-x_0))=0$ and  $\cos(\frac{3}{2}\mathrm{arg}(x-x_0))<0$.  Thus, we have shown that (\ref{eq:expsmall}) switches on as we cross the Stokes line $\mathrm{arg}(x-x_0)=-2\pi/3$.

\subsection{Matching in inner region to determine $\Lambda$}\label{sec:expasympt_match}

Finally, we need to determine the Stokes constant $\Lambda$, which we do by matching into the inner region near $x=x_0$.  To begin, we employ the inner variable $\xi$, defined in (\ref{eq:xi}), and construct the inner limit of the outer expansion using
\begin{equation}
t^nu_n\sim \frac{\Lambda t^n (-3^{3/2}/2)^{2n+3/2}
\Gamma(2n+\sfrac{3}{2})}{(x-x_0)^{3n+2}}
=\Lambda \left(\frac{-3^{3/2}}{2}\right)^{3/2}
\left(\frac{3}{2}\right)^{2n}
\frac{\Gamma(2n+\sfrac{3}{2})}{(3t)^{2/3}\xi^{3n+2}}
\label{eq:innerouterspecial}
\end{equation}
as $n\rightarrow\infty$.  However, we also have from the inner region that
$$
u\sim \frac{1}{(3t)^{2/3}}f_0(\xi)
\quad\mbox{as}\quad t\rightarrow 0^+,
$$
where
\begin{equation}
f_0\sim \sum_{n=0}^\infty \frac{a_n}{\xi^{3n+2}}
\quad\mbox{as}\quad \xi\rightarrow -\mathrm{i}\infty,
\quad
a_0=A_0.
\label{eq:seriesf0}
\end{equation}
Matching these gives
\begin{equation}
\Lambda=\left(\frac{-2}{3^{3/2}}\right)^{3/2}
\lim_{n\rightarrow\infty}
\frac{(2/3)^{2n}a_n}{\Gamma(2n+\sfrac{3}{2})}.
\label{eq:lambda1}
\end{equation}
It remains to consider the coefficients $a_n$ in order to take the limit in (\ref{eq:lambda1}).

One way to determine the $a_n$ is to use the relationship between $f_0$ and $F$, given by (\ref{eq:PIIreduction}), and then relying on (\ref{eq:divergent}) and (\ref{eq:its}), recalling that $\alpha$ and $A_0$ are related by (\ref{eq:BA02}).  This approach gives
$$
a_n=-\alpha(3n+1)b_n-\alpha^2\sum_{j=0}^n b_jb_{n-j},
$$
where the $b_n$ are given by (\ref{eq:its}).  Now, using the large $n$ behaviour of the $b_n$ in (\ref{eq:largebn}), we find
$$
a_n\sim -\alpha(3n+1)b_n \sim -\alpha(3n+1)\frac{\sin\pi\alpha}{\alpha\pi^{3/2}}
\left(\frac{3}{2}\right)^{2n+1/2}
\Gamma\left(2n+\sfrac{1}{2}\right),
$$
which implies that
$$
\frac{(2/3)^{2n}a_n}{\Gamma(2n+\sfrac{3}{2})}\sim
-\left(\frac{3}{2\pi}\right)^{3/2}\sin \pi\alpha
\quad\mbox{as}\quad n\rightarrow\infty.
$$
Upon using the relationship between $\alpha$ and $A_0$ in (\ref{eq:BA02}), we find from (\ref{eq:lambda1}) that $\Lambda$ is given by (\ref{eq:Lambda}).

Another method is to substitute (\ref{eq:seriesf0}) into the governing equation (\ref{eq:odef0}) directly to give
$$
-2\sum_{n=0}^\infty \frac{a_n}{\xi^{3n+2}}
+
\sum_{n=0}^\infty \frac{(3n+2)a_n}{\xi^{3n+2}}
-6
\sum_{n=0}^\infty \frac{a_n}{\xi^{3n+2}}
\sum_{m=0}^\infty \frac{(3m+2)a_m}{\xi^{3m+3}}
-
\sum_{n=0}^\infty \frac{(3n+2)(3n+3)(3n+4)a_n}{\xi^{3n+5}}=0,
$$
which implies that
$$
-2a_n+(3n+2)a_n
-6\sum_{m=0}^{n-1}(3m+2)a_m a_{n-m-1}
-(3n-1)(3n)(3n+1)a_{n-1}=0,
\quad n\geq 1.
$$
As a result,
\begin{equation}
a_n= (3n-1)(3n+1)a_{n-1}+
\frac{2}{n}\sum_{m=0}^{n-1}(3m+2)a_m a_{n-m-1}.
\label{eq:an}
\end{equation}
As a check, we have generated $a_n$ for large $n$ and then estimated the limit in (\ref{eq:lambda1}) for various values of $A_0$ using repeated Shanks transformations.  Combined with (\ref{eq:lambda1}), this numerical exercise confirms (\ref{eq:Lambda}).

\subsection{Special case $A_0=-N(N+1)$}\label{sec:expasympt_special}

For an initial condition with double poles whose local behaviour is given by (\ref{eq:ICA}), our leading-order prediction (\ref{eq:expdispersive}) with (\ref{eq:Lambda}) suggests that the amplitude of the dispersive waves vanishes for triangular values $A_0=-N(N+1)$.  However, this absence of dispersive waves for certain values of $A_0$ is known to hold only for the exceptional $\mathrm{sech}^2$-type initial conditions (\ref{eq:ICsech}).  Otherwise, when $A_0=-N(N+1)$ we need to revisit our analysis and consider higher-order contributions, the details of which we summarise here.

To proceed, we will need some results from appendix~\ref{sec:appendinneragain} below.  First, we will need to consider the first two terms in (\ref{eq:ICAmore}), which are characterised by $A_0$ and the residue $A_1$.  We shall also need the first two terms in the expansion for $f(\xi)$, given by (\ref{eq:innerexpansion}), where now the correction term $f_1$ satisfies the ode (\ref{eq:odef1}).  Using the far-field behaviour (\ref{eq:farfieldf0more})--(\ref{eq:farfieldf1more}), we can integrate the equation for $f_1$ to give the second-order linear equation (\ref{eq:f1reduced}).  For our purposes, we shall write out the full series for $f_1$ in (\ref{eq:farfieldf1more}) as
\begin{equation}
f_1\sim A_1\sum_{n=0}^\infty \frac{c_n}{\xi^{3n+1}}
\quad\mbox{as}\quad \xi\rightarrow -\mathrm{i}\infty,
\label{eq:f1expand}
\end{equation}
where $c_0=1$.

A key point is that, in general, both of the series in (\ref{eq:farfieldf0more}) and (\ref{eq:farfieldf1more}) are divergent; however, for the special case $A_0=-N(N+1)$ the series (\ref{eq:farfieldf0more}) converges (for sufficiently large $|\xi|$) to the appropriate rational solution described in subsection~\ref{sec:exactsolns} (the first four of which are listed in (\ref{eq:B1f0})--(\ref{eq:B4f0})). Therefore, when $A_0=-N(N+1)$, the crucial divergence that is driving the exponentially small terms (that are turned on across Stokes lines) comes from $f_1$ and not $f_0$.  Thus, our inner expansion for $A_0=-N(N+1)$ can be interpreted as being
\begin{equation}
u(x,t)\sim \frac{1}{(3t)^{2/3}}\,\left[\mbox{rational soln}\,f_0(\xi)\right]
+\frac{1}{(3t)^{1/3}}f_1(\xi)
\quad\mbox{as}\quad t\rightarrow 0^+,
\label{eq:rationalplusseries}
\end{equation}
where $f_1$ has the expansion (\ref{eq:f1expand}).

Returning to our outer expansion, we found that our late-order terms are of the form (\ref{eq:lateorder}).  To be consistent with (\ref{eq:rationalplusseries}), we must now choose $\gamma=1/2$ (and not $\gamma=3/2$), in which case the late-order terms behave as
$$
t^n u_n\sim \frac{\Lambda t^n(-3^{3/2}/2)^{2n+1/2}
\Gamma(2n+\sfrac{1}{2})}
{(x-x_0)^{3n+1}}
=\Lambda \left(\frac{-3^{3/2}}{2}\right)^{1/2}
\left(\frac{3}{2}\right)^{2n}
\frac{\Gamma(2n+\sfrac{1}{2})}{(3t)^{1/3}\xi^{3n+1}}
$$
as $n\rightarrow\infty$ (and not (\ref{eq:innerouterspecial})).  Matching with the inner region (\ref{eq:f1expand})--(\ref{eq:rationalplusseries}), we find
\begin{equation}
\Lambda=\left(\frac{-2}{3^{3/2}}\right)^{1/2}A_1
\lim_{n\rightarrow\infty}
\frac{(2/3)^{2n}c_n}{\Gamma(2n+\sfrac{1}{2})}.
\label{eq:lambdalimitspecial}
\end{equation}
To evaluate this limit, we need to consider $c_n$ in more detail.

We can obtain a recurrence relation for $c_n$ by substituting (\ref{eq:f1expand}) into (\ref{eq:f1reduced})
$$
\sum_{n=0}^\infty \frac{A_1(3n+1)(3n+2)c_n}{\xi^{3n+3}}
+
6\sum_{n=0}^\infty \frac{a_n}{\xi^{3n+2}} \sum_{m=0}^\infty \frac{A_1c_n}{\xi^{3n+1}}
-
\sum_{n=0}^\infty \frac{A_1c_n}{\xi^{3n}}+A_1=0,
$$
where recall we have set $c_0=1$.  Therefore,
$$
c_n=(3n-2)(3n-1)c_{n-1}+6\sum_{m=0}^{n-1}a_mc_{n-m-1},
\quad n\geq 1.
$$
In the most simple case, $A_0=-2$, we have $a_0=-2$ and $a_n=0$ for $n\geq 1$.  Therefore,
$$
c_n=(3n-5)(3n+2)c_{n-1},
$$
which has the exact solution
$$
c_n=-\frac{3^{2n+1/2}}{2\pi}\Gamma\left(n+\sfrac{5}{3}\right)
\Gamma\left(n-\sfrac{2}{3}\right)
\sim -\sqrt{3}\left(\frac{3n}{\mathrm{e}}\right)^{2n}
\quad\mbox{as}\quad n\rightarrow\infty.
$$
Thus, for $A_0=-2$ we have from (\ref{eq:lambdalimitspecial}) that
$$
\Lambda = \left(\frac{-2}{3^{3/2}}\right)^{1/2}A_1 \left(\frac{-\sqrt{3}}{\sqrt{2\pi}}\right)
=-\frac{\mathrm{i}A_1}{3^{1/4}\pi^{1/2}}
\approx -0.42869138\,\mathrm{i}\,A_1.
$$
For other values of $A_0=-N(N+1)$, there is no obvious exact solution for $c_n$ and therefore $\Lambda$ may be approximated numerically.  Careful numerical simulations suggest that
$$
\Lambda = (-1)^N\frac{\mathrm{i}A_1}{3^{1/4}\pi^{1/2}}.
$$
That is, it appears that $\Lambda$ takes the same absolute value regardless of $N$, with the sign alternating with $N$.

In summary, for the special cases $A_0=-N(N+1)$, where $N\in\mathbb N$, instead of the  exponentially small contribution (\ref{eq:expsmall}) being switched on across a Stokes line $\mathrm{arg}(x-x_0)=-2\pi/3$, we have
$$
\pi\mathrm{i}\,\mathcal{A}\, t^{-1/4}\mathrm{e}^{-\chi/t^{1/2}}
=
(-1)^{N+1}
\frac{\sqrt{\pi}A_1}{3^{1/4}}
(x-x_0)^{-1/4}
t^{-1/4}
\mathrm{e}^{2(x-x_0)^{3/2}/3(3t)^{1/2}}.
$$
Together with the analogous contribution from the lower-half plane, the result is that, on the real line, instead of (\ref{eq:expdispersive}), we have
\begin{align}
U_{\mbox{\footnotesize dis}}\sim
&
\,\,\pi\mathrm{i}\Lambda t^{-1/4}\left(
(x-\mathrm{i}y_0)^{-1/4}
\mathrm{e}^{2(x-\mathrm{i}y_0)^{3/2}/3(3t)^{1/2}}
+
(x+\mathrm{i}y_0)^{-1/4}
\mathrm{e}^{2(x+\mathrm{i}y_0)^{3/2}/3(3t)^{1/2}}
\right)
\nonumber \\
=
&
(-1)^{N+1}
\frac{2\sqrt{\pi}A_1}{3^{1/4}}
\,t^{-1/4}
(x^2+y_0^2)^{-1/8}
\mathrm{e}^{
2(x^2+y_0^2)^{3/4}\cos (3\phi/2)/(3(3t)^{1/2})
}
\nonumber \\
& \times
\cos\left(
\frac{2}{3(3t)^{1/2}}
(x^2+y_0^2)^{3/4}
\sin \sfrac{3}{2}\phi -\sfrac{1}{4}\phi
\right),
\nonumber
\end{align}
$x<-y_0/\sqrt{3}$, $t\rightarrow 0^+$, where $\phi=\arctan(-x)$.
For large negative $x$, we have
$$
U_{\mbox{\footnotesize dis}}\sim
(-1)^{N+1}
\frac{\sqrt{\pi}A_1}{3^{1/4}}
\frac{\mathrm{e}^{-y_0(-x/3t)^{1/2}}}{(-xt)^{1/4}}
\,\cos\left(\frac{2(-x)^{3/2}}{3(3t)^{1/2}}+\frac{\pi}{4}
\right)
$$
as $x\rightarrow -\infty$, $t\rightarrow 0^+$.  Compared with the generic case (\ref{eq:farfielddispersivewaves}), the dispersive waves are algebraically smaller by a scaling of $(-x/t)^{1/2}$ due to the term $(-1/xt)^{1/4}$ compared to $(-x/t^3)^{1/4}$.

\section{Higher-order terms in matched asymptotic expansions and special cases}\label{sec:higherorder}

In this appendix, we consider higher-order terms in our expansions in order to demonstrate how further terms in (\ref{eq:scalingfors}) could be generated.  Further, we use these results to clarify a couple of special cases, including $A_1=0$ and $A_0=-2$.

\subsection{Outer and inner expansions with more terms}\label{sec:appendinneragain}

Here we shall consider more terms in our expansions, starting with our initial condition and then our inner problem.  As such, we shall revisit the outer expansion, but this time including more details.  To progress, we extend the expansion for our class of initial conditions (\ref{eq:ICA}) to be (\ref{eq:ICAmore}).
The first three terms of the expansion (\ref{eq:powerseries}) are
\begin{equation}
u\sim u_0(x)+tu_1(x)+t^2u_2(x)
\quad\mbox{as}\quad t\rightarrow 0^+,
\label{eq:outermore}
\end{equation}
which gives
$$
u_1=-6u_0u_0'-u_0''', \quad 2u_2=-6u_0u_1'-6u_0'u_1-u_1'''.
$$
Therefore, we have the local behaviour
\begin{equation}
u_1\sim \frac{12A_0(A_0+2)}{(x-x_0)^5}+\frac{6A_1(3A_0+1)}{(x-x_0)^4}+
\frac{6(A_1^2+2A_0A_2)}{(x-x_0)^3}
\quad\mbox{as}
\quad x\rightarrow x_0,
\label{eq:local1}
\end{equation}
\begin{equation}
u_2\sim \frac{252A_0(A_0+2)(A_0+5)}{(x-x_0)^8}+
\frac{180A_1(3A_0^2+9A_0+2)}{(x-x_0)^7}
\quad\mbox{as}
\quad x\rightarrow x_0.
\label{eq:local2}
\end{equation}
We now combine (\ref{eq:outermore}) and (\ref{eq:local2}) to give
\begin{align*}
u \sim & \frac{1}{(3t)^{2/3}}\left(
\frac{A_0}{\xi^2}+\frac{4A_0(A_0+2)}{\xi^5}+\frac{28A_0(A_0+2)(A_0+5)}{\xi^8}+\ldots
\right)
\\
& +
\frac{1}{(3t)^{1/3}}\left(
\frac{A_1}{\xi}+\frac{2A_1(3A_0+1)}{\xi^4}+
\frac{20A_1(3A_0^2+9A_0+2)}{\xi^7}+\ldots
\right)
\\
& +
\left(
A_2+\frac{2(A_1^2+2A_0A_2)}{\xi^3}+\frac{40A_0A_2(A_0+2)+10A_1^2(4A_0+3)}{\xi^6}
+\ldots
\right)+\ldots.
\end{align*}
Therefore, for the inner problem (\ref{eq:inner}) with pde (\ref{eq:pdef}), we can write
\begin{equation}
f\sim f_0(\xi)+(3t)^{1/3}f_1(\xi)+(3t)^{2/3}f_2(\xi)
\quad\mbox{as}\quad t\rightarrow 0^+,
\label{eq:innerexpansion}
\end{equation}
so that $f_0$ satisfies (\ref{eq:odef0}), while $f_1$ and $f_2$ satisfy
\begin{equation}
-f_1-\xi \frac{\mathrm{d}f_1}{\mathrm{d}\xi}
+6\left(f_0\frac{\mathrm{d}f_1}{\mathrm{d}\xi}
+\frac{\mathrm{d}f_0}{\mathrm{d}\xi}f_1\right)
+\frac{\mathrm{d}^3f_1}{\mathrm{d}\xi^3}=0,
\label{eq:odef1}
\end{equation}
\begin{equation}
-\xi \frac{\mathrm{d}f_2}{\mathrm{d}\xi}
+6\left(f_0\frac{\mathrm{d}f_2}{\mathrm{d}\xi}
+f_1\frac{\mathrm{d}f_1}{\mathrm{d}\xi}
+\frac{\mathrm{d}f_0}{\mathrm{d}\xi}f_2\right)
+\frac{\mathrm{d}^3f_2}{\mathrm{d}\xi^3}=0,
\label{eq:odef2}
\end{equation}
with far-field conditions
\begin{align}
f_0 & \sim \frac{A_0}{\xi^2}+\frac{4A_0(A_0+2)}{\xi^5}
+\frac{28A_0(A_0+2)(A_0+5)}{\xi^8}
\quad\mbox{as}\quad \xi\rightarrow -\mathrm{i}\infty,
\label{eq:farfieldf0more}
\\
f_1 & \sim
\frac{A_1}{\xi}+\frac{2A_1(3A_0+1)}{\xi^4}+
\frac{20A_1(3A_0^2+9A_0+2)}{\xi^7}
\quad\mbox{as}\quad \xi\rightarrow -\mathrm{i}\infty,
\label{eq:farfieldf1more}
\\
f_2 & \sim
A_2+\frac{2(A_1^2+2A_0A_2)}{\xi^3}+\frac{40A_0A_2(A_0+2)+10A_1^2(4A_0+3)}{\xi^6}
\quad\mbox{as}\quad \xi\rightarrow -\mathrm{i}\infty,
\label{eq:farfieldf2more}
\end{align}
together with other conditions suppressing exponentially growing correction terms, which we will discuss below where appropriate.  (Note that the terms in (\ref{eq:farfieldf0more})--(\ref{eq:farfieldf2more}) could also be determined directly from (\ref{eq:odef0}), (\ref{eq:odef1}) or (\ref{eq:odef2}); for example, (\ref{eq:farfieldf0more}) contains the first few terms of (\ref{eq:seriesf0}) together with the recurrence relation (\ref{eq:an}).)

In general, the solutions for $f_0$ will be related to our decreasing tritronqu\'{e}e solutions $F$ of P$_{\mathrm{II}}$ via (\ref{eq:PIIreduction}). Thus, $f_0$ will generically have an infinite number of double poles at points that we label $\xi=\xi_0$.  As we have already noted in (\ref{eq:scalingfors}), this means that for each $\xi=\xi_0$, there is a corresponding singularity of our KdV solution that behaves like $s(t)\sim x_0+(3t)^{1/3}\xi_0$ as $t\rightarrow 0^+$.

By considering (\ref{eq:odef0}), we find $f_0$ has the local behaviour about each of its singularities of the form
\begin{equation}
f_0\sim \frac{-2}{(\xi-\xi_0)^2}+\sfrac{1}{6}\xi_0+\mathcal{O}\left((\xi-\xi_0)^2
\right)
\quad\mbox{as}\quad \xi\rightarrow\xi_0.
\label{eq:f0nearxi0}
\end{equation}
Turning to the problem for $f_1$, we can integrate (\ref{eq:odef1}) together with (\ref{eq:farfieldf1more}) to give
\begin{equation}
\frac{\mathrm{d}^2f_1}{\mathrm{d}\xi^2}
+(6f_0-\xi)f_1+A_1=0.
\label{eq:f1reduced}
\end{equation}
As $f_1$ satisfies a linear equation, it will also be singular at each of the points $\xi=\xi_0$.  Regardless of the precise form of the solution for $f_1$, the governing equation (\ref{eq:f1reduced}) can be used to show $f_1$ will behave as
\begin{equation}
f_1\sim \frac{\mu_1}{(\xi-\xi_0)^3}-\sfrac{1}{12}\mu_1 + \mathcal{O}(\xi-\xi_0)
\quad\mbox{as}\quad \xi\rightarrow\xi_0,
\label{eq:f1nearxi0}
\end{equation}
where $\mu_1$ is some constant that depends on $A_1$ (to compute $\mu_1$, the full solution $f_1$ would need to be calculated).  Finally, directly from (\ref{eq:odef2}) we find
\begin{equation}
f_2\sim -\frac{3\mu_1^2}{8(\xi-\xi_0)^4}+\frac{\mu_2}{(\xi-\xi_0)^3}
\quad\mbox{as}\quad \xi\rightarrow\xi_0,
\label{eq:f2nearxi0}
\end{equation}
where $\mu_2$ is a constant that depends on $A_2$ (and can only be determined by the full solution for $f_2$).

\subsection{Higher-order terms for singularity location $x=s(t)$}

In order to derive further terms in our expansion for $x=s(t)$, the location of each singularity of our KdV solution, we first note that including the next nonzero term in the expansion (\ref{eq:doublepoleleadingorder}) gives
\begin{equation}
u\sim -\frac{2}{(x-s(t))^2}+\frac{\dot{s}}{6}\quad\mbox{as}\quad x\rightarrow s(t).
\label{eq:doublepolecorrection}
\end{equation}
Thus, if we write
\begin{equation}
s(t)\sim x_0+(3t)^{1/3}\xi_0+(3t)^{2/3}\xi_1+(3t)\xi_2
\quad\mbox{as}\quad t\rightarrow 0^+,
\label{eq:higherordersingularity}
\end{equation}
then we have the limiting behaviour
\begin{align}
u\sim \frac{1}{(3t)^{2/3}}
&
\left[
\frac{-2}{(\xi-\xi_0)^2}+\frac{1}{6}\xi_0+\ldots
+(3t)^{1/3}\left(-\frac{4\xi_1}{(\xi-\xi_0)^3}+\frac{1}{3}\xi_1+\ldots
\right)
\right.
\nonumber \\
& \left.
+(3t)^{2/3}\left(-\frac{6\xi_1^2}{(\xi-\xi_0)^4}
-\frac{4\xi_2}{(\xi-\xi_0)^3}+\ldots
\right)+\ldots
\right]
\quad\mbox{as}\quad x\rightarrow x_0,\quad t\rightarrow 0^+,
\label{eq:doublelimit}
\end{align}
where we have been careful not to include terms in (\ref{eq:doublelimit}) that are neglected in (\ref{eq:doublepolecorrection}).

Now, given the expansion (\ref{eq:innerexpansion}), we can infer from (\ref{eq:doublelimit}) that
\begin{align}
f_0\sim & \frac{-2}{(\xi-\xi_0)^2}+\frac{1}{6}\xi_0+\ldots,
\\
f_1\sim & -\frac{4\xi_1}{(\xi-\xi_0)^3}+\frac{1}{3}\xi_1+\ldots,
\\
f_2\sim & -\frac{6\xi_1^2}{(\xi-\xi_0)^4}
-\frac{4\xi_2}{(\xi-\xi_0)^3}+\ldots
\quad\mbox{as}\quad x\rightarrow x_0.
\end{align}
Thus, by comparing with (\ref{eq:f1nearxi0})--(\ref{eq:f2nearxi0}), we find
\begin{equation}
\xi_1=-\sfrac{1}{4}\mu_1,
\quad
\xi_2=-\sfrac{1}{4}\mu_2.
\label{eq:xi1xi2}
\end{equation}
Putting it together, we see that singularities of KdV solutions move as
\begin{equation}
s(t)\sim x_0+(3t)^{1/3}\xi_0 -\sfrac{1}{4}(3t)^{2/3}\mu_1 -\sfrac{1}{4}(3t)\mu_2
\quad\mbox{as}\quad t\rightarrow 0^+,
\label{eq:3rdscalingfors}
\end{equation}
remembering that $\mu_1$ and $\mu_2$ depend on $A_1$ and $A_2$ and are found by solving for $f_1$ and $f_2$, respectively.  Even higher-order contributions could in principle be calculated in a similar way by formulating problems for $f_3$ and so on.

\subsection{Special case $A_1=0$}\label{eq:A0neqm2A1eq0}

Very briefly, if $A_1=0$, then the solution for $f_1$ is simply $f_1=0$, so that (\ref{eq:f1nearxi0}) implies $\mu_1=0$.  Further, the ode for $f_2$, (\ref{eq:odef2}), becomes
\begin{equation}
-\xi \frac{\mathrm{d}f_2}{\mathrm{d}\xi}
+6\left(f_0\frac{\mathrm{d}f_2}{\mathrm{d}\xi}
+\frac{\mathrm{d}f_0}{\mathrm{d}\xi}f_2\right)
+\frac{\mathrm{d}^3f_2}{\mathrm{d}\xi^3}=0,
\label{eq:odef2ifA10}
\end{equation}
and the local behaviour near $\xi=\xi_0$, namely (\ref{eq:f2nearxi0}), simplifies to
\begin{equation}
f_2\sim \frac{\mu_2}{(\xi-\xi_0)^3}
\quad\mbox{as}\quad \xi\rightarrow\xi_0.
\label{eq:f2nearxi0_A10}
\end{equation}
As a consequence, when $A_1=0$, each singularity of the KdV solution that emerges from $x=x_0$ evolves according to
$$
s(t)\sim x_0+(3t)^{1/3}\xi_0 -\sfrac{1}{4}(3t)\mu_2
\quad\mbox{as}\quad t\rightarrow 0^+,
$$
where the value $\mu_2$ comes from solving (\ref{eq:odef2ifA10}) subject to
\begin{equation}
f_2 \sim A_2+\frac{4A_0A_2}{\xi^3}+\frac{40A_0A_2(A_0+2)}{\xi^6}
\quad\mbox{as}\quad \xi\rightarrow -\mathrm{i}\infty
\label{eq:farfieldf2more_A10}
\end{equation}
and a further condition that controls exponential terms in the far-field as appropriate.

An interesting check on these arguments is with $A_0=-6$, $A_1=0$ and $A_2=2$, which is what we have for the initial condition $u_0=6 \,\mathrm{sech}^2 x$ and the corresponding $2$-soliton solution (\ref{eq:2soliton}).  In that case, we have
$$
f_0=-\frac{6\xi(\xi^3-8)}{(\xi^3+4)^2},
\quad
f_1=0,
\quad
f_2=\frac{2\xi^9-24\xi^6+1440\xi^3-640}{(\xi^3+4)^3},
$$
as can be checked by substitution into (\ref{eq:odef0}) with (\ref{eq:farfieldf0}) and
(\ref{eq:odef2ifA10}) with (\ref{eq:farfieldf2more_A10}) using $A_0=-6$ and $A_2=2$.  In this exceptional case, there are only three singularities of $f_0$, which are given by (\ref{eq:xi0forB2}).  For each of these, the near-singularity behaviour for $f_2$ is
$$
f_2\sim -\frac{16}{(\xi-\xi_0)^3}
\quad\mbox{as}\quad \xi\rightarrow\xi_0,
$$
which, by comparison with (\ref{eq:f2nearxi0_A10}), means that $\mu_2=4$ and the three singularities emerge out of $x=x_0$ according to
\begin{equation}
s(t)\sim x_0+(3t)^{1/3}\xi_0+4(3t)
\quad\mbox{as}\quad t\rightarrow 0^+.
\label{eq:predictA0min6A2equal2}
\end{equation}
This result agrees with the asymptotics for the exact $2$-soliton solution considered in subsection~\ref{sec:2soliton}.

Note the special case $A_0=-12$, $A_1=0$, $A_2=4$ is relevant for the initial condition $u_0=12 \,\mathrm{sech}^2 x$, which leads to the $3$-soliton (\ref{eq:3soliton}) described in subsection~\ref{sec:3soliton}.  In this case, the solutions of (\ref{eq:odef0}) with (\ref{eq:farfieldf0}) and (\ref{eq:odef2ifA10}) with (\ref{eq:farfieldf2more_A10}) are
$$
f_0= -\frac{12\xi(\xi^9+600\xi^3+1600)}{(\xi^6+20\xi^3-80)^2},
\quad
f_1=0,
$$
$$
f_2=\frac{4(\xi^{18} + 12\xi^{15} + 2880\xi^{12} - 136000\xi^9 - 1075200\xi^6 - 8064000\xi^3 - 3584000)}{(\xi^6+20\xi^3-80)^3}.
$$
For each of the six singularities $x=x_0$ of $f_0$, we have
$$
f_2\sim -\frac{32}{(\xi-\xi_0)^3}
\quad\mbox{as}\quad \xi\rightarrow\xi_0,
$$
which means that $\mu_2=8$.  Thus, while each of these six singularities have a different $\xi_0$, their small-time behaviour has the same $\mathcal{O}(t)$ part, namely
\begin{equation}
s(t)\sim x_0+(3t)^{1/3}\xi_0+8(3t)
\quad\mbox{as}\quad t\rightarrow 0^+.
\label{eq:predictA0min12A2equal4}
\end{equation}
The asymptotic prediction (\ref{eq:predictA0min12A2equal4}) agrees with the exact results for the $3$-soliton solution in subsection~\ref{sec:2soliton}.

\subsection{Special case $A_0=-2$}\label{eq:A0m2}

In terms of our original problem (\ref{eq:kdv}) with (\ref{eq:ICA}), the case $A_0=-2$ is subtly different because this is when the initial condition has a double pole that has the same strength exhibited by the double poles that time-dependent solutions to KdV have.  We therefore expect the temporal scaling to be different for $A_0=-2$, with any singularity moving more slowly away from $x=x_0$ than those that are born for $A_0\neq -2$.

To help clarify these cases for $A_0=-2$, we observe that (\ref{eq:doublepolecorrection}) means that singularities of solutions of KdV are doubles poles with zero residue.  Thus we have two broad cases to consider.  First, if $A_0=-2$ but $A_1\neq 0$, then the initial condition has the same leading-order behaviour near $x=x_0$ as time-dependent solutions of KdV have near each singularity $x=s(t)$, but with a different first-order correction.  It turns out, in this case, that singularities are modified at $x=x_0$ but move with initial speed $\mathcal{O}(t^{-1/3})$ (instead of $\mathcal{O}(t^{-2/3})$, which is what happens if $A_0\neq -2$).  Second, if $A_0=-2$ and $A_1=0$, then the initial condition has the same leading-order {\em and} correction term near its singularity as the full time-dependent solution does.  This would happen, of course, if we take any time-dependent solution of (\ref{eq:kdv}) (that has evolved for any nonzero time) and treat it as an initial condition.  Here the singularity present at $x_0$ propagates with speed $\mathcal{O}(1)$.

Before going further, we note briefly that a cursory glance suggests the arguments in subsections~\ref{sec:outerexpansion} and \ref{sec:innernearx0} should fail for $A_0=-2$.  This is because the main scalings for the inner problem, namely (\ref{eq:xi}) and (\ref{eq:inner}), depend on the balance between $u_0$ and $tu_1$ (using (\ref{eq:local})) happening when $x-x_0=\mathcal{O}(t^{1/3})$, whereas for $A_0=-2$ this balance is different (the balance between $u_0$ and $tu_1$ for $A_0=-2$ happens when $x-x_0=\mathcal{O}(t^{1/2})$).  However, as we see from the higher-order terms (\ref{eq:local1})--(\ref{eq:local2}), for $A_0=-2$ we have
$$
u_1\sim -\frac{30A_1}{(x-x_0)^4},
\quad
u_2\sim -\frac{720A_1}{(x-x_0)^7},
\quad\mbox{as}\quad
x\rightarrow x_0,
$$
which suggests the crucial balance is again $x-x_0=\mathcal{O}(t^{1/3})$.
Therefore, the scalings used in subsections~\ref{sec:outerexpansion} and \ref{sec:innernearx0} are still correct for $A_0=-2$.

\subsubsection{Special case $A_0=-2$, $A_1\neq 0$}\label{sec:innerspecialcase}

In the special case $A_0=-2$, the leading-order problem (\ref{eq:odef0})--(\ref{eq:farfieldf0}) (see also (\ref{eq:farfieldf0more})) has the exact solution $f_0=-2/\xi^2$ and so $\xi_0=0$ and we immediately see from (\ref{eq:3rdscalingfors}) that the speed of the singularities is $\mathcal{O}(t^{-1/3})$ instead of $\mathcal{O}(t^{-2/3})$.  For $A_0=-2$, $A_1\neq 0$, the problem  (\ref{eq:odef1}), (\ref{eq:farfieldf1more}), becomes
\begin{align}
& \frac{\mathrm{d}^2f_1}{\mathrm{d}\xi^2}
-\left(\frac{12}{\xi^2}+\xi\right)f_1+A_1=0,
\label{eq:f1reduced}
\\
& f_1\sim A_1\left(\frac{1}{\xi}-\frac{10}{\xi^4}-
\frac{80}{\xi^7}\right)
\quad\mbox{as}\quad \xi\rightarrow -\mathrm{i}\infty,
\label{eq:farfieldf1special}
\end{align}
which is a second-order linear ode problem.

In addition to (\ref{eq:farfieldf1special}), we are after a solution that does not grow exponentially in the lower half plane.  To make that slightly more precise, we note that the homogenous part of (\ref{eq:f1reduced}) is approximated by Airy's equation in the far field, with linearly independent solutions
$$
\mathrm{Ai}(\xi)\sim \frac{\mathrm{e}^{-2\xi^{3/2}/3}}{2\sqrt{\pi}\xi^{1/4}},
\quad
\mathrm{Ai}(\xi\mathrm{e}^{2\pi\mathrm{i}/3})\sim \frac{\mathrm{e}^{2\xi^{3/2}/3}}{2\sqrt{\pi}
\mathrm{e}^{\pi\mathrm{i}/6 }\xi^{1/4}},
\quad\mbox{as}\quad \xi\rightarrow -\mathrm{i}\infty.
$$
As exponential growth is inconsistent with (\ref{eq:farfieldf1special}), we must eliminate that option.  As such, we may rewrite (\ref{eq:farfieldf1special}) as
\begin{equation}
f_1\sim A_1\left(\frac{1}{\xi}-\frac{10}{\xi^4}-
\frac{80}{\xi^7}+\ldots\right)+\frac{K_1}{\xi^{1/4}}\,\mathrm{e}^{2\xi^{3/2}/3}
\quad\mbox{as}\quad \xi\rightarrow -\mathrm{i}\infty,
\label{eq:farfieldf1special2}
\end{equation}
where the ellipsis denotes higher-order terms in a divergent expansion.
Further, as with our discussion for P$_{\mathrm{II}}$ in subsection~\ref{sec:WKB}, we note that, while the exponential $\xi^{-1/4}\mathrm{e}^{2\xi^{3/2}/3}$ decays as $\xi\rightarrow -\mathrm{i}\infty$, it becomes of $\mathcal{O}(1)$ on the anti-Stokes line $\theta=-\pi/3$ and becomes exponentially large for $\theta>-\pi/3$.  Therefore, in order to exclude any exponential growth in the lower half $\xi$-plane, we also supplement (\ref{eq:f1reduced})--(\ref{eq:farfieldf1special2}) with
\begin{equation}
K_1=0.
\label{eq:farfieldf1special3}
\end{equation}
With this in mind, a more complete description of the problem for $f_1$ is (\ref{eq:f1reduced}), (\ref{eq:farfieldf1special2})--(\ref{eq:farfieldf1special3}).

In practice, it is easiest to apply (\ref{eq:farfieldf1special2}) along the anti-Stokes line $\theta=-\pi/2$ instead of the negative imaginary axis, as setting $K_1=0$ then kills off both exponentials, as required.  Thus we can change variables via
$$
f_1=\mathrm{e}^{\pi\mathrm{i}/3}g_1(\eta),
\quad
\eta=\xi\mathrm{e}^{\pi\mathrm{i}/3},
$$
so that
\begin{equation}
\frac{\mathrm{d}^2g_1}{\mathrm{d}\xi^2}
+\left(-\frac{12}{\eta^2}+\eta\right)g_1=A_1,
\label{eq:g1reduced}
\end{equation}
\begin{equation}
g_1\sim A_1\left(\frac{1}{\eta}+\frac{10}{\eta^4}-
\frac{80}{\eta^7}+\ldots\right)+\frac{K_1}{\eta^{1/4}}\,\mathrm{e}^{-2\mathrm{i}
\eta^{3/2}/3+\pi\mathrm{i}/12}
\quad\mbox{as}\quad \eta\rightarrow +\infty
\quad\mbox{with}
\quad K_1=0.
\label{eq:farfieldg1special2}
\end{equation}
The ode (\ref{eq:g1reduced}) is an inhomogeneous version of Bessel's equation, with exact solution
$$
g_1=C_1\eta^{1/2}J_{7/3}(\sfrac{2}{3}\eta^{3/2})
+C_2\eta^{1/2}Y_{7/3}(\sfrac{2}{3}\eta^{3/2})+A_1g_{1p},
$$
where $g_{1p}$ is a particular solution that involves Anger and Weber functions \cite{DLMF}.
The near- and far-field behaviours are
$$
g_1\sim (3C_1+C_2)\frac{3^{2/3}\Gamma(\sfrac{2}{3})}{56\pi}
\left(\eta^4-\frac{\eta^7}{30}+\ldots\right)
-C_2\frac{8}{3^{1/6}\Gamma(\sfrac{2}{3})}
\left(\frac{1}{\eta^3}+\frac{1}{12}+\frac{\eta^3}{72}-\frac{\eta^6}{1296}+\ldots
\right)
$$
$$
+A_1\left(-\frac{\eta^2}{10}+\frac{\eta^5}{80}-\frac{\eta^8}{3520}+\ldots\right),
$$
as $\eta\rightarrow 0$ and
$$
g_1\sim \left(-C_1\sqrt{\frac{3}{\pi}}-A_1\frac{\sqrt{\pi}}{3}\right)
\frac{1}{\eta^{1/4}}\sin\left(\sfrac{2}{3}\eta^{3/2}+\sfrac{1}{12}\pi
\right)
+\left(C_2\sqrt{\frac{3}{\pi}}+A_1\sqrt{\frac{\pi}{3}}\right)
\frac{1}{\eta^{1/4}}\cos\left(\sfrac{2}{3}\eta^{3/2}+\sfrac{1}{12}\pi
\right)+\frac{A_1}{\eta}
$$
as $\eta\rightarrow +\infty$.  That latter limit implies that, in order to kill both exponentials, we need to choose
$$
C_1=-\frac{\pi}{3\sqrt{3}}A_1,
\quad
C_2=-\frac{\pi}{3}A_1,
$$
which is enough information for what follows.

Returning to variables $f_1$ and $\xi$, this solution for $f_1$ has the following behaviour near $\xi=\xi_0=0$:
$$
f_1\sim \frac{\mu_1}{\xi^3}
\quad\mbox{as}\quad \xi\rightarrow 0,
\quad\mbox{where}\quad
\mu_1 =
-\mathrm{e}^{\pi\mathrm{i}/3}\frac{8\pi A_1}{3^{7/6}\Gamma(\sfrac{2}{3})}.
$$
Thus, by using the result $\xi_1=-\mu_1/4$ (from (\ref{eq:xi1xi2})),
we conclude the initial double pole at $x=x_0$ moves as
\begin{equation}
s(t)\sim x_0 + (3t)^{2/3} \frac{2\pi A_1\mathrm{e}^{\pi\mathrm{i}/3}}{3^{7/6}\Gamma(\sfrac{2}{3})}
\quad\mbox{as}\quad t\rightarrow 0^+.
\label{eq:finalresultA1neq0}
\end{equation}
The initial speed $\mathcal{O}(t^{-1/3})$ implies the singularity moves more slowly than the case $A_0\neq -2$, for which the initial speed is $\mathcal{O}(t^{-2/3})$.

\subsubsection{Special case $A_0=-2$, $A_1=0$, $A_2\neq 0$}\label{sec:evenmorespecial}

An even more special case, which happens to be relevant for the 1-soliton solution (or, indeed, if we take any existing solution of KdV (\ref{eq:kdv}) and interpret it as an initial condition), is when $A_0=-2$ but $A_1=0$.  Here, we have $f_0=-2/\xi^2$ ($\xi_0=0$) and $f_1=0$ ($\xi_1=0$), and we need to go to the next order to solve for $f_2$.  With these solutions for $f_0$ and $f_1$, (\ref{eq:odef2}) and (\ref{eq:farfieldf2more}) become
$$
\frac{24}{\xi^3}f_2-\left(\frac{12}{\xi^2}+\xi\right)
\frac{\mathrm{d}f_2}{\mathrm{d}\xi}
+\frac{\mathrm{d}^3f_2}{\mathrm{d}\xi^3}=0,
$$
$$
f_2\sim A_2-\frac{8A_2}{\xi^3}+\frac{0}{\xi^6}
\quad\mbox{as}\quad \xi\rightarrow -\mathrm{i}\infty.
$$
The solution that satisfies this far-field condition is simply $f_2=A_2(1-8/\xi^3)$.

Thus, given that $f_2\sim \mu_2/\xi^3$ as $\xi\rightarrow 0$, where $\mu_2=-8A_2$, we find from the result $\xi_2=-\mu_2/4$ (in (\ref{eq:xi1xi2})) that $\xi_2=2A_2$.  As a consequence, for this special case with $A_0=-2$, $A_1=0$, there is one singularity that emerges from $x_0$ and propagates as
\begin{equation}
s(t)\sim x_0+6A_2 t\quad\mbox{as}\quad t\rightarrow 0^+.
\label{eq:veryspecialcase}
\end{equation}
For example, the case $A_2=2/3$ is discussed in subsection~\ref{eq:1soliton}.

\subsection{Summary}\label{sec:summary}

In summary, by considering higher-order terms in our asymptotic expansions, we are able to describe the small-time motion of each of the singularities of the KdV solution as (\ref{eq:higherordersingularity}), namely $s(t)\sim x_0+(3t)^{1/3}\xi_0+(3t)^{2/3}\xi_1+(3t)\xi_2$.  Each term in (\ref{eq:higherordersingularity}) comes from determining the location of singularities of a cascade of problem as follows:
\begin{myitemize}

\item The value $x_0$ represents the location of each singularity $x=x_0$ of the initial condition $u_0(x)$.  Here we are mostly interested in problems for which $u_0$ is real on the real axis, whereby each of these singularities will come as a complex-conjugate pair, it being sufficient to consider only those in the upper half plane.

\item The value $\xi_0$ represents the location of each singularity $\xi=\xi_0$ of the function $f_0(\xi)$, which is the leading-order term for the inner expansion near $x=x_0$, where $\xi=(x-x_0)/(3t)^{1/3}$.  These singularities are determined as the poles of our P$_{\mathrm{II}}$ problem that have a residue $+1$.  Their precise location will depend on the parameter $\alpha$, which itself depends on $A_0$, the strength of the double pole of the initial condition.  Note that generically there are infinitely many of these singularities $\xi=\xi_0$, which means for each $x_0$ in the upper half plane, there are (generically) infinitely many singularities, with (\ref{eq:higherordersingularity}).

\item The values $\xi_1$ and $\xi_2$ come from solving linear problems for correction terms $f_1$ and $f_2$ (which depend on constants $A_1$ and $A_2$), determining the near $\xi_0$ behaviour $f_1\sim \mu_1/(\xi-\xi_0)^3$, $f_2\sim -(3/8)\mu_1^2(\xi-\xi_0)^4+\mu_2/(\xi-\xi_0)^3$ and then reading off $\xi_1=-\mu_1/4$ and $\xi_2=-\mu_2/4$, respectively.

\item In principle, further terms in (\ref{eq:higherordersingularity}) could be computed; these would depend on $A_3$, $A_4$, and so on.

\end{myitemize}

\section{Further details on transseries expansion for Painlev\'{e}}
\label{sec:transappendix}

We include here further details of the analysis in subsection~\ref{sec:trans} to locate singularities of our P$_{\mathrm{II}}$ solutions using transseries.

\subsection{Careful stocktake of terms in transseries}\label{sec:stocktake}

A starting point for applying our transseries expansion is to extend (\ref{eq:farfieldFantiS})--(\ref{eq:K1noequals0}) to be
$$
F\sim \sum_{n=0}^\infty \sigma_1^n\,\mathrm{e}^{2n\xi^{3/2}/3}F^{(n)}(\xi)
\quad\mbox{as}\quad |\xi|\rightarrow \infty, \quad -\pi<\mathrm{arg}(\xi)<-2\pi/3,
$$
so that, after substituting into P$_{\mathrm{II}}$ ($F''-\xi F+\alpha = 2F^3$), we find the $F^{(n)}$ satisfy
$$
\sum_{n=0}^\infty \sigma_1^n
\left(
\frac{\mathrm{d}^2F^{(n)}}{\mathrm{d}\xi^2}
+2n\xi^{1/2}\frac{\mathrm{d}F^{(n)}}{\mathrm{d}\xi}
+\left((n^2-1)\xi+\sfrac{1}{2}n\xi^{-1/2}\right)F^{(n)}
\right)\, \mathrm{e}^{2n\xi^{3/2}/3}+\alpha
$$
$$
=2\sum_{n=0}^\infty \sum_{j=0}^n \sum_{\ell=0}^{n-j}
\sigma_1^n\, \mathrm{e}^{2n\xi^{3/2}/3}
F^{(j)}F^{(\ell)}F^{(n-j-\ell)}.
$$
To appreciate the pattern, it is worth considering the first four of these, namely
\begin{align}
&
\frac{\mathrm{d}^2F^{(0)}}{\mathrm{d}\xi^2}-\xi F^{(0)} + \alpha = 2(F^{(0)})^3
\nonumber
\\
&
\frac{\mathrm{d}^2F^{(1)}}{\mathrm{d}\xi^2}
+2\xi^{1/2}
\frac{\mathrm{d}F^{(1)}}{\mathrm{d}\xi}
+\sfrac{1}{2}\xi^{-1/2} F^{(1)} = 6(F^{(0)})^2F^{(1)}
\nonumber
\\
&
\frac{\mathrm{d}^2F^{(2)}}{\mathrm{d}\xi^2}
+4\xi^{1/2}
\frac{\mathrm{d}F^{(2)}}{\mathrm{d}\xi}
+(3\xi+\xi^{-1/2}) F^{(2)} = 6(F^{(0)})^2F^{(2)} + 6F^{(0)}(F^{(1)})^2
\nonumber
\\
&
\frac{\mathrm{d}^2F^{(3)}}{\mathrm{d}\xi^2}
+6\xi^{1/2}
\frac{\mathrm{d}F^{(3)}}{\mathrm{d}\xi}
+(8\xi+\sfrac{3}{2}\xi^{-1/2}) F^{(3)} = 6(F^{(0)})^2F^{(3)} + 12 F^{(0)}F^{(1)}F^{(2)}
+ 2(F^{(1)})^3.
\nonumber
\end{align}
By expanding out the solution for each $F^{(n)}$ as a power series in the limit $|\xi|\rightarrow\infty$, we find
\begin{align}
F\sim & \, F^{(0)} + \sigma_1\,\mathrm{e}^{2\xi^{3/2}/3}F^{(1)}
+ \sigma_1^2\,\mathrm{e}^{4\xi^{3/2}/3}F^{(2)}
+ \sigma_1^3\,\mathrm{e}^{6\xi^{3/2}/3}F^{(3)}
+ \sigma_1^4\,\mathrm{e}^{8\xi^{3/2}/3}F^{(4)}+\ldots
\nonumber \\
= & \left(\frac{0}{\xi^{-1/2}}+
\frac{\alpha}{\xi}
+\frac{0}{\xi^{5/2}}
-\frac{2\alpha(\alpha^2-1)}{\xi^4}
+\frac{0}{\xi^{11/2}}
+\frac{4\alpha(\alpha^2-1)(3\alpha^2-10)}{\xi^7}
+\ldots \right)
\nonumber \\
& +\sigma_1\,\mathrm{e}^{2\xi^{3/2}/3}
\left(\frac{1}{\xi^{1/4}}
-\frac{96\alpha^2-5}{48\xi^{7/4}}
+\frac{(96\alpha^2-5)(96\alpha^2-77)}{4608\xi^{13/4}}
+\ldots \right)
\nonumber \\
& + \sigma_1^2\,\mathrm{e}^{4\xi^{3/2}/3}
\left(\frac{0}{\xi}
+\frac{2\alpha}{\xi^{5/2}}
-\frac{\alpha(96\alpha^2-77)}{12\xi^4}
+\frac{\alpha(9216\alpha^4-27456\alpha^2+17629)}{576\xi^{11/2}}
+\ldots \right)
\nonumber \\
& + \sigma_1^3\,\mathrm{e}^{6\xi^{3/2}/3}
\left(\frac{1}{4\xi^{7/4}}-\frac{96\alpha^2-23}{64\xi^{13/4}}
+\frac{9216\alpha^4-3648\alpha^2+1493}{2048\xi^{19/4}}
+\ldots \right)
\nonumber \\
&
+ \sigma_1^4\,\mathrm{e}^{8\xi^{3/2}/3}
\left(\frac{0}{\xi^{5/2}}
+\frac{\alpha}{\xi^4}-\frac{\alpha(192\alpha^2-121)}{24\xi^{11/2}}
+\frac{\alpha(9216\alpha^4-16224\alpha^2+7699)}{288\xi^7}
+\ldots \right)
+\ldots
\nonumber
\end{align}
In this way, by including terms with zero on the numerator, the pattern is clearer, suggesting we introduce the notation
$$
F^{(n)}\sim \frac{1}{\xi^{-1/2+3n/4}}
\sum_{m=0}^\infty \frac{F_m^{(n)}}{\xi^{3m/2}},
$$
provided  $F_{2m}^{(0)}=0$, $F_0^{(2n)}=0$, $F_0^{(1)}=1$.  This argument justifies (\ref{eq:transseries1}).

\subsection{Solving for $G_m(\tau)$}\label{sec:termsGm}

We need to substitute the asymptotic expansion (\ref{eq:transseries3}) into P$_{\mathrm{II}}$, $F''=2F^3+\xi F-\alpha$, to derive ode problems for the $G_m(\tau)$, $m\geq 0$.  By carefully applying the chain rule, we find
\begin{align}
\xi F\sim & \,\, \xi^{3/2}G_0+G_1+
\,\sum_{m=0}^\infty \frac{G_{m+2}}{\xi^{3/2+3m/2}},
\nonumber
\\
\frac{\mathrm{d}^2F}{\mathrm{d}\xi^2}
\sim & \,\,\xi^{3/2}\left(\tau^2
\frac{\mathrm{d}^2G_0}{\mathrm{d}\tau^2}
+ \tau \frac{\mathrm{d}G_0}{\mathrm{d}\tau}
\right)
+\tau^2
\frac{\mathrm{d}^2G_1}{\mathrm{d}\tau^2}
+\tau \frac{\mathrm{d}G_1}{\mathrm{d}\tau}
-\sfrac{3}{2}\tau^2
\frac{\mathrm{d}^2G_0}{\mathrm{d}\tau^2}
\nonumber
\\
& \,+\sum_{m=0}^\infty \frac{1}{\xi^{3/2+3m/2}}
\bigg(\tau^2
\frac{\mathrm{d}^2G_{m+2}}{\mathrm{d}\tau^2}
+\tau \frac{\mathrm{d}G_{m+2}}{\mathrm{d}\tau}
-\sfrac{3}{2}\tau^2
\frac{\mathrm{d}^2G_{m+1}}{\mathrm{d}\tau^2}
-3(1+m)\tau
\frac{\mathrm{d}G_{m+1}}{\mathrm{d}\tau}
\nonumber
\\
& \hspace{10ex} +\sfrac{9}{16}\tau^2
\frac{\mathrm{d}^2G_{m}}{\mathrm{d}\tau^2}
+\sfrac{9}{16}(1+4m)\tau
\frac{\mathrm{d}G_{m}}{\mathrm{d}\tau}
+\frac{9m^2-1}{4}G_m  \bigg)
\nonumber
\\
F^3\sim & \,\, \xi^{3/2}G_0^{3}+3G_0^2G_1
+ \sum_{m=0}^\infty \frac{1}{\xi^{3/2+3m/2}}
\sum_{i=0}^{m+2}\sum_{j=0}^{m+2-i}G_iG_jG_{m+2-i-j},
\end{align}
therefore the governing odes for $G_0$ and $G_1$ are
\begin{equation}
\tau^2
\frac{\mathrm{d}^2G_0}{\mathrm{d}\tau^2}
+ \tau
\frac{\mathrm{d}G_0}{\mathrm{d}\tau}
= 2G_0^{3}+G_0,
\label{eq:G0}
\end{equation}
\begin{equation}
\tau^2
\frac{\mathrm{d}^2G_1}{\mathrm{d}\tau^2}
+\tau
\frac{\mathrm{d}G_1}{\mathrm{d}\tau}
-\sfrac{3}{2}\tau^2
\frac{\mathrm{d}^2G_0}{\mathrm{d}\tau^2}
= 6G_0^2G_1+G_1-\alpha,
\label{eq:G1}
\end{equation}
while analogous (but more complicated) odes for $G_m$, $m\geq 2$, can be also be derived using the information above.  For boundary conditions, we apply (\ref{eq:Gmmatching}) for $m=0$ and $1$, to give
\begin{align}
G_0 \sim & \,\,
0+\tau+0\,\tau^2+\sfrac{1}{4}\tau^3+
0\,\tau^4+\ldots
\label{eq:G0bc}
\\
G_1\sim & \,\,
\alpha -\frac{96\alpha^2-5}{48}\tau
+2\alpha\tau^2 -\frac{96\alpha^2-23}{64}\tau^3
+\tau^4+\ldots.
\label{eq:G1bc}
\end{align}
Again, boundary conditions for other $G_m$ can be taken from (\ref{eq:Gmmatching}).

Leaving out the details, we can solve (\ref{eq:G0})--(\ref{eq:G1bc}) using maple (or equivalent) to give
\begin{equation}
G_0=\frac{4\tau}{4-\tau^2},
\quad
G_1=\frac{3\tau^5+24\alpha\tau^4+(118-192\alpha^2)\tau^3
+576\alpha\tau^2+(40-768\alpha^2)\tau+384\alpha}
{24(\tau^2-4)^2}.
\label{eq:G0G1}
\end{equation}
Thus, we have a better approximation than (\ref{eq:firstapprox}) for $F$ in the neighbourhood of the negative $\xi$-axis in the far field, namely $F\sim \xi^{1/2}G_0(\tau)+G_1(\tau)/\xi$ with (\ref{eq:G0G1}) and $\tau$ given by (\ref{eq:transvariable}).
For the next subsection, we shall use the results
\begin{equation}
G_0\sim \frac{-2}{\tau+2},
\quad
G_1\sim \frac{8\alpha(\alpha+1)-35/12}{(\tau+2)^2}
\quad\mbox{as}\quad \tau\rightarrow -2,
\label{eq:G0singular_v1}
\end{equation}
\begin{equation}
G_0\sim \frac{-2}{\tau-2},
\quad
G_1\sim \frac{-8\alpha(\alpha-1)+35/12}{(\tau-2)^2}
\quad\mbox{as}\quad \tau\rightarrow 2.
\label{eq:G0singular_v2}
\end{equation}

\subsection{Singularities near anti-Stokes lines and higher-order corrections}
\label{sec:C3}

We shall use the notation $\tau=\tau_0$ for each singularity of $F$ and write
\begin{equation}
\tau_0\sim \tau_0^{(0)} + \frac{\tau_0^{(1)}}{\xi^{3/2}} + \frac{\tau_0^{(2)}}{\xi^{3}}
+\ldots\quad\mbox{as}\quad |\xi|\rightarrow\infty.
\label{eq:taunotation}
\end{equation}
From (\ref{eq:G0G1}), there are two options, namely
$$
\tau_0^{(0)}=\pm 2,
$$
associated with poles of $F$ with residues $-1$ and $+1$, respectively.  To go further, suppose we have $F/\xi^{1/2}\sim G_0+G_1\xi^{-3/2}+\ldots$, where
$$
G_0\sim \frac{k_0}{\tau - \tau_0^{(0)}},
\quad
G_1\sim \frac{k_1}{(\tau - \tau_0^{(0)})^2}
\quad\mbox{as}\quad \tau \rightarrow \tau_0^{(0)},
$$
then we have that
$$
\frac{F}{\xi^{1/2}}\sim \frac{k_0}{\tau - \tau_0}
=\frac{k_0}{\tau - (\tau_0^{(0)}+\tau_0^{(1)}\xi^{-3/2}+\ldots)}
\quad\mbox{as}\quad \tau \rightarrow \tau_0,
$$
provided $\tau_0^{(1)}=k_1/k_0$.  Thus, comparing with (\ref{eq:G0singular_v1})--(\ref{eq:G0singular_v2}), we find the two-term approximations for the singularities in terms of $\tau$ come from (\ref{eq:taunotation}) with either
$$
\tau_0^{(0)}=-2,
\quad
\tau_0^{(1)}=-4\alpha(\alpha+1)+35/24
\quad\mbox{for poles with residue +1, or}
$$
$$
\tau_0^{(0)}=2,
\quad
\tau_0^{(1)}=4\alpha(\alpha-1)-35/24
\quad\mbox{for poles with residue -1}.
$$

With these correction terms we can provide a more accurate approximation for the location of the singularities than (\ref{eq:xi0largen}).  The updated version of (\ref{eq:transcend}) becomes
$$
\sigma_1\xi_0^{-3/4}\,\mathrm{e}^{2\xi_0^{3/2}/3}=\tau_0^{(0)}
+ \frac{\tau_0^{(1)}}{\xi^{3/2}} + \ldots,
$$
which can be solved asymptotically to give
\begin{align}
\xi_0\sim & -(3\pi n)^{2/3}+\frac{1}{(3\pi n)^{1/3}}
\left(
\frac{3\pi}{4}+\frac{\mathrm{i}}{2}\ln (3\pi n)
+\mathrm{i}\log\left(\frac{\tau_0^{(0)}}{\sigma_1}\right)
\right)
\nonumber
\\
& +\frac{1}{(3\pi n)^{4/3}}
\left(
-\frac{1}{4}
\left(
\frac{3\pi}{4}+\frac{\mathrm{i}}{2}\ln (3\pi n)
+\mathrm{i}\log\left(\frac{\tau_0^{(0)}}{\sigma_1}\right)
+\frac{3\mathrm{i}}{2}\right)^2
-\frac{9}{16}+\frac{\tau_0^{(1)}}{\tau_0^{(0)}}
\right)
\quad\mbox{as}\quad n\rightarrow\infty,
\end{align}
where the values of $\tau_0^{(0)}$ and $\tau_0^{(1)}$ depend on whether the poles have residue $+1$ or $-1$, as summarised above.

\end{appendix}

\end{document}